\documentclass[12pt]{article}

\usepackage{graphicx,psfrag,epsf,color}
\usepackage{amsmath,amssymb,amsfonts}
\usepackage{array}
\usepackage{cite}
\usepackage{multirow}
\usepackage{rotating}

\renewcommand{\arraystretch}{1.25}

\def\slash#1{#1 \hskip-0.45em /}

\newcommand{\bff}[1]{\mbox{\boldmath ${#1}$}}

\begin{document}

\allowdisplaybreaks

\begin{titlepage}

\begin{flushright}
{\small
TUM-HEP-954/14\\
SFB/CPP-14-69 \\
TTK-14-21 \\
IFIC/14-58 \\[0.2cm]

November 19, 2014
}
\end{flushright}

\vskip1cm
\begin{center}
\Large\bf\boldmath
Non-relativistic pair annihilation of nearly mass degenerate neutralinos 
and charginos III. Computation of the Sommerfeld enhancements 
\end{center}

\vspace{0.8cm}
\begin{center}
{\sc M.~Beneke$^{a,b}$, C.~Hellmann$^{a,b}$} and 
{\sc P. Ruiz-Femen\'\i a$^{c}$}\\[5mm]
{\it ${}^a$Physik Department T31,\\
James-Franck-Stra\ss e, 
Technische Universit\"at M\"unchen,\\
D--85748 Garching, Germany\\
\vspace{0.3cm}
${}^b$Institut f\"ur Theoretische Teilchenphysik und 
Kosmologie,\\
RWTH Aachen University, D--52056 Aachen, Germany\\
\vspace{0.3cm}
${}^c$Instituto de F\'\i sica Corpuscular (IFIC), 
CSIC-Universitat de Val\`encia \\
Apdo. Correos 22085, E-46071 Valencia, Spain}\\[0.3cm]
\end{center}

\vspace{1cm}
\begin{abstract}
\vskip0.2cm\noindent
This paper concludes the presentation of the non-relativistic effective 
field theory formalism designed to calculate the radiative corrections 
that enhance the pair-annihilation cross sections of slowly moving
neutralinos and charginos within the general minimal supersymmetric 
standard model (MSSM). While papers I and II 
focused on the computation of the tree-level annihilation rates 
that feed into the short-distance part, here we describe in detail 
the method to obtain the Sommerfeld factors that contain the 
enhanced long-distance corrections. This includes the computation
of the potential interactions in the MSSM, which are provided in 
compact analytic form, and a novel solution
of the multi-state Schr\"odinger equation that is free from
the numerical instabilities generated by large mass splittings
between the scattering states. Our results allow for a precise
computation of the MSSM neutralino dark matter relic abundance 
and pair-annihilation rates in the present Universe, when Sommerfeld 
enhancements are important.
\end{abstract}
\end{titlepage}




\section{Introduction}
\label{sec:introduction}

For heavy neutralino dark matter (DM) there is class of radiative 
corrections to the tree-level pair annihilation cross section, which can 
become larger than naively expected by the electroweak nature of the 
interaction, and even exceed the lowest-order cross section by orders 
of magnitude. This so-called Sommerfeld effect arises when an attractive  
interaction between the non-relativistic DM particles significantly 
distorts their wave function, such that they have a larger
probability to undergo annihilation. In terms of Feynman
diagrams the effect arises from the exchange of the electroweak gauge bosons
between the DM particles, which contributes a factor $g^2 M_{\rm DM}/M_W$, 
such that an additional particle exchange
is not suppressed by the electroweak coupling 
$g^2$ when the DM mass is much larger than the 
mediator mass. The Sommerfeld effect is non-perturbative in the sense 
that a resummation of diagrams to all orders in $g^2$ is needed in order to
calculate the annihilation cross section. The relevance of the Sommerfeld 
effect was first pointed out for the annihilation cross section of 
(wino- or higgsino-like) neutralino DM into two photons~\cite{Hisano:2004ds},
although it was not until 2008, when an anomalous positron excess was
measured by PAMELA, that Sommerfeld enhanced DM models attracted more 
attention as a mechanism to boost the DM annihilation 
rates~\cite{ArkaniHamed:2008qn}. For these models, the larger
velocities around DM freeze-out in the early Universe $(v \sim 0.3c)$ 
can yield an annihilation rate in accordance with the observed relic density, 
whereas the much smaller neutralino velocities today $(v \sim 10^{-3}c)$ 
enhance the annihilation rate into electrons and positrons observed as 
cosmic rays. Since then the Sommerfeld enhancements in cosmic ray 
signatures and in the thermal relic abundance have been discussed 
extensively for relevant MSSM scenarios 
with neutralino DM~\cite{Hisano:2006nn,Hryczuk:2010zi,Drees:2009gt,Hryczuk:2011tq,Hryczuk:2011vi,Fan:2013faa,Cohen:2013ama,Hryczuk:2014hpa}, 
but also for generic multi-state 
dark matter models~\cite{Cirelli:2007xd,Cirelli:2010nh,Finkbeiner:2010sm}. 

Here, and in our previous papers~\cite{Beneke:2012tg} (from now on referred 
to as paper I) and~\cite{Hellmann:2013jxa} (paper II), we develop a 
formalism that improves the calculation of the DM annihilation cross section
in the general minimal supersymmetric standard model (MSSM) with 
neutralino DM by including the Sommerfeld corrections.
By allowing the lightest supersymmetric particle (LSP) to be 
an arbitrary admixture of the electroweak gauginos and higgsinos,
the framework can be used to study regions of the MSSM parameter space
where the Sommerfeld correction may no longer be an order one effect, but 
still constitute the dominant radiative correction.
Our method builds upon the non-relativistic nature of the pair of 
annihilating particles and separates the short-distance annihilation 
process (taking place at distances ${\cal O}(1/m_{\rm LSP})$)
from the long-distance interactions characterized by the Bohr radius 
of order $1/(m_{\rm LSP} g^2)$ or $1/m_W$, 
responsible for the Sommerfeld effect, in analogy to the 
NRQCD treatment of quarkonium annihilation. However, in the MSSM
co-annihilation effects of the LSP with heavier neutralino and chargino 
species have to be accounted for in regions where mass degeneracies are 
generic. Dealing with many nearly mass-degenerate scattering states 
(channels) requires an extension of the conventional 
NRQCD setup, which is provided in this work to accommodate any number of 
neutralino and chargino species. Previous works on the Sommerfeld
enhancement focused on the wino and higgsino limits of the MSSM 
with at most three neutralino-chargino two-particle states in the case 
of the neutral sector of higgsino-like scenarios. Further aspects where the
formalism we present extends previous approaches are:
\begin{itemize}
\item The total annihilation cross section for a given co-annihilating state $\chi_i\chi_j$
is obtained from the imaginary part of the forward scattering amplitude 
$\chi_i\chi_j\to \dots\to \chi_{e_1} \chi_{e_2} \to X_A X_B \to \chi_{e_4} \chi_{e_3} \to \dots \chi_i\chi_j$
where  $X_A X_B$ denotes a pair of SM or light Higgs particles. The transitions of $\chi_i\chi_j$ to 
other two-particle states through
electroweak gauge boson exchange prior to the short-distance annihilation 
$\chi_{e_1} \chi_{e_2} \to X_A X_B \to \chi_{e_4} \chi_{e_3}$ are 
accounted for by off-diagonal terms in the potential interactions, which 
are provided in analytic form in this
paper. Likewise, off-diagonal entries in the space of 
two-particle states are required to describe the short-distance part. The 
(off)-diagonal annihilation matrix entries
in the general MSSM were obtained in analytic form in papers I and II. 
The off-diagonal annihilation terms cannot be 
be obtained directly from the numerical codes that compute the MSSM 
tree-level annihilation rates, and were consequently not considered 
previously except in the strict wino and higgsino limits. 
\item Our multi-channel approach implies that an incoming  state 
$\chi_i\chi_j$ can scatter to heavier two-particle states that are 
kinematically closed for the available center-of-mass (cms) energy.
Eventually, when some mass-splittings become larger than $M_W^2/M_{\rm DM}$, 
numerical instabilities in the solution of the 
matrix-Schr\"odinger equation for the scattering wave functions prevents 
one from obtaining accurate results for the Sommerfeld factors. We 
circumvent this problem by reformulating the Schr\"odinger problem 
directly for the entries of the relevant matrix that 
provides the Sommerfeld factors, instead of solving for the wave functions. 
Leaving aside limitations related to the CPU time needed to solve a 
system of many coupled differential equations, this method allows to 
compute the Sommerfeld factors reliably also when many non-degenerate 
two-particle channels are present.
\item The short-distance annihilation rates are obtained in the 
non-relativistic limit including corrections of 
${\cal O}(v_\text{rel}^2)$, where $v_\text{rel}=\vert\vec{v}_1-\vec{v}_2\vert$ 
denotes the relative velocity of the annihilating 
particles in their center-of-mass frame, {\it i.e.}
\begin{align}
\label{eq:rates}
\Gamma_{\chi_{e_1} \chi_{e_2} \to X_A X_B \to \chi_{e_4} \chi_{e_3}}
 \, \sim \,a + b\,v_\text{rel}^2 \,.
\end{align}
While the leading order describes only $S$-wave annihilations, the 
subleading term $b$ contains both $S$- and $P$-wave contributions, that 
get multiplied by different Sommerfeld factors when building the full 
annihilation cross section from the rates above.
A separation of the two partial-wave components of $b$ is thus needed, which
is automatically achieved in our EFT framework, since the 
Sommerfeld factors in question arise from different non-relativistic 
operator matrix elements. 
\end{itemize}

In the EFT approach the short-distance contributions to the neutralino 
and chargino pair-annihilation processes 
are encoded into the  Wilson coefficients of local four-fermion 
operators, whereas the absorptive part of the matrix element of these 
four-fermion operators gives the full annihilation rates, including the 
Sommerfeld corrections. In papers I and II we have written down the 
dimension-6 and dimension-8 four-fermion
operators that mediate the short-distance annihilation rates 
at leading and next-to-next-to leading order in $v_\text{rel}$,
respectively, and presented the complete results for the
corresponding Wilson coefficients in the general MSSM. In the present paper 
we turn to the long-distance part of the annihilation process and provide 
all the technical details required for the computation of the matrix elements
of the four-fermion operators. Apart from the annihilation rates, the other
model-specific ingredient, the non-relativistic 
potentials generated by electroweak gauge boson and Higgs boson 
exchange in the MSSM, are also provided in compact analytic form in 
this work. The contents of papers I-III therefore
allow us to compute the full Sommerfeld-enhanced neutralino and chargino 
co-annihilation rates for an arbitrary MSSM parameter-space point and 
the corresponding relic abundance. 
A detailed investigation and discussion of Sommerfeld
enhancements in the relic abundance calculation in some popular MSSM scenarios
is the subject of an accompanying paper~\cite{paperIV}.
The general phenomenological study of Sommerfeld enhancements in 
the MSSM parameter space is postponed for a future publication.

A number of limitations needs to be mentioned. The formalism in its 
present form cannot be applied to DM annihilation through a narrow-width  
$s$-channel resonance, since in this case the annihilation process 
is no longer short-distance. Furthermore, we computed the pair 
annihilation and potentials of neutralinos and charginos, but not 
of neutralinos and sfermions, which excludes MSSM scenarios where 
neutralino-sfermion co-annihilation is important. Both, resonant 
annihilation and neutralino-sfermion co-annihilation scenarios require 
the accidental coincidence of some MSSM parameters and are in this 
sense less generic than neutralino-chargino co-annihilations, 
especially when the LSP is heavy ($m_{\rm LSP} \gg m_W$), and 
therefore a member of an approximate electroweak multiplet. On 
the technical side, we presently neglect the scale dependence of 
the electroweak couplings as well as electroweak double logarithms 
of the Sudakov type. The latter, of course, are only important when 
the formalism is applied to exclusive final states, as is relevant 
to indirect detection signals of DM. EFT methods have been proposed 
recently for summing up the Sudakov logarithms in 
multi-TeV DM annihilation~\cite{Baumgart:2014vma,Bauer:2014ula,Ovanesyan:2014fwa}.
More important to relic-density 
computations is the fact that for heavy dark matter the freeze-out 
occurs at temperatures where the temperature-dependence of the 
electroweak gauge-bosons masses can be relevant, as can be the 
temperature-dependence of the gaugino masses. While in general the 
thermal corrections to DM freeze-out are tiny \cite{Beneke:2014gla}, 
the case of Sommerfeld enhancements is special, since they depend 
sensitively on the range of the potential and the small mass splittings 
between the co-annihilating particles. In the wino and Higgsino limits 
some of the above effects have already been studied in the context 
of Sommerfeld enhancement \cite{Cirelli:2007xd, Hryczuk:2010zi}. For 
the general case, this is left for future work.

The contents of the paper are the following. The theoretical framework 
is summarized at some length in Sec.~\ref{sec:eft}. We discuss the 
structure of the EFT Lagrangian and derive the master formula for the 
Sommerfeld-corrected cross section in Subsec.~\ref{sec:eft-approach}. 
The standard formalism to obtain the thermal relic abundance
is reviewed in~\ref{subsec:relic-abundance}. The computation of the 
potentials for neutralino and chargino scattering in the MSSM is one 
of the main results of this paper, and we have devoted 
Sec.~\ref{sec:potMSSM} to outline the details.
The computation of the Sommerfeld factors is the subject 
of Sec.~\ref{sec:Sommerfeld}. In the first part of this section we
relate the four-fermion matrix elements with the scattering wave functions that
are determined by solving a multi-channel Schr\"odinger equation with MSSM
potentials. Subsec.~\ref{sec:schrsolution} describes the standard method to solve
the Schr\"odinger equation and points out to the numerical instabilities 
that are caused by kinematically closed channels. 
The improved method that solves this problem 
is then explained in Subsec.~\ref{sec:schrsolutionimp}.
As in NRQCD, the matrix element of second-derivative operators 
can be related to the leading order ones, though here it
becomes more involved due to the presence of
Yukawa-like interactions generated by massive gauge bosons; the exact
relation is derived in Subsec.~\ref{sec:derivops}.
In the final part of  Sec.~\ref{sec:Sommerfeld}
we give an approximation to the treatment of heavier channels which
can be used to reduce the number of channels to be treated exactly in the 
multi-channel Schr\"odinger equation, and thus the time required for its
numerical solution. A summary of our main results is given 
in Sec.~\ref{sec:summary}. Finally, we collect the analytic expressions 
for the potentials in the MSSM in Appendix~\ref{sec:appendixpot}, and we 
show the equivalence between the two different basis of
two-particle states that can be used to evaluate the Sommerfeld factors in
Appendix~\ref{sec:appendixmethodIvsII}. As mentioned above, the present 
papers concentrates on technical aspects of the framework, which is 
rather general and applicable to any model of heavy dark matter with 
electroweak interactions. A non-technical discussion of results is 
presented in an accompanying paper \cite{paperIV}.


\section{Theoretical framework}
\label{sec:eft}

Together with previous results given in \cite{Beneke:2012tg, Hellmann:2013jxa},
the formalism that we present in this work allows to describe
Sommerfeld-enhanced neutralino and chargino pair-annihilation rates within
generic R-parity conserving MSSM scenarios, including the most general
form of flavor mixing in the squark and slepton sector.
After electroweak symmetry breaking the four neutralino states
$\chi^0_i$, $i=1,\ldots,4$ result as combinations of the 
four $SU(2)_L\times U(1)_Y$ eigenstates bino, the electrically neutral 
wino and the two electrically neutral higgsinos. The  
chargino states $\chi^\pm_j$, $j=1,2$ are a 
mixture of the charged wino and higgsino electroweak eigenstates.
While the bino and the winos are related to the soft SUSY breaking
mass-parameters $M_1$ and $M_2$, respectively, the higgsino states are
associated with the mass-parameter $\mu$ in the underlying SUSY Lagrangian.

A MSSM scenario can be obtained with publicly available MSSM spectrum
generators, for example
\cite{Djouadi:2002ze,Allanach:2009bv,Porod:2011nf},
where the parameters $M_1$, $M_2$ and $\mu$ among other required SUSY
parameter inputs have to be specified. 
In constrained MSSM scenarios, as for instance models with grand
unification of gauge couplings, certain relations among the input SUSY
parameters are assumed. Our setup is not
restricted to such cases, but allows to analyze Sommerfeld enhancements in
$\chi^0$, $\chi^\pm$ co-annihilations in the most general MSSM models.
Generically we require for our calculations a (SLHA formatted) MSSM spectrum
including mass parameters, mixing matrices and angles, typically provided as
output of a MSSM spectrum calculator.
From this spectrum we determine the corresponding Sommerfeld-enhanced
$\chi^0$, $\chi^\pm$ co-annihilation rates as well as the $\chi^0_1$ relic
abundance. Our formalism requires positive mass parameters of the neutralino
and chargino states, which we automatically account for by an appropriate
rotation of the neutralino and chargino mixing matrices, as explained in
Appendix A of paper I~\cite{Beneke:2012tg}.

A rigorous analysis of Sommerfeld-enhanced $\chi^0$, $\chi^\pm$ 
co-annihilation processes in a given model should refer to the on-shell 
mass spectrum of the neutralino and chargino states, instead of to the 
$\overline{\text{DR}}$-parameters that are provided by most spectrum 
calculators. Furthermore, the mass splittings between the co-annihilating 
states play an essential role in the precise calculation of Sommerfeld 
enhancements, requiring one-loop on-shell renormalized masses in some 
cases. Results on the one-loop on-shell renormalized $\chi^0$, $\chi^\pm$ 
sector of the MSSM are available
\cite{Heinemeyer:2011gk, Bharucha:2012re, Bharucha:2012nx, Fritzsche:2013fta},
but have not yet been implemented in public spectrum generators.

We perform our calculation in the framework of an effective field theory (EFT)
of non-relativistic neutralinos and charginos, which generically covers 
the case of models with $n_0\leq4$ neutralino-states $\chi^0_i$, 
$i=1,\ldots,n_0$ and $n_+\leq2$ chargino-states $\chi^\pm_j$, 
$j=1,\ldots,n_+$, that are nearly mass-degenerate with the $\chi^0_1$.
The description of our EFT framework is the purpose of Sec.~\ref{sec:eft-approach}, starting with
the discussion of the Lagrangian in the effective theory in
Sec.~\ref{subsec:lagrangian}.
In Sec.~\ref{subsec:basis} we recall the notation for the four-fermion
operators that reproduce the short-distance annihilation reactions
and that were already introduced in \cite{Beneke:2012tg, Hellmann:2013jxa}.
The relevant formulae for the 
Sommerfeld-enhanced annihilation matrix elements, which determine
the co-annihilation cross sections entering the $\chi^0_1$ relic abundance
calculation, are given in Sec.~\ref{subsec:SExsec}.
Finally the $\chi^0_1$ relic abundance calculation is summarized in
Sec.~\ref{subsec:relic-abundance}.
Our EFT framework does not cover scenarios where co-annihilations with
nearly mass-degenerate sfermion such as the $\tilde t_1$ or $\tilde \tau_1$ are
important in the relic abundance calculation. 
These cases require a straightforward extension of our EFT setup but are beyond
the scope of this work. For the time being we exclude MSSM scenarios
from our analysis, where other than $\chi^0/\chi^\pm$ co-annihilations are
important in the $\chi^0_1$ relic abundance calculation.

\subsection{Effective theory approach}
\label{sec:eft-approach}

\subsubsection{Lagrangian}
\label{subsec:lagrangian}

In \cite{Beneke:2012tg} we have introduced an effective field theory (EFT),
the non-relativistic MSSM (NRMSSM), 
designed to describe the dynamics of charginos and neutralinos
which are off-shell by an amount of the order of $(m_{\rm LSP} v)^2$, where 
$m_{\rm LSP}$ is the mass of the lightest neutralino and 
$v$ its small velocity. The framework allows us to compute
the inclusive annihilation rates of pairs of charginos and neutralinos 
moving at small velocities in a systematic 
expansion in the coupling constant and the velocity. The NRMSSM shares
many similarities with the non-relativistic EFT of 
QCD~\cite{Bodwin:1994jh}, that has
been employed for computing heavy quarkonium properties and
radiative corrections to heavy quark production with remarkable accuracy,
and can be considered as an extension of the latter in two aspects. 
First, the NRMSSM can account for several non-relativistic particle species, 
namely those neutralinos and charginos whose masses
are nearly degenerate with $m_{\rm LSP}$, and second, it includes potential
interactions generated by massive gauge bosons ({\it i.e.} 
Yukawa potentials) and not just Coulomb potentials.

The structure of the EFT Lagrangian has already been discussed 
in~\cite{Beneke:2012tg}. It consists of three parts:
\begin{align}
 \mathcal L^\text{NRMSSM}
	=
   \mathcal L_\text{kin} + \mathcal L_\text{pot}
 + \delta\mathcal L_\text{ann} + \dots
 \ ,
\end{align}
where the dots stand for terms of higher order in the non-relativistic 
expansion, that are not required for the present calculation of the Sommerfeld-enhanced annihilation rates.
$\mathcal L_\text{kin}$ contains the bilinear terms in 
the two-component spinor fields $\xi_i$ and  $\psi_j=\eta_j, \zeta_j$ 
that represent the non-relativistic neutralinos ($\chi^0_i$) and charginos ($\chi^-_j$ and
$\chi^+_j$), respectively.
For  $n_0\le 4$ non-relativistic
neutralino species and  $n_+\le 2$ non-relativistic chargino species,
$\mathcal L_\text{kin}$ is 
given by
\begin{eqnarray}
\mathcal L_\text{kin}&= &
 \sum\limits_{i=1}^{n_0}
  \xi^\dagger_i \left( i\partial_t - (m_i - m_{\text{LSP}}) + 
\frac{\vec\partial^{\,2}}{2 m_{\mathrm{LSP}} } \right) \xi_i
\nonumber\\
&& + \,
 \sum_{\psi=\eta,\zeta} \sum\limits_{j=1}^{n_+}
   \psi^\dagger_j \left( i\partial_t - (m_j - m_{\text{LSP}}) + 
\frac{\vec\partial^{\,2}}{2 m_{\mathrm{LSP}}} \right) \psi_j\,.
\label{eq:kin}
\end{eqnarray}
In order to have a consistent power-counting in the amplitudes 
describing transitions between two-particle states formed from the neutralino and 
chargino species included in the EFT we need that the mass differences
$(m_i - m_{\text{LSP}})$ are formally considered 
of order $m_{\rm LSP}v^2$~\cite{Beneke:2012tg},\footnote{As discussed 
in~\cite{Beneke:2012tg}, a straightforward extension of the EFT  
covers the case where the non-relativistic
particle species are nearly mass-degenerate with respect to two well-separated
mass scales $m$ and $\overline{m}$;
within this modified set-up  pair annihilations of a set of
hydrogen-like two-particle states can be also considered.} 
of the same order as the time-derivative and kinetic-energy term in the 
Lagrangian. This implies that heavier neutralinos and charginos (as well 
as further heavy SUSY particles and
higher mass Higgs) are not among the degrees of freedom of the effective 
theory, and their virtual effects can only appear as short-distance 
corrections to the operators in $\mathcal L^\text{NRMSSM}$. In the same way, 
the hard modes associated to the SM and light Higgs-particle
produced in neutralino and chargino pair-annihilations are encoded in
the Wilson coefficients of four-fermion operators in $\delta\mathcal L_\text{ann}$,
which are local in space-time because the annihilation takes place
at short-distances of ${\cal O}(1/m_{\rm LSP})$, as compared to the 
characteristic range ${\cal O}(1/m_{\rm LSP}v)$ or ${\cal O}(1/m_W)$ 
(whichever is smaller) of the non-relativistic  interactions between 
the charginos and neutralinos. The explicit form of the
operators in $\delta\mathcal L_\text{ann}$, which are relevant for this work,
and the details on the associated Wilson coefficients are given below in 
Sec.~\ref{subsec:basis}. 

\begin{table}[t]
\centering
\begin{tabular}{|c|c|c|}
\hline
   neutral reactions
 & single-charged reactions
 & double-charged reactions
\\
\hline\hline
   $\chi^0 \chi^0 \to \chi^0 \chi^0 $
 & $\chi^0 \chi^+ \to \chi^0 \chi^+$
 & $\chi^+ \chi^+ \to \chi^+ \chi^+$
\\
   $\chi^0 \chi^0 \to \chi^- \chi^+$
 & $\chi^- \chi^0 \to \chi^- \chi^0$
 & $\chi^- \chi^- \to \chi^- \chi^-$
\\
   $\chi^- \chi^+ \to \chi^0 \chi^0$
 &
 &
\\
   $\chi^- \chi^+ \to \chi^- \chi^+$
 &
 &
\\
\hline
\end{tabular}
\caption{Possible $\chi_{e_1} \chi_{e_2} \to \chi_{e_4} \chi_{e_3}$
         scattering reactions classified according to the total charge.
         The labels $e_i$ on the fields $\chi_{e_i}$ are
         suppressed in the above table. If $\chi_{e_i}$ represents a field
         $\chi^0_{e_i}$, the label $e_i$ can range over $e_i = 1, \ldots, n_0$,
         whereas $e_i = 1, \ldots, n_+$ for the case of a $\chi^\pm_{e_i}$ field.
         Redundant reactions where the type of particle in position 1 and 2 
         and/or 3 and 4 are interchanged ({\it e.g.} the reaction $\chi^+ \chi^- \to \chi^0 \chi^0$) 
         are not explicitly written.}
\label{tab:scattering_reactions}
\end{table}

The term $\mathcal L_\text{pot}$ accounts for the exchange of SM gauge 
bosons and Higgs particles between the two-particle states 
$\chi_{e_1}\chi_{e_2}$ and $\chi_{e_4}\chi_{e_3}$ with
non-relativistic relative velocity. In the non-relativistic limit, 
such interactions become instantaneous but spatially non-local, and 
are described in the EFT by four-fermion operators whose matching 
coefficients are Yukawa- and Coulomb potentials depending on the 
relative distance $\vec{r}=\vec{x}{\,^\prime}-\vec{x}$ 
($r\equiv |\vec{r}\,|$) in the two-body system:
\begin{align}
\mathcal L_\text{pot} = 
- \sum\limits_{ \chi \chi \rightarrow \chi \chi}
  \int d^3 \vec{r} ~ V^{\chi\chi \to \chi\chi}_{ \lbrace e_1 e_2\rbrace 
  \lbrace e_4 e_3\rbrace }(r) 
  \, \chi_{e_4}^\dagger (t, \vec{x})   
  \chi_{e_3}^\dagger (t, \vec{x}+\vec{r}\,)  
   \chi_{e_1} (t, \vec{x})   \chi_{e_2} (t, \vec{x}+\vec{r}\,)  \,.
\label{eq:pot}
\end{align}
The sum in (\ref{eq:pot}) is taken over all $\chi_{e_1}\chi_{e_2}\to\chi_{e_4}\chi_{e_3}$ neutral, single-charged
and double-charged reactions with $\chi_{e_i}=\chi^0_{e_i},\,\chi^\pm_{e_i}$, which have been
summarized in Table~\ref{tab:scattering_reactions}. The particular particle species ($\chi^0$ or $\chi^\pm$)
participating in the reaction is indicated by the label $\chi\chi\to\chi\chi$ in the potentials. 
The explicit form of the terms in (\ref{eq:pot}) are generated by
replacing the generic fields $\chi_{e_i}$ by
the field symbols  $\xi_{e_i}, \eta_{e_i}$ or $\zeta_{e_i}$, corresponding to particle species $\chi^0_{e_i}$, $\chi^-_{e_i}$ and $\chi^+_{e_i}$
in all possible ways compatible with charge-conservation in the reaction. 
At leading order in the non-relativistic expansion, the potentials depend only on the total spin of the two-particle states,
which is thus conserved. An individual potential contribution from the exchange of a gauge boson or Higgs particle with mass $m_{\phi}$
has, at leading order, the generic form
\begin{align}
V^{\chi\chi \to \chi\chi}_{ \lbrace e_1 e_2\rbrace \lbrace e_4 e_3\rbrace }(r) = 
  \left( A_{e_1 e_2 e_4 e_3}\delta_{\alpha_4 \alpha_1}\delta_{\alpha_3\alpha_2}
  +  B_{e_1 e_2 e_4 e_3} \big( \vec{S}^2 \big)_{\alpha_4 \alpha_1,\alpha_3\alpha_2} \right)
  \frac{e^{-m_\phi r}}{r}\,,
\label{eq:potgeneric}
\end{align}
where we have written explicitly the spin indices $\alpha_i$ omitted 
in (\ref{eq:pot}), which are contracted with the corresponding
indices in the field operators $\chi_{e_i}$.
The total spin operator $\vec{S}$ is built from the spin operators acting on the particles interacting
at points $\vec{x}_1$ and $\vec{x}_2$ as
$\vec{S}_{\alpha_4 \alpha_1,\alpha_3\alpha_2}=
 \vec{\sigma}_{\alpha_4\alpha_1}/2\,\delta_{\alpha_3\alpha_2} 
+\delta_{\alpha_4 \alpha_1}\vec{\sigma}_{\alpha_3\alpha_1}/2 
\equiv 1/2 \, ( \vec{\sigma} \otimes  \mathbf{1} +   \mathbf{1}  \otimes \vec{\sigma} )$.
Since we shall decompose  the two-particle states $\chi_{e_1}\chi_{e_2}$ and $\chi_{e_4}\chi_{e_3}$
undergoing potential interactions into ${}^{2S+1}L_J$ partial-wave states with defined spin $S=0,1$, we can
drop the spin indices in the potentials in what follows and replace the $\vec{S}^2$ operator acting on 
$\chi_{e_1}\chi_{e_2}$,$\chi_{e_4}\chi_{e_3}$ by its eigenvalue $S(S+1)=2S$ for $S=0,1$. 

In this work we account for ${\cal O}(v^2)$ effects in the 
(co-)annihilation of neutralino and chargino pairs coming from the 
short-distance part of the annihilation but ignore ${\cal O}(v^2)$ 
contributions from the long-range part. Consequently, only the 
leading-order Coulomb- and Yukawa potential interactions
need to be considered in $\mathcal L_\text{pot}$.
Details on the calculation of the potentials at ${\cal O}(g_2^2)$ in the MSSM,
where $g_i$ are the generic $SU(2)_L \otimes U(1)_Y$ gauge couplings,
will be given in Sec.~\ref{sec:potMSSM}. 

\subsubsection{Annihilation matrices}
\label{subsec:basis}

The short-distance annihilation of the chargino and neutralino pairs
into SM and light Higgs final states is reproduced in the EFT by local 
four-fermion operators contained in $\delta \mathcal L_\text{ann}$.
The Wilson coefficients of these operators can be determined by matching 
the MSSM amplitudes for the process 
$\chi_{e1}\chi_{e_2} \to X \to \chi_{e_4}\chi_{e_3}$
with SM and light Higgs intermediate states to the tree-level matrix 
element of the four-fermion operators for the same incoming and 
outgoing states. For the computation of
the neutralino and chargino inclusive annihilation rates, only the 
absorptive part of
these Wilson coefficients are required, see~(\ref{eq:SommerSigma}), and
consequently the matching can be done for the absorptive part
of the amplitude only. At lowest order in the electroweak gauge couplings
$g_i$, the contributions to the Wilson
coefficients arise from the absorptive part of $\chi_{e1}\chi_{e_2}\to 
\chi_{e_4}\chi_{e_3}$
one-loop scattering diagrams with two SM or light Higgs particles in the 
intermediate state, $X_A X_B$, and are of ${\cal O}(\alpha_i^2)$, where
$\alpha_i=g_i^2/4\pi$. Note that the annihilation rates include 
the absorptive parts of off-diagonal amplitudes in the space of 
two particle states $\chi_e\chi_{e^\prime}$, since the exchange of 
gauge and Higgs bosons before the short-distance annihilation may transform 
one two-particle state into another.

The leading-order contributions to $\delta \mathcal L_\text{ann}$ are given by
dimension-6 four-fermion operators, that describe leading-order $S$-wave 
neutralino and chargino scattering processes 
$\chi_{e1}\chi_{e_2}\to \chi_{e_4}\chi_{e_3}$. 
They read~\cite{Beneke:2012tg}
\begin{align}
\delta \mathcal L_\text{ann}^{d = 6}  = 
\sum\limits_{ \chi \chi \rightarrow \chi \chi}
\sum\limits_{S = 0,1 }~\frac{1}{4}~
f^{ \chi \chi \to \chi \chi }_{ \lbrace e_1 e_2\rbrace \lbrace e_4 e_3\rbrace }
       \left( {}^{2S+1}S_S \right) \
\mathcal O^{\chi \chi \to \chi \chi }_{ \lbrace e_4 e_3\rbrace 
\lbrace e_2 e_1\rbrace }
\left( {}^{2S+1}S_S \right) \,,
\label{eq:deltaL4fermion}
\end{align}
where $f^{ \chi \chi \to \chi \chi }_{ \lbrace e_1 e_2\rbrace \lbrace e_4 e_3\rbrace } \left( {}^{2S+1}S_J \right)$ 
are the corresponding Wilson coefficients, which will be often abbreviated as $f\left( {}^{2S+1}S_J \right)$.
The explicit form of the dimension-6 $S$-wave operators with $S=0,1$ is
\begin{align}
\mathcal O^{\chi \chi \to \chi \chi }_{ \lbrace e_4 e_3\rbrace \lbrace e_2 e_1\rbrace }\left( {}^{1}S_0 \right)
 \, = & \
    \chi^\dagger_{e_4} \chi^c_{e_3}
    \ \chi^{c \dagger}_{e_2}\chi^{}_{e_1}
    \ ,
\\
\mathcal O^{\chi \chi \to \chi \chi }_{ \lbrace e_4 e_3\rbrace \lbrace e_2 e_1\rbrace }\left( {}^{3}S_1 \right)
 \, = & \
    \chi^\dagger_{e_4} \sigma^i \chi^c_{e_3}
    \  \chi^{c \dagger}_{e_2} \sigma^i \chi^{}_{e_1}
    \ .
\label{eq:LOSwave_ops}
\end{align}
The first sum in (\ref{eq:deltaL4fermion}) is taken over all neutralino 
and chargino scattering reactions $\chi_{e_1} \chi_{e_2} \to 
\chi_{e_4} \chi_{e_3}$ specified
in Tab.~\ref{tab:scattering_reactions}, including redundant ones where 
the particle species at the first and second and/or third and fourth position
are interchanged. This redundancy implies that several operators describe one
specific process with a $\chi_{e_1}$ and $\chi_{e_2}$ ($\chi_{e_3}$ and $\chi_{e_4}$)
particle in the initial (final) state, and as a consequence their respective Wilson coefficients 
obey certain symmetry relations under the exchange of the labels $e_1 \leftrightarrow e_2$
and/or $e_3 \leftrightarrow e_4$~\cite{Beneke:2012tg}, see~(\ref{eq:WCsymmetries}) below. 
The absorptive parts of the Wilson coefficients,
$\hat f \left( {}^{2S+1}S_J \right)$, 
have been calculated in~\cite{Beneke:2012tg}
in the MSSM at $\mathcal O(\alpha_i^2)$. A master formula and necessary ingredients
to obtain the contributions to  $\hat f \left( {}^{2S+1}S_J \right)$ from 
individual states $X_A X_B$ in analytic form can be found therein. 

At ${\cal O}(v^2)$ in the non-relativistic expansion in momenta
and mass differences, dimen\-sion-8 four-fermion operators
contribute to $\delta \mathcal L_\text{ann}$:\footnote{In a parity-violating 
theory such as the MSSM, there exist also  ${\cal O}(v)$ contributions to
$\delta \mathcal L_\text{ann}$, corresponding to dimension-7 four-fermion
operators which describe $^1S_0 - {}^3P_0$, $^3S_1 - {}^1P_1$ and 
$^3S_1 - {}^3P_1$ transitions where the spin and/or the orbital-angular 
momentum of the $\chi\chi$ pair is changed.
They are not considered here because they contribute to the 
annihilation rates only in conjunction with ${\cal O}(v)$-suppressed potential
interactions in ${\cal L}_{\rm pot}$, which we neglect. For the same 
reason, dimension-8 four-fermion operators in $\delta \mathcal L_\text{ann}$ 
causing 
${}^3P_1 \to {}^1P_1$ and ${}^3S_1 \to {}^3D_1$ transitions
are ignored.}
\begin{align}
\nonumber
 \delta \mathcal L ^{d=8}_\text{ann}
  \ =& \
   \sum\limits_{\chi\chi \to \chi\chi} 
   \frac{1}{4M^2} ~
   f^{\chi\chi \to \chi\chi}_{\lbrace e_1 e_2 \rbrace \lbrace e_4 e_3 \rbrace}\left( ^1P_1 \right)
   ~
   \mathcal O^{\chi\chi \to \chi\chi}_{\lbrace e_4 e_3 \rbrace \lbrace e_2 e_1 \rbrace}\left( ^1P_1 \right)
 \\\nonumber
 & +
  \sum\limits_{\chi\chi \to \chi\chi} \, \sum\limits_{J=0,1,2} ~
   \frac{1}{4M^2} ~
   f^{\chi\chi \to \chi\chi}_{\lbrace e_1 e_2 \rbrace \lbrace e_4 e_3 \rbrace}\left( ^3P_J \right)
   ~
   \mathcal O^{\chi\chi \to \chi\chi}_{\lbrace e_4 e_3 \rbrace \lbrace e_2 e_1 \rbrace}\left( ^3P_J \right)
 \\\nonumber
 & +
  \sum\limits_{\chi\chi \to \chi\chi} \, \sum\limits_{s=0,1} ~
   \frac{1}{4M^2} ~
   g^{\chi\chi \to \chi\chi}_{\lbrace e_1 e_2 \rbrace \lbrace e_4 e_3 \rbrace}\left( ^{2S+1}S_S \right)
   ~
   \mathcal P^{\chi\chi \to \chi\chi}_{\lbrace e_4 e_3 \rbrace \lbrace e_2 e_1 \rbrace}\left( ^{2S+1}S_S \right)
 \\
 & +
  \sum\limits_{\chi\chi \to \chi\chi} \, \sum\limits_{s=0,1}\, \sum\limits_{i=1,2} ~
   \frac{1}{4M^2} ~
   h^{\chi\chi \to \chi\chi}_{i\,\lbrace e_1 e_2 \rbrace \lbrace e_4 e_3 \rbrace}\left( ^{2S+1}S_S \right)
   ~
   \mathcal Q^{\chi\chi \to \chi\chi}_{i\,\lbrace e_4 e_3 \rbrace \lbrace e_2 e_1 \rbrace}\left( ^{2S+1}S_S \right) \ .
\label{eq:NNLO_ops}
\end{align}
The operators $\mathcal O\left( ^1P_1 \right)$ and $\mathcal 
O\left( ^3P_J \right)$, $J=0,1,2$, contain one derivative acting on each 
of the initial and final bilinear operators 
($\chi^{c \dagger}_{e_2}\chi^{}_{e_1}$ and $\chi^\dagger_{e_4} \chi^c_{e_3}$, respectively),
while in the $\mathcal P\left( ^{2S+1}S_S \right)$
operators the two derivatives act either on the initial or in the final state. 
The explicit form of these  
operators can be read off from Tab.~1 in~\cite{Hellmann:2013jxa}.
The remaining next-to-next-to leading $S$-wave operators
$\mathcal Q_{i}\left( ^{2S+1}S_S \right)$, $i=1,2$,
are the same as the dimension-6 operators $\mathcal O\left( ^{2S+1}S_S \right)$
written in (\ref{eq:LOSwave_ops}) up to a factor $(\delta m\, M)$ in the case $i=1$
and $(\delta \overline{m}\,M)$ for $i=2$, where
\begin{align}
\label{eq:M}
 M \ = \ \frac{1}{2} \, \sum_{k=1}^4 \, m_{e_k} \, ,
\end{align}
with $m_{e_k}$ the masses of the $\chi_{e_k}$ particles involved in the 
reaction $\chi_{e_1} \chi_{e_2} \to \chi_{e_4} \chi_{e_3}$, and 
\begin{align}
 \delta m \ =& \ \frac{m_{e_4} - m_{e_1}}{2} \, , \hspace{7ex}
 \delta \overline m \ = \ \frac{m_{e_3} - m_{e_2}}{2} \, .
\label{eq:deltam}
\end{align}
The mass scale $M$ and mass differences $\delta m$, $\delta \overline m$
are process-specific quantities, their value being determined  
by the curly bracket labels of the short-distance coefficients 
$f$, $g$, $h_i$ multiplying them. Since $\delta m=\delta \overline m=0$ for 
diagonal annihilation reactions $\chi_{e_1}\chi_{e_2} \to \chi_{e_1}\chi_{e_2}$
(where the absorptive part of the amplitudes is related to
the corresponding annihilation cross section), the $\mathcal Q_i
\left( ^{2S+1}S_S \right)$ are only relevant for the computation
of the off-diagonal rates. The convergence of the non-relativistic expansion 
for these off-diagonal terms requires that the mass differences are 
considered as ${\cal O}(v^2)$ effects~\cite{Beneke:2012tg}.
Therefore, in scenarios where the off-diagonal annihilation terms 
can be relevant, the non-degeneracies among the particle species whose 
long-distance interactions are described within the EFT formalism 
are limited to $\delta m \ll m_{\text{LSP}}$. 
Particles with masses 
such that $\delta m \sim m_{\text{LSP}}$ or larger should be decoupled 
explicitly and integrated out -- they cause small modifications of 
short-distance coefficients in the effective Lagrangian, which are not 
relevant to the relic density computation.

Analytic results for the absorptive parts of the Wilson coefficients 
appearing in $\delta \mathcal L ^{d=8}_\text{ann}$ in the general MSSM
can be extracted from the expressions given in~\cite{Hellmann:2013jxa}.
For the $P$- and ${\cal O}(v^2)$ $S$-wave Wilson coefficients in 
(\ref{eq:NNLO_ops}) there are symmetry relations under the exchange of 
the particle labels analogous to those for the leading-order
$S$-wave Wilson coefficients~\cite{Hellmann:2013jxa}.
We reproduce them here for later reference:
\begin{align}
\nonumber
   k_{ \lbrace e_2 e_1\rbrace \lbrace e_4 e_3\rbrace }^{ \chi\chi \to \chi \chi}
                                              \left( {}^{2S+1}L_J \right)
	=
 (-1)^{S+L} \  k_{ \lbrace e_1  e_2\rbrace \lbrace e_4 e_3\rbrace }^{ \chi \chi \to \chi \chi }
                                               \left( {}^{2S+1}L_J \right)
\ ,
\\
   k_{ \lbrace e_1 e_2\rbrace \lbrace e_3 e_4\rbrace }^{ \chi \chi \to \chi \chi }
                                              \left( {}^{2S+1}L_J \right)
	=
 (-1)^{S+L} \  k_{ \lbrace e_1  e_2\rbrace \lbrace e_4 e_3\rbrace }^{ \chi \chi \to \chi \chi }
                                               \left( {}^{2S+1}L_J \right)
\ ,
\label{eq:WCsymmetries}
\end{align}
where $k$ refers to any of the Wilson coefficients, $f, g$ and $h_i$ in (\ref{eq:deltaL4fermion})
and (\ref{eq:NNLO_ops}).

\subsubsection{Sommerfeld-corrected cross section}
\label{subsec:SExsec}

\begin{figure}[t]
\begin{center}
\includegraphics[width=0.95\textwidth]{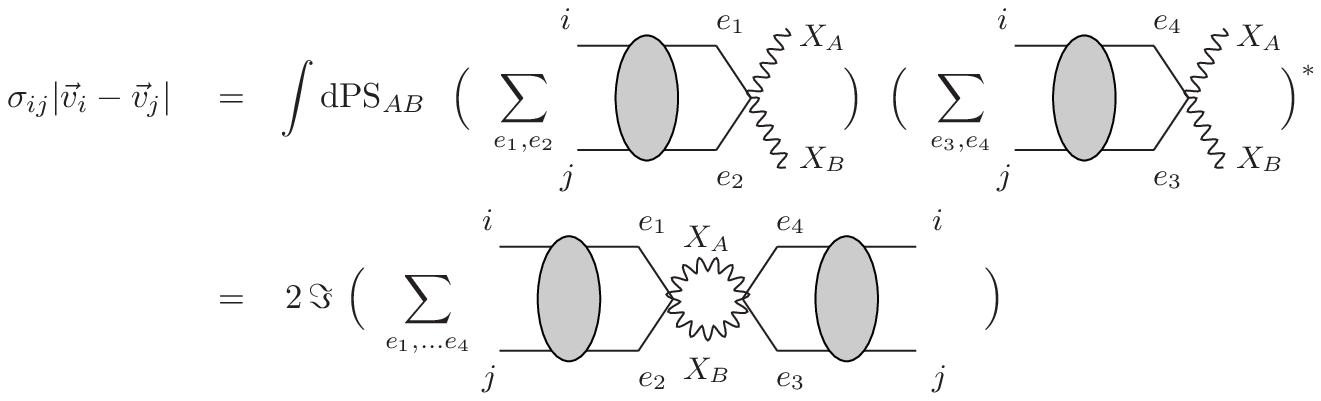}
\caption{ Diagrammatic picture for the relation among
          the annihilation amplitude and the absorptive
          part of the corresponding forward scattering amplitude
          in presence of long-range potential interactions.}
\label{fig:genericdiagram}
\end{center}
\end{figure}

The spin-averaged center-of-mass frame $\chi_{i} \chi_{j}$
annihilation cross section summed over all accessible light final states 
is given by the imaginary part of 
the forward-scattering amplitude $\chi_{i} \chi_{j} \to \chi_{i} \chi_{j}$ by virtue of unitarity,
see Fig.~\ref{fig:genericdiagram}.  In the non-relativistic effective theory, including up
to ${\cal O}(\vec{p}^{\,2})$ corrections, this observable is obtained as
\begin{align}
& \sigma^{\chi_{i} \chi_{j} \to \,{\rm light}} \, v_\text{rel} \ = \  
   \bigg( \frac{1}{4}\sum\limits_{ s_i, s_j} \bigg) \,2\, {\rm Im} \ \langle \chi_i \chi_j | \  \delta {\mathcal L}_\text{ann}  \ | \chi_i \chi_j \rangle \
\nonumber\\
& = \,\frac{1}{8} \sum\limits_{ s_i, s_j}   
\bigg\{   
    \,\bigg( \hat  f(^{1}S_0) + \frac{\delta m}{M} \, \hat h_1(^{1}S_0) + \frac{\delta \overline{m} }{M} \,\hat  h_2(^{1}S_0) \bigg)
    \, \langle \chi_i \chi_j | \, \mathcal O(^{1}S_0) \, | \chi_i \chi_j \rangle \ 
\nonumber\\
    & \qquad \qquad \;\; +\bigg(  \hat f(^{3}S_1) + \frac{\delta m}{M} \, \hat h_1(^{3}S_1) + \frac{\delta \overline{m} }{M} \, \hat h_2(^{3}S_1) \bigg) 
    \,   \langle \chi_i \chi_j | \, \mathcal O(^{3}S_1) \, | \chi_i \chi_j \rangle \
\nonumber\\
    & \qquad \qquad \;\;  + \,\frac{\hat  g(^{1}S_0)}{M^2}  \, \langle \chi_i \chi_j | \,  \mathcal P(^{1}S_0) \, | \chi_i \chi_j \rangle \
    + \frac{\hat g(^{3}S_1)}{M^2} \,  \langle \chi_i \chi_j | \, \mathcal P(^{3}S_1) \, | \chi_i \chi_j \rangle \ 
\nonumber\\
    & \qquad \qquad \;\; + \, \frac{\hat f(^{1}P_1)}{M^2}   \, \langle \chi_i \chi_j | \,  \mathcal O(^{1}P_1)\, | \chi_i \chi_j \rangle \ 
\nonumber\\
    & \qquad \qquad \;\;  + \frac{1}{M^2}\,\Big( \hat f(^{3}P_0) + 3 \, \hat f(^{3}P_1) + 5 \, \hat f(^{3}P_2) \Big)   \,\langle \chi_i \chi_j | \,
       \, \mathcal O(^{3}P_0) \, | \chi_i \chi_j \rangle \
\bigg\}\,,
\label{eq:SommerSigma}
\end{align}
with $v_{\rm rel} = \vert \vec v_{i} - \vec v_{j}\vert$  the
relative velocity of the
annihilating particles in the center-of-mass frame,\footnote{To make 
contact with commonly used notation in quarkonium annihilation,
we abuse notation when writing in~(\ref{eq:SommerSigma}) the matrix elements of operators $\delta {\cal L}_\text{ann}$, ${\cal O}(^{2S+1}L_J)$
and  ${\cal P}(^{2S+1}L_J)$ instead of the 
corresponding forward scattering amplitudes; the former contain an additional factor
$(2\pi)^4\delta^4(p_{\rm f}-p_{\rm i})$ which should not be included in the relation~(\ref{eq:SommerSigma}).} and assuming
the non-relativistic normalization $\langle \vec{p} |\vec{p}^{\,\prime} \rangle = (2\pi)^3 \delta^{(3)}(\vec{p}-\vec{p}^{\,\prime})$
for the incoming chargino and neutralino one-particle states.
In order to make the notation simpler we have omitted in (\ref{eq:SommerSigma}) 
the sum symbol over the intermediate states 
$\chi_{e_1} \chi_{e_2}$ and $\chi_{e_4} \chi_{e_3}$ that undergo annihilation
and
the labels in the Wilson coefficients
and operators (as well as in the quantities  $M,\,\delta m$ and $\delta \overline{m}$).
This means that, for instance, the first term in (\ref{eq:SommerSigma}) in full form
reads
\begin{align}
\frac{1}{8} \sum\limits_{ s_i, s_j}  \sum\limits_{ \chi \chi \rightarrow \chi \chi}
\,
\hat f^{ \chi \chi \to \chi \chi }_{ \lbrace e_1 e_2\rbrace \lbrace e_4 e_3\rbrace }
       \left( {}^{1}S_0 \right) \
       \, \langle \chi_i \chi_j | \, \mathcal O^{\chi \chi \to \chi \chi }_{ \lbrace e_4 e_3\rbrace 
\lbrace e_2 e_1\rbrace }(^{1}S_0) \, | \chi_i \chi_j \rangle \ 
\,,
\label{eq:notation}
\end{align}
and the sum extends over all $\chi_{e_1}\chi_{e_2}\to\chi_{e_4}\chi_{e_3}$ 
annihilation reactions where the states $\chi_{e_1}\chi_{e_2}\,,\chi_{e_4}\chi_{e_3}$ 
have the same charge as the incoming $\chi_i\chi_j$ pair.
In what follows we will always omit the symbol $\sum_{ \chi \chi \rightarrow \chi \chi}$,
and when repeated indices $e_i$ appear in an expression 
a summation over the particle species will be implied. 
To obtain the last equality in (\ref{eq:SommerSigma}) we have used that  
\begin{equation}
{\rm Im} \, \Big\{f(^{2S+1}L_J) \, \langle \chi_i \chi_j | \, 
\mathcal O(^{2S+1}L_J) \, | \chi_i \chi_j \rangle \Big\} \,
=  \hat f(^{2S+1}L_J) \,
\langle \chi_i \chi_j | \, \mathcal O(^{2S+1}L_J) \, | \chi_i \chi_j \rangle 
\, ,
\label{eq:opticaltheorem}
\end{equation}
following from the definition of the absorptive part of the Wilson 
coefficients $\hat{f}$~\cite{Beneke:2012tg}, which 
implies\footnote{It follows directly from the definition 
of $\hat f$ that $\hat f^{\chi\chi \to \chi \chi}_{\lbrace e_1 e_2 \rbrace 
\lbrace e_4 e_3 \rbrace}(^{2S+1}L_J)
= \big[ \, \hat f^{\chi\chi \to \chi \chi}_{\lbrace e_4 e_3 \rbrace 
\lbrace e_1 e_2 \rbrace}(^{2S+1}L_J) \,\big]^*$. In papers I and II this 
relation was incorrectly written for the Wilson coefficients themselves 
and not for their absorptive parts; we take the opportunity to correct it 
here.}
\begin{align}
-i \, \Big( \, f^{\chi\chi \to \chi \chi}_{\lbrace e_1 e_2 \rbrace \lbrace e_4 e_3 \rbrace}(^{2S+1}L_J)
  -  \big[ \, f^{\chi\chi \to \chi \chi}_{\lbrace e_4 e_3 \rbrace \lbrace e_1 e_2 \rbrace}(^{2S+1}L_J) \,\big]^* \,\Big)
  = 2 \,\hat f^{\chi\chi \to \chi \chi}_{\lbrace e_1 e_2 \rbrace \lbrace e_4 e_3 \rbrace}(^{2S+1}L_J)
  \ ,
\label{eq:fabs}
\end{align}
and the fact that the adjoint of the four-fermion operators in $\delta {\mathcal L}_\text{ann}$ satisfy 
\begin{align}
\mathcal O_{\lbrace e_4 e_3 \rbrace \lbrace e_2 e_1 \rbrace}(^{2S+1}L_J)=\mathcal O^{\,\dagger}_{\lbrace e_1 e_2 \rbrace \lbrace e_3 e_4 \rbrace}(^{2S+1}L_J) \ .
\label{eq:adjoint}
\end{align}
Note also that we have related the spin-average of the matrix element of the operators $\mathcal O(^{3}P_J)$, $J=1,2$,
with that of $\mathcal O(^{3}P_0)$  in the last line in~(\ref{eq:SommerSigma}).

%
%
 
The matrix-elements of four-fermion operators in~(\ref{eq:SommerSigma}) 
account for
the long-distance interactions between the annihilating pair, while the
short-distance annihilation into light particles is described
by the Wilson coefficients. The matching calculation of the
absorptive part of the Wilson coefficients can be performed 
for exclusive two-particle final states
$X_A X_B$ at the tree-level~\cite{Beneke:2012tg} ({\it i.e.} at 
${\cal O}(\alpha_i^2)$), since infrared divergences
are absent at that order. Therefore~(\ref{eq:SommerSigma})
also applies separately for  
every final state $X_A X_B$ to yield the exclusive annihilation
rates $\sigma_{\chi_{i} \chi_{j} \to X_A X_B} ~ v_{\rm rel}$ 
with ${\cal O}(\alpha_i^2)$ short-distance corrections.

It is well-known from quarkonium physics that matrix-elements of 
four-fermion operators analogous to those in~(\ref{eq:SommerSigma}) 
can be expressed in terms of non-relativistic
wave functions and their derivatives evaluated at the origin.
This relation becomes clear, if we explicitly insert the
operator $|0\rangle \, \langle0|$ that projects onto the Fock space
with no neutralino and chargino states into the 
four-fermion operators, which is exact at the level of terms 
included here in $\mathcal L_\text{kin} + \mathcal L_\text{pot}$. 
For instance, the matrix element
of the operator ${\cal O}^{ \chi_{e_1} \chi_{e_2} \rightarrow \chi_{e_4} \chi_{e_3}}({}^1S_0)$ can be written as
\begin{align}
\langle \chi_i \chi_j | \, \mathcal O^{ \chi \chi \rightarrow \chi \chi}_{\lbrace e_4 e_3 \rbrace \lbrace e_2 e_1 \rbrace}(^{1}S_0) \, | \chi_i \chi_j \rangle &= 
\langle \chi_i \chi_j | \, 
    \chi^\dagger_{e_4} \chi^c_{e_3} |0 \rangle \, \langle 0| \chi^{c \dagger}_{e_2}\chi^{}_{e_1}
\, | \chi_i \chi_j \rangle
\nonumber\\
 = \Big[\,\langle \xi^{c \dagger}_{j} \xi_{i} \rangle \, 
        \big( \psi^{(0,0)}_{e_4 e_3,\,ij} \, + \, & \psi^{(0,0)}_{e_3 e_4,\,ij} \big) \Big]^*
    \,\langle \xi^{c \dagger}_{j} \xi_{i} \rangle \,
          \big( \psi^{(0,0)}_{e_1 e_2, \,ij} \, + \,  \psi^{(0,0)}_{e_2 e_1, \,ij} \big)
\ ,
\label{eq:wave}
\end{align}
where $\psi^{(L,S)}_{e_1 e_2,\,ij}$ is the $\chi_{e_1} \chi_{e_2}$-component 
of the scattering wave function for an incoming $\chi_{i} \chi_{j}$ state 
with center-of-mass energy $\sqrt{s}$, orbital quantum number $L$ and 
total spin $S$, evaluated for zero relative distance and
normalized to the free scattering solution.
The symbols $\xi_{i},\,\xi_j$ in the second
line of~(\ref{eq:wave}) denote the Pauli spinor of the incoming particles $\chi_{i}$ and $\chi_j$,
and $\langle \dots \rangle$ stands for the trace in spin space.
The multi-component wave function $\vec{\psi}^{\,(L,S)}_{ij}$  accounts for the potential interactions of the incoming 
$\chi_{i} \chi_{j}$ state with all possible intermediate
two-body chargino and neutralino states with the same charge and identical spin and partial-wave
configuration. Recall from Section~\ref{subsec:lagrangian} that we consider only leading-order potential interactions
in ${\cal L}_{\rm pot}$, which cannot change the spin and orbital angular momentum of the two-particle
states. 
Both wave-function components $e_1 e_2$ and $e_2 e_1$ are generated by the matrix-element of operator  
$\chi^{c \dagger}_{e_2}\chi^{}_{e_1}$; for an operator with quantum numbers $L$ and $S$, there is a 
relative sign $(-1)^{L+S}$ between the two components, see~(\ref{eq:medef}).
The precise definition of  $\vec{\psi}^{\,(L,S)}_{ij}$ is postponed to Section~\ref{sec:Sommerfeld}.
Let us just mention here that the lowest-order perturbative result for the matrix-elements of four-fermion operators
is obtained by replacing  
$\psi^{(L,S)}_{e_a e_b,\,ij}\to \delta_{e_a i}\,\delta_{e_b j}$,
as can be easily checked by explicit computation.

In terms of the non-relativistic wave functions the spin-averaged 
annihilation cross section~(\ref{eq:SommerSigma}) takes the form 
\begin{align}
& \sigma^{\chi_{i} \chi_{j} \to \,{\rm light}} \, v_\text{rel} \
\nonumber\\
& = \,\left[ \psi^{(0,0)}_{e_4 e_3,\,ij}\right]^*
    \bigg( \hat  f(^{1}S_0) + \frac{\delta m}{M} \, \hat h_1(^{1}S_0) + \frac{\delta \overline{m} }{M} \,\hat  h_2(^{1}S_0) 
           +   \hat  g_{\kappa}(^{1}S_0) \bigg)
    \, \psi^{(0,0)}_{e_1 e_2, \, ij}
\nonumber\\
    & \quad \;+\, 3\, \left[ \psi^{(0,1)}_{e_4 e_3,\,ij} \right]^*
    \bigg(  \hat f(^{3}S_1) + \frac{\delta m}{M} \, \hat h_1(^{3}S_1) + \frac{\delta \overline{m} }{M} \, \hat h_2(^{3}S_1) 
          +   \hat  g_{\kappa}(^{3}S_1) \bigg) 
    \,  \psi^{(0,1)}_{e_1 e_2, \,ij}
\nonumber\\
    & \quad\; +\,\vec{p}_{ij}^{\,2} \,\left[ \psi^{(1,0)}_{e_4 e_3,\,ij} \right]^* \,\frac{\hat f(^{1}P_1)}{M^2} \,
    \psi^{(1,0)}_{e_1 e_2,\,ij} 
\nonumber\\
    & \quad \; +\, \vec{p}_{ij}^{\,2} \,\left[ \psi^{(1,1)}_{e_4 e_3,\,ij} \right]^*
      \,\frac{1}{M^2}\bigg( \, \frac{1}{3} \, \hat f(^{3}P_0) + \, \hat f(^{3}P_1) + \frac{5}{3} \, \hat f(^{3}P_2) \bigg)   
       \, \psi^{(1,1)}_{e_1 e_2, \,ij}
\ ,
\label{eq:SigmaWave}
\end{align}
where we have used the symmetry relations~(\ref{eq:WCsymmetries}) for the 
Wilson coefficients, and the spin sums
\begin{equation}
\frac{1}{2} \, \sum\limits_{ s_i, s_j}\,  \langle \xi^{c \dagger}_{j} \xi_{i} \rangle \, \langle \xi^{c \dagger}_{j} \xi_{i} \rangle^*  = 1 \,,
\qquad 
\frac{1}{2} \, \sum\limits_{ s_i, s_j}\,  \langle \xi^{c \dagger}_{j} \sigma^k \xi_{i} \rangle \, 
\langle \xi^{c \dagger}_{j} \sigma^\ell \xi_{i} \rangle^* = \delta_{k\ell} 
\,.
\label{eq:spintraces}
\end{equation}
For a given two-particle state, the relative momentum of particles 
$\chi_i$, $\chi_j$ in their center-of-mass frame is related to the 
center-of-mass energy of the collision by  
$\vec{p}_{ij}^{\,2}=2\mu_{ij}(\sqrt{s}-M_{ij})+
{\cal O}(\vec{p}_{ij}^{\,4})$, where
$M_{ij}$  and  $\mu_{ij}$ are the total and reduced mass, respectively, 
of the two-particle system.
In addition we have used the relation 
\begin{align}
\frac{\hat  g(^{2S+1}S_S)}{M^2} \, \langle \chi_i \chi_j | \, 
\mathcal P(^{2S+1}S_S) \, | \chi_i \chi_j \rangle &= 
\, \hat g_\kappa(^{2S+1}S_S) \, \langle \chi_i \chi_j | \, 
\mathcal O(^{2S+1}S_S) \, | \chi_i \chi_j \rangle 
\label{eq:p2Swaveeom}
\end{align}
between the matrix-elements
of operators ${\cal O}(^{2S+1}S_S)$ and ${\cal P}(^{2S+1}S_S)$, where
\begin{align}
\hat  g_{\kappa \,\lbrace e_1^\prime e_2^\prime \rbrace \lbrace e_4^\prime e_3^\prime \rbrace}(^{2S+1}S_S) =  
\frac{\hat g_{\lbrace e_1 e_2 \rbrace \lbrace e_4 e_3 \rbrace}(^{2S+1}S_S)}{2M^2} 
\, \Big( \kappa^*_{ e_1 e_2, e_1^\prime e_2^\prime} \, 
\delta_{e_4 e_3,e_4^\prime e_3^\prime}
+  \delta_{e_1 e_2, e_1^\prime e_2^\prime} \, \kappa_{  e_4 e_3 , e_4^\prime e_3^\prime } \,\Big)
\,,
\label{eq:gkappa}
\end{align}
and
\begin{eqnarray}
&&\nonumber\\[-1cm]
&&\kappa_{\,  e_1 e_2 , e_1^\prime e_2^\prime } = \vec{p}^{\,2}_{e_1 e_2}\, 
\delta_{e_1 e_2,e_1^\prime e_2^\prime}
+ 2 \, \mu_{e_1 e_2}\alpha_2\, \sum_a  \,  m_{\phi_a}  \,  c^{(a)}_{ e_1 e_2,  e_1^\prime e_2^\prime    } 
\ .
\label{eq:kappa}
\end{eqnarray}
In (\ref{eq:gkappa}), (\ref{eq:kappa}) the value of $M^2$ is determined 
by the particle labels of the short-distance coefficient in the numerator 
of the fraction.
The sum in the second term in (\ref{eq:kappa}) extends over 
all potential interactions $\chi_{e_1}\chi_{e_1}\to\chi_{e_1^\prime}\chi_{e_2^\prime}$
arising from $\phi_a$-boson exchange. The coefficients of the potentials, $c^{(a)}_{  e_1 e_2  ,e_1^\prime e_2^\prime }$, are given
in Tab.~\ref{tab:potentials} of Appendix~\ref{sec:appendixpot}. 
The derivation of (\ref{eq:p2Swaveeom}) is postponed to 
Sec.~\ref{sec:Sommerfeld}. It is interesting to note that 
when we reduce the spectrum of two-particle states to just one state and  the 
exchanged bosons are massless,  the relation (\ref{eq:p2Swaveeom}) between 
the leading and the $v^2$-suppressed $S$-wave
operators acquires the simpler form familiar from the NRQCD applications to heavy quarkonium,
$ \langle \chi \chi | \, \mathcal P(^{2S+1}S_S) \, | \chi \chi \rangle = 
\vec{p}^{\,2} \, \langle \chi \chi | \, \mathcal O(^{2S+1}S_S) \, | \chi \chi \rangle$.

We define for an incoming state $\chi_i\chi_j$  with center-of-mass energy $\sqrt{s}$ 
the Sommerfeld enhancement factor  associated to a generic Wilson coefficient or combination of
Wilson coefficients, $\hat f$, describing the short-distance annihilation of $\chi\chi$ states
with spin $S$ and orbital-momentum $L$
as the ratio 
\begin{equation}
S_{ij}[\hat f(^{2S+1}L_J)] = 
\frac{ 
 \left[ \psi^{(L,S)}_{e_4 e_3,\,ij}\right]^*
     \hat f^{\chi\chi \to \chi \chi}_{\lbrace e_1 e_2 \rbrace \lbrace e_4 e_3 \rbrace}(^{2S+1}L_J) 
    \, \psi^{(L,S)}_{e_1 e_2, \,ij}}
{\hat  f^{\chi\chi\to \chi\chi}_{\lbrace i j \rbrace \lbrace i j \rbrace}(^{2S+1}L_J)|_{\rm LO}
}
\ .
\label{eq:SFdef}
\end{equation}
The subscript ``LO'' in the denominator of (\ref{eq:SFdef}) means that only the leading order
in $\alpha_2$ of the Wilson coefficient $\hat f(^{2S+1}L_J)$ should be kept in the denominator.
For our purposes,\footnote{In general, the definition (\ref{eq:SFdef}) also 
allows to incorporate
the higher-order (hard) radiative corrections to the short-distance part of the annihilation into
the Sommerfeld factor.} this is only relevant for the case of
$S_{ij}[\hat  g_\kappa(^{2S+1}S_S)]$, where we have to set the $\alpha_2$ term in $\kappa$ to zero,
so that
$\hat  g^{\chi\chi\to \chi\chi}_{\kappa \lbrace i j \rbrace \lbrace i j \rbrace}(^{2S+1}S_S)|_{\rm LO}
= \vec{p}_{ij}^{\;2}/M_{ij}^2\,\hat  g^{\chi\chi\to \chi\chi}_{\lbrace i j \rbrace \lbrace i j \rbrace}(^{2S+1}S_S)$.
The Sommerfeld factors are functions of $\sqrt{s}$ or, equivalently, of the relative
velocity $v_{\rm rel}$ of the incoming state.
They allow us to parametrize the long-distance corrections to the annihilation rate
of the state $\chi_i\chi_j$ in a convenient way. Indeed, in terms of the Sommerfeld factors, the 
spin-averaged annihilation cross section~(\ref{eq:SommerSigma}) acquires 
the simple form
\begin{eqnarray}
\sigma^{\chi_{i} \chi_{j} \to \,{\rm light}} \, v_\text{rel} 
& =& \, S_{ij} [\hat f_h(^{1}S_0)] 
     \; \hat  f^{\chi\chi \to \chi \chi}_{\lbrace i j \rbrace \lbrace i j \rbrace}(^{1}S_0)
 + \, S_{ij}[\hat f_h(^{3}S_1)] 
     \; 3 \,\hat  f^{\chi\chi \to \chi \chi}_{\lbrace i j \rbrace \lbrace i j \rbrace}(^{3}S_1)
\nonumber\\
&&  \hspace*{-2cm} + \, \frac{\vec{p}_{ij}^{\,2}}{M_{ij}^2} \, 
     \Big( \, S_{ij} [\hat g_\kappa(^{1}S_0)]  \; \hat  g^{\chi\chi \to \chi \chi}_{\lbrace i j \rbrace \lbrace i j \rbrace}(^{1}S_0)
          +  S_{ij}[\hat g_\kappa(^{3}S_1)] \; 3 \, \hat  g^{\chi\chi \to \chi \chi}_{\lbrace i j \rbrace \lbrace i j \rbrace}(^{3}S_1)
\nonumber\\
&&  \hspace*{-0.5cm}
         + \,S_{ij} \Big[\frac{\hat f(^{1}P_1)}{M^2}\Big] \; \hat  f^{\chi\chi \to \chi \chi}_{\lbrace i j \rbrace \lbrace i j \rbrace}(^{1}P_1)
         + S_{ij} \Big[\frac{\hat f({}^3P_{\cal J})}{M^2} \Big] \; \hat f^{\chi\chi \to \chi \chi}_{\lbrace i j \rbrace \lbrace i j \rbrace}(^{3}P_{\cal J})
     \Big)
\ ,\qquad
\label{eq:SFenhancedsigma}
\end{eqnarray}
where we have introduced the combinations of Wilson coefficients
\begin{align}
\hat  f_h(^{1}S_0)  & = \hat  f(^{1}S_0) + \frac{\delta m}{M} \, \hat h_1(^{1}S_0) + \frac{\delta \overline{m} }{M} \,\hat  h_2(^{1}S_0) \ ,
\nonumber\\
\hat  f_h(^{3}S_1)  & = \hat  f(^{3}S_1) + \frac{\delta m}{M} \, \hat h_1(^{3}S_1) + \frac{\delta \overline{m} }{M} \,\hat  h_2(^{3}S_1) \ ,
\nonumber\\
\hat f(^3P_{\cal J}) & = \, \frac{1}{3} \hat f(^3P_0) +  \hat f(^3P_1) + \frac{5}{3} \hat f(^3P_2) \ .
\label{eq:effWilsoncoeff}
\end{align}
The last relation in~(\ref{eq:effWilsoncoeff}) reminds us that the knowledge 
of the spin-weighted sum over the three
different $^3P_0$, $^3P_1$ and $^3P_2$ partial-wave Wilson coefficients
is sufficient, since the leading-order potential interactions while being 
spin-dependent, do not discriminate among the three spin-1
$P$-wave states $^3P_J$ with different total angular momentum $J = 0,1,2$.
The pure tree-level annihilation rate with no long-distance corrections is 
readily recovered  by setting all the Sommerfeld factors 
in~(\ref{eq:SFenhancedsigma}) to one. The 
tree-level annihilation cross section thus obtained depends only on the
diagonal entry of the Wilson coefficients corresponding to channel $\chi_i\chi_j$. Note that
we have written in~(\ref{eq:SFenhancedsigma})
$f^{\chi_i\chi_j \to \chi_i \chi_j}_{\lbrace i j \rbrace \lbrace i j \rbrace}(^{2S+1}S_S)$ 
and not $f^{\chi_i\chi_j \to \chi_i \chi_j}_{h\,\lbrace i j \rbrace \lbrace i j \rbrace}(^{2S+1}S_S)$ because
for the diagonal entries $\delta m=\delta \overline{m}=0$.

\subsection{Relic abundance}
\label{subsec:relic-abundance}

The thermal relic abundance of neutralino dark matter can be obtained by 
solving the Boltzmann equation
\begin{align}
\frac{dn}{dt} + 3 H n = - \langle \sigma_{\rm eff} v \rangle 
( n^2 - n_{\rm eq}^2 )\,,
\label{eq:boltzeq}
\end{align}
that describes the time evolution of $n=\sum_{i=1}^N n_i$, the sum of the 
particle number densities of all supersymmetric particles $\chi_i$ taking 
part in the relevant annihilation reactions $\chi_i\chi_j\to X$, 
which change the lightest neutralino number density $n_1$, either directly 
or indirectly through the later decay  into $\chi_1$ of
the species involved in those reactions. Note that one can solve 
(\ref{eq:boltzeq}) for $n$
to obtain $n_1$ in the present Universe because $R$-parity
conservation implies that all other $\chi_i$ must decay into the 
lightest one by today. In this work the supersymmetric particles $\chi_i$ 
considered in the relic density calculation are the neutralinos and 
charginos. For SUSY models where other particle species (staus for example)
may have important co-annihilation effects with the neutralino, an extension 
of the EFT framework presented here to a larger set of two-particle states 
would be needed. The co-annihilation rates enter the 
Boltzmann equation through the thermal average of the effective cross section, $\langle \sigma_{\rm eff} v \rangle$,
whose specific form is given below. The other quantities entering (\ref{eq:boltzeq}) are the 
Hubble parameter $H$ and $n_{\rm eq}$, the sum of the equilibrium number densities of each particle 
$n_{i,\,\rm eq}$. Eq.~(\ref{eq:boltzeq}) is derived using Maxwell-Boltzmann 
statistics for all species in thermal equilibrium instead of Fermi-Dirac for fermions and Bose-Einstein for 
bosons, which is a very good approximation for the temperatures relevant for the relic 
density calculation of heavy dark matter ($T \lesssim T_{f}$ with $T_f\sim m_{\chi_0}/20$
the typical temperature where the departure from equilibrium takes place). 
Furthermore, the evolution equation~(\ref{eq:boltzeq}) is valid under the assumption that
the particle's phase-space distributions are proportional to those in 
equilibrium, $f_i$,  with
a factor of proportionality that depends on $T$ but not on the energy. As argued in~\cite{Bernstein:1985th},
this is true if the $\chi_i$ particles are maintained in kinetic equilibrium through 
elastic scattering with the much more abundant standard model
particles in the thermal plasma during all of their evolution, even after they leave chemical equilibrium. In this work we assume that dark matter is kept in kinetic equilibrium after chemical decoupling sufficiently long until freeze-out is completed. The possibility that the elastic scattering processes stop to be effective and the consequences of early kinetic
decoupling in the context of Sommerfeld-enhancement 
have been investigated in~\cite{Bringmann:2006mu,Dent:2009bv,Feng:2010zp,vandenAarssen:2012ag}.

The quantity $\langle \sigma_{\rm eff} v \rangle$ that feeds into the 
equation for the dark matter relic density reads
\begin{align}
\langle \sigma_{\rm eff} v \rangle = \sum_{i,j=1}^{N}  \langle \sigma_{ij} v_{ij} \rangle \,
\frac{n_{i,\,\rm eq}\,n_{j,\,\rm eq}}{n_{\rm eq}^2}
\,,
\label{eq:sigmaeffv}
\end{align}
where  $\langle \sigma_{ij} v_{ij} \rangle$ is the thermal average of the co-annihilation cross section
of $\chi_i\chi_j$ into standard model particles  and $v_{ij}$ is the so-called M\o ller velocity of
particles $\chi_i$ and $\chi_j$.\footnote{$v_{ij}$ is defined by $v_{ij}=\sqrt{(p_i\cdot p_j)^2-m_i^2\,m_j^2}/(E_iE_j)$
with $p_i$ and $E_i$ the four-momentum and energy of particle $i$, and is equal to the relative
velocity of particles $i$ and $j$ in any frame where the two particles move collinearly.} In the cosmic comoving frame,
{\it i.e.} the frame where the 
plasma is at rest as a whole,
the thermal average $\langle \sigma_{ij} v_{ij} \rangle$  can be calculated using $f_{i} = e^{-E_i/T}$
for the Maxwell-Boltzmann equilibrium distributions. A general expression for 
$\langle \sigma_{\rm eff} v \rangle$ written in terms of Lorentz invariants was first
derived in~\cite{Gondolo:1990dk} and further generalized to include co-annihilations in~\cite{Edsjo:1997bg}. It reads
\begin{align}
\langle \sigma_{\rm eff} v \rangle = 
\frac{1}{n_{\rm eq}^2}\, \displaystyle\sum_{i,j=1}^N \frac{g_i g_j}{4\pi^4} \, T \int_{(m_i+m_j)^2}^{\infty} ds \, 
\frac{(p_i\cdot p_j)^2-m_i^2m_j^2}{2\sqrt{s}} \, \sigma_{ij}
\,K_1 \big( \sqrt{s}/{T} \big)
\,,
\label{eq:invsigmaeffv}
\end{align}
\begin{align}
n_{\rm eq} = \frac{T}{2\pi^2} \sum_{i=1}^N g_i m_i^2 K_2\big( m_i/T \big)
\,,
\label{eq:neq}
\end{align}
with $g_i$ the internal degrees of freedom of particle species $\chi_i$, which for
neutralinos and charginos is $g_i=2$,
and $K_{n}$ the modified Bessel function of the second kind of order $n$.
The integrand in~(\ref{eq:invsigmaeffv}) evaluated in the center-of-mass
frame allow us to use the expressions 
obtained in Sec.~\ref{subsec:SExsec} for $(\sigma_{ij}\,v_{ij})$ (there denoted as 
$\sigma^{\chi_{i} \chi_{j} \to \,{\rm light}} \, v_\text{rel}$)
in the non-relativistic limit including up to ${\cal O}(v_\text{rel}^2)$ corrections.
Note that the thermal average of a heavy co-annihilation
channel is typically
suppressed by a factor $e^{-(m_i+m_j-2m_1)/T}$ with respect 
to $\langle \sigma_{11}v_{11}\rangle$, which arises from the asymptotic expansion of the Bessel 
function $K_1(\sqrt{s}/T)$ for large $\sqrt{s}/T > 2m_1/T \gg 1$.  

The efficiency of dark matter production and annihilation processes to 
maintain chemical equilibrium with the plasma in an expanding Universe can 
be better understood when the Boltzmann
equation is written in terms of yield $Y\equiv n/s$, defined as the ratio of the particle density 
to the entropy density in the comoving cosmic frame. 
This is because,
assuming that the entropy per comoving volume is conserved, 
the change of $n$ and $s$ due to the expansion of the Universe is 
the same, namely $ds/dt= -3Hs$, and gets scaled out from the Boltzmann equation.
Since $\langle \sigma_{\rm eff} v \rangle$ is
computed as a function of the temperature rather than 
time, it is also convenient to trade the independent variable $t$ in the evolution equation
by $T$. In terms of the yield, and defining  $x=m_1/T$, the Boltzmann equation
(\ref{eq:boltzeq}) takes the form 
\begin{align}
\frac{dY}{dx} = \frac{1}{3H} \, \frac{ds}{dx} \langle \sigma_{\rm eff} v \rangle ( Y^2 - Y_{\rm eq}^2 )
\,,
\label{eq:boltzeqY1}
\end{align}
where $Y_{\rm eq}=n_{\rm eq}/s$. The final step requires the Friedmann equation that relates the expansion
of the Universe (Hubble rate $H$) to its energy density. In a radiation-dominated universe at early times, the
equation for the evolution of $Y$ can be finally written as~\cite{Gondolo:1990dk}
\begin{align}
\frac{dY}{dx} = -\sqrt{\frac{\pi}{45G}}\,\frac{g_*^{1/2}\,m_1}{x^2} 
\, \langle \sigma_{\rm eff} v \rangle ( Y^2 - Y_{\rm eq}^2 )
\,,
\label{eq:boltzeqY2}
\end{align}
with $G$ the gravitational constant. The parameter $g_*^{1/2}$ is defined as
\begin{align}
g_*^{1/2} = \frac{h_{\rm eff}}{g_{\rm eff}^{1/2}} 
\bigg( 1+ \frac{T}{3h_{\rm eff}} \, \frac{dh_{\rm eff}}{dT} \bigg)
\,
\label{eq:gstar}
\end{align}
in terms of the effective degrees of freedom 
$g_{\rm eff}$ and $h_{\rm eff}$ of the energy and entropy densities:
\begin{align}
\rho = g_{\rm eff}(T) \frac{\pi^2}{30}\,T^4 
\quad,\quad 
s= h_{\rm eff}(T) \frac{2\pi^2}{45}\,T^3 
\,.
\label{eq:rhos}
\end{align}
The effective degrees of freedom are slowly-varying functions of the 
temperature except at the QCD quark-hadron phase transition 
($T\sim 150-400$~MeV), where there is a sharp 
increase~\cite{Srednicki:1988ce,Gondolo:1990dk}.
The equilibrium yield $Y_{\rm eq}$ can  be obtained using the formula 
for $n_{\rm eq}$ given in~(\ref{eq:neq}) and the parametrization of the 
entropy density in (\ref{eq:rhos}) above. 

In the form~(\ref{eq:boltzeqY2}), we see that the change in the yield  
tries to compensate for deviations from the equilibrium, such that if $Y$ 
was initially
close to $Y_{\rm eq}$ at high temperatures (low $x$) it will continue to 
track $Y_{\rm eq}$ as the temperature decreases. At a given temperature 
the prefactor $\langle \sigma_{\rm eff} v \rangle/x^2$ (take $g_*^{1/2}$
constant) is not large enough to provide the change in $Y$ needed to
keep up with the rapid exponential drop of $Y_{\rm eq}$,
and chemical decoupling of the massive particle from the plasma takes place. 
The yield then remains constant as the universe cools down 
(phenomenon known as ``sudden freeze-out'') or, if
the annihilation cross section is further increased for lower temperatures
the yield continues to be reduced; that is the case
when the annihilation rates get enhanced by Sommerfeld corrections.

In order to obtain the relic density we solve~(\ref{eq:boltzeqY2}) 
numerically from $x=1$, where one can safely assume that 
the yield still tracks the equilibrium value and take $Y=Y_{\rm eq}$ as initial condition,
up to $x_0=m_1/T_0$, where $T_0$ is the current photon temperature in the Universe. 
Once the present value of the yield, $Y_0\equiv Y(x_0)$, is known, 
the neutralino relic density today in units of the critical density 
is determined by the relation
\begin{align}
\Omega = \rho_{0}^{\rm DM} /\rho_{\rm crit} = m_1 s_0 Y_0  /\rho_{\rm crit}
\,,
\label{eq:Omegah2}
\end{align}
with $\rho_{\rm crit}=3H^2/8\pi G$  the critical density and $s_0$ the entropy
density of the present universe, that can be calculated from~(\ref{eq:rhos}).
It is customary to provide the relic density as $\Omega h^2$, where
$h$ is the value of the Hubble parameter determined at present in units
of 100~km~s${}^{-1}$~Mpc${}^{-1}$.

In the presence of Sommerfeld corrections the effective annihilation 
cross section $\sigma_{\rm eff} v$ acquires a complicated velocity 
dependence that can no longer be parameterized in the simple form 
$a+b v^2$. For the results that we present in \cite{paperIV},
$Y_0$ has been obtained by solving (\ref{eq:boltzeq}) using numerical 
routines for differential equations built in {\sc Mathematica}.
Due to the stiffness of the evolution equation, the 
numerical integration of the equation is done by splitting the region 
from $x=1$ to $x=10^8$ into several pieces and adapting the starting and 
maximum step sizes in each of them. For $g_{*}^{1/2}(T)$ and $h_{\rm eff}(T)$ 
we have used the values derived in~\cite{Gondolo:1990dk}, 
which can be found conveniently
tabulated as a function of temperature among the
package files of the automated programs {\sf DarkSUSY}~\cite{Gondolo:2004sc} 
and {\tt micrOMEGAs}~\cite{Belanger:2010gh,Belanger:2013oya}. 
Other numerical values needed for the computation of the relic density are 
$T_0=2.7255$~K and $\rho_{\rm crit}=1.05368\times 10^{-5} h^2$~GeV~cm${}^{-3}$,
both taken from~\cite{Beringer:1900zz}.


\section{MSSM potentials}
\label{sec:potMSSM}

Sommerfeld enhancement occurs when the interaction between the incoming DM 
particles significantly distorts their two-body wave function away 
from the initial plane wave.
In terms of Feynman diagrams the effect arises from the
exchange of electroweak gauge (and to a lesser extent, Higgs) bosons 
between the neutralinos and charginos. In the non-relativistic limit the 
dominant contribution arises from the potential loop momentum region 
in ladder box graphs, which have to be summed to all orders in $\alpha_{\rm EW}$, as is shown in Sec.~\ref{sec:Sommerfeld} below. A necessary input to 
perform such resummation are the potential interactions generated by 
electroweak- and Higgs-boson exchange, which are given in this section. 
 
%
\begin{figure}[t]
\begin{center}
\includegraphics[width=0.5\textwidth]{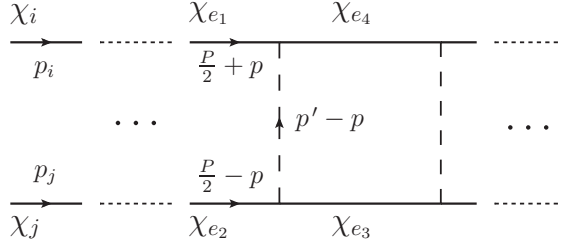}
\caption{Box subgraph of a characteristic diagram with multiple 
ladder-like exchanges of vector and Higgs bosons among intermediate 
$\chi\chi$ states, which contributes to the set of radiative corrections 
to the annihilation of the incoming pair $\chi_i\chi_j$. Arrows in this 
picture indicate the direction of the labelled momenta.}
\label{fig:potential}
\end{center}
\end{figure}
%

In the center-of-mass system of the incoming $\chi_i\chi_j$ pair,
$p_i+p_j=P=(\sqrt{s},\vec{0}\,)$,
the potential loop momentum $p^\prime$ running inside a box graph with 
internal fermions $\chi_{e_4}\chi_{e_3}$
can be chosen to be the relative momentum of the $\chi_{e_4}\chi_{e_3}$ pair, {\it i.e.}
$p^\prime=(p_{4}-p_{3})/2$; the momentum carried by the exchanged boson is then equal to the
difference between the relative momenta in the $\chi\chi$ pair before and 
after the interaction (see Fig.~\ref{fig:potential}). 
The potential loop-momentum routed in this way is characterized by the 
scaling  $p^{\prime\,0}\sim \vec{p}^{\,\prime\,2}/m_{\rm LSP}\ll m_{\rm LSP}$ (idem for $p$), and fermion and boson propagators can be expanded 
accordingly. To leading order in the potential-region expansion,
the denominator of the boson propagator, $D=[(p^{\,\prime}-p)^2-M^2]^{-1}$ becomes energy independent,
$D_{\rm pot}=-[(\vec{p}^{\,\prime}-\vec{p}\,)^2+M^2]^{-1}$ and thus represents an instantaneous
interaction between the $\chi_{e_1}\chi_{e_2}$ and $\chi_{e_3}\chi_{e_4}$ pairs, 
that in the non-relativistic EFT is taken into account by the potentials in
${\cal L}_{\rm pot}$. We note that the $(p^{\prime\,0}-p^0)^2$ term dropped in the boson propagator
has a term proportional to the mass difference squared $(m_{4}-m_{1})^2$, which we
neglect consistently given that we are dealing with a set of nearly mass-degenerate
with the neutralino LSP and mass differences are assumed to be at most of order of
$p_{ij}^2/m_{\rm LSP}\sim {\cal O}(m_{\rm LSP} v^2)$, where $p_{ij}=(p_i-p_j)/2$ is the
incoming relative momentum.

\begin{figure}[t]
\begin{center}
\includegraphics[width=0.8\textwidth]{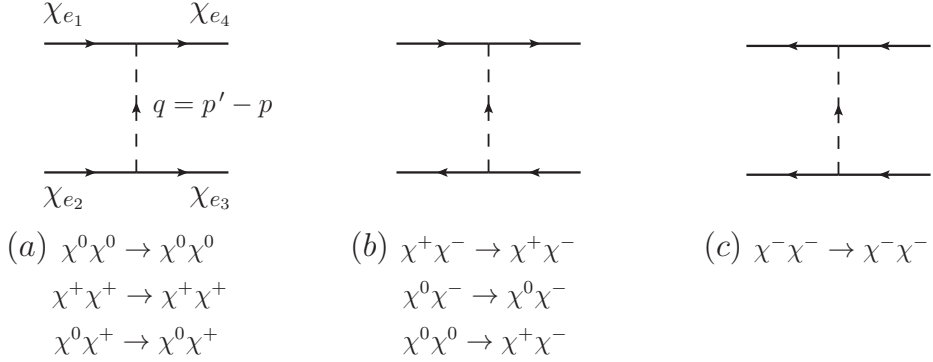}
\caption{Tree-level diagrams with $t$-channel boson exchange that generate the 
leading-order potential among non-relativistic neutralinos and charginos. 
The arrows in the neutralino/chargino lines indicate the fermion flow, 
such that 
each diagram contributes only to the scattering processes written below.}
\label{fig:pot_matching}
\end{center}
\end{figure}

The leading-order potential interactions in the EFT can be obtained by 
expanding the full-theory tree-level scattering amplitude 
$\chi_{e_1}\chi_{e_2} \to\chi_{e_4}\chi_{e_3}$ for on-shell
charginos or neutralinos exchanging a gauge or Higgs boson. 
Three tree diagrams which differ in the direction of the fermion flow
have to be considered
depending on the particle nature of the scattering particles, see Fig.~\ref{fig:pot_matching}.
By means of an example, $Z$-boson exchange, we
illustrate in the following the essential steps to perform this tree-level matching. 

Let us denote the vector ($v$), axial-vector ($a$) interaction vertices of charginos and neutralinos 
with gauge bosons in the MSSM generically as
\begin{align}
g_2\,\overline{\chi}_{e_i}  \, [ \, v_{ij} \gamma^\mu + a_{ij} \gamma^\mu \gamma_5 \, ] \, \chi_{e_j} A_\mu^V
\,,
\label{eq:gaugecoup}
\end{align}
where the spin-1 field $A^V_\mu$ stands for either the $Z$-boson (then
$\overline{\chi}_{e_i} \chi_{e_j}=\overline{\chi}^0_{e_i} \chi^0_{e_j},\,
\overline{\chi}^+_{e_i} \chi^+_{e_j}$), the photon field 
($\overline{\chi}_{e_i} \chi_{e_j}=\overline{\chi}^+_{e_i} \chi^+_{e_j}$),
or the $W^+$-field (which sets 
$\overline{\chi}_{e_i} \chi_{e_j}=\overline{\chi}^+_{e_i} \chi^0_{e_j}$)\footnote{
We adopt the convention that the $\chi^+$-field annihilates positive charginos. 
Therefore, the arrow on the fermion line for a chargino refers to the direction of $\chi^+$ flow.
For consistency, the $W^+$-field in the interaction vertex (\ref{eq:gaugecoup}) 
must annihilate positively charged $W$ bosons.}.
In the latter case, the hermitian conjugate of (\ref{eq:gaugecoup}) has to be 
added to describe the charge-conjugated interaction, {\it i.e.}
\begin{align}
g_2\,\overline{\chi}_{e_i}^0  \, [ \, (v^{W\dagger})_{ij} \gamma^\mu 
 + (a^{W\dagger})_{ij} \gamma^\mu \gamma_5 \, ] \, \chi_{e_j}^+ \, [W_\mu^+]^\dagger
\,.
\label{eq:hc}
\end{align}
Likewise, the interaction vertices of charginos/neutralinos with Higgs particles
has the generic form
\begin{align}
g_2\,\overline{\chi}_{e_i}  \, [ \, s_{ij}  + p_{ij}  \gamma_5 \, ] \, \chi_{e_j} \, \phi
\,,
\label{eq:higgscoup}
\end{align}
where the scalar field $\phi$ can be any of the $CP$-even ($H^0_m,\,m=1,2$) or 
$CP$-odd ($A^0_m,\,m=1,2$) Higgs bosons, in which case 
$\overline{\chi}_{e_i} \chi_{e_j}=\overline{\chi}^0_{e_i} \chi^0_{e_j},\,
\overline{\chi}^+_{e_i} \chi^+_{e_j}$, or
a charged Higgs boson ($H^\pm_m,\, m=1,2$), then
$\overline{\chi}_{e_i} \chi_{e_j}=\overline{\chi}^+_{e_i} \chi^0_{e_j}$.
Note, however, that our results apply as well to the $CP$-violating MSSM 
where the neutral Higgs particles are no longer $CP$ eigenstates. 
For charged Higgs-boson exchange the hermitian conjugate of (\ref{eq:higgscoup})
has to be considered as well, which introduces the hermitian conjugates of the 
coupling matrices, $(s^{H_m^+\dagger})_{ij}$ and $(p^{H_m^+\dagger})_{ij}$.
Explicit expressions for all the $v_{ij},\,a_{ij},\,s_{ij},\,p_{ij}$
couplings are given in the Appendix~\ref{sec:appendixpot}. 

%
%

The amplitude of Fig.~\ref{fig:pot_matching}a with $Z$-boson exchange 
is obtained at leading order in the non-relativistic expansion by 
taking the limit of small relative momenta,
$\vec{p},\vec{p}^{\,\prime}\sim m_{\rm LSP} v_\text{rel}\sim M_Z$
and $p^0,p^{\prime\,0}\sim m_{\rm LSP} v_\text{rel}^2$,
in the particle spinors and $Z$-boson propagator,
\begin{align}
\frac{-i}{q^2-M_Z^2}\left( g_{\mu\nu} - \frac{q_\mu q_\nu}{M_Z^2} \right) + 
\frac{q_\mu q_\nu}{M_Z^2} \, \frac{-i}{q^2-\xi M_Z^2} 
\,,
\label{eq:Zprop}
\end{align}
here defined in the $R_\xi$-gauge. 
At leading-order, the non-relativistic expansion of the product of Dirac bilinears in
the amplitude of diagram~\ref{fig:pot_matching}a,
$\bar{u}(p_4)\Gamma^\mu u(p_1)\,
\bar{u}(p_3)\tilde{\Gamma}_{\mu} u(p_2)$,
is equivalent to the replacements ($q=p_4-p_1=p_2-p_3$)
\begin{gather}
  \gamma^\mu \otimes \gamma_\mu \to \mathbf{1} \otimes  \mathbf{1} \ \ , \ \
  \gamma^\mu \gamma_5 \otimes \gamma_\mu \gamma_5\to  - \,\sigma^i \otimes  \sigma^i 
\ , \nonumber \\*
  \gamma^\mu \gamma_5 \otimes \gamma_\mu \,,\, \gamma^\mu  \otimes \gamma_\mu \gamma_5\to 0
\ , \nonumber \\*
  \slash{q} \otimes \slash{q} \to -(m_{e_4}-m_{e_1})(m_{e_3}-m_{e_2}) \, \mathbf{1} \otimes  \mathbf{1} 
\ ,
\nonumber \\*
  \slash{q} \otimes \slash{q} \gamma_5 \,,\, \slash{q}\gamma_5 \otimes \slash{q} \,,\,
  \slash{q} \gamma_5\otimes \slash{q} \gamma_5 \to 0
\,,
\label{eq:NRbilinears}
\end{gather}
in the full-theory amplitude, where the right-hand-side of these relations 
should be understood as the matrices acting on the two-component Pauli spinors of the non-relativistic neutralinos and 
charginos at the upper and lower interaction vertices of Fig.~\ref{fig:pot_matching}a.
Written in components, the spin operators above read 
$\sigma^i \otimes  \sigma^i \equiv \sigma^i_{\alpha_4 \alpha_1} \sigma^i_{\alpha_3 \alpha_2}$
and  $\mathbf{1} \otimes  \mathbf{1} \equiv \delta_{\alpha_4 \alpha_1} \delta_{\alpha_3 \alpha_2}$.
The use of the non-relativistic normalization for the relativistic spinors,
$u^\dagger(p)u(p)=1$, is implied in the replacements (\ref{eq:NRbilinears}).
The relation in the third line of (\ref{eq:NRbilinears}) can be obtained using the equation of motion for the 
relativistic spinors; we comment on the non-relativistic scaling of this term 
below. Using (\ref{eq:NRbilinears}) we thus obtain 
\begin{align}
({\rm Fig. 3a})^Z  \, = \ &  
\frac{-ig_2^2}{\vec{q}^{\,2}+M_Z^2} \, 
\left[ \, \Big( 1+ \frac{\delta m_{e_4 e_1}\,\delta m_{e_3 e_2}}{M_Z^2} \Big)
      v^Z_{e_4 e_1} v^Z_{e_3 e_2} \, \mathbf{1} \otimes  \mathbf{1} 
     -a^Z_{e_4 e_1} a^Z_{e_3 e_2} \, \sigma^i \otimes  \sigma^i  \right]
\nonumber\\
& + \frac{ig_2^2}{\vec{q}^{\,2}+\xi M_Z^2} \, \frac{\delta m_{e_4 e_1}\,\delta m_{e_3 e_2}}{M_Z^2} 
\, v^Z_{e_4 e_1} v^Z_{e_3 e_2} \, \mathbf{1} \otimes  \mathbf{1} 
\; 
\label{eq:potfiga}
\end{align}
for the leading-order term 
in the expansion in $\vec{p},\vec{p}^{\,\prime}\sim m_{\rm LSP} 
v_\text{rel}\sim M_Z$ and $p^0,p^{\prime\,0}\sim m_{\rm LSP} v_\text{rel}^2$
of the amplitude in Fig.~\ref{fig:pot_matching}a
when the exchanged particle is a $Z$-boson. For simplicity we have considered the case where the couplings in both vertices are equal,
which applies to $\chi^0\chi^0\to \chi^0\chi^0$ and $\chi^+\chi^+\to \chi^+\chi^+$ processes.

The term in the second line of (\ref{eq:potfiga}) arises from the $\xi$-dependent term of the $Z$-boson propagator (\ref{eq:Zprop}). 
Gauge invariance requires that this term
is cancelled against the contribution from the exchange of the Goldstone boson $A_2^0\equiv G^0$,
which has a mass $\xi M_Z^2$. The cancellation holds because certain 
conditions among the vector and scalar couplings of the neutralinos and charginos are fulfilled
in the MSSM. To see this, write the leading-order contribution 
corresponding to diagram \ref{fig:pot_matching}a with scalar-boson $\phi$ exchange by expanding 
the scalar propagator, $i/(q^2-m_\phi^2)$, and the Dirac bilinear in the non-relativistic 
limit; the result  reads
\begin{align}
({\rm Fig. 3a})^\phi  \, = \ &  
\frac{ig_2^2}{\vec{q}^{\,2}+m_\phi^2} \, 
      s^\phi_{e_4 e_1} s^\phi_{e_3 e_2} \, \mathbf{1} \otimes  \mathbf{1} 
\; .
\label{eq:potfigaphi}
\end{align}
The pseudoscalar interactions do not survive in (\ref{eq:potfigaphi}) because  
$\mathbf{1}\otimes \gamma_5 \,,\, \gamma_5\otimes \mathbf{1}\,,\,\gamma_5\otimes\gamma_5\to 0$
at the leading order.
Adding (\ref{eq:potfigaphi}) for $\phi=G^0$ ($m_{G^0}=\xi M_Z$) to  
(\ref{eq:potfiga}) one obtains the condition 
\begin{align}
\frac{\delta m_{e_4 e_1}\,\delta m_{e_3 e_2}}{M_Z^2} \, v^Z_{e_4 e_1} v^Z_{e_3 e_2}
+
s^{G^0}_{e_4 e_1} s^{G^0}_{e_3 e_2} = 0
\label{eq:gaugecancel}
\end{align}
for the cancellation of the $\xi$-dependent terms. Eq.~(\ref{eq:gaugecancel}) 
holds provided that $v^Z_{ij}\,(m_i-m_j)/M_Z = \pm i s^{G^0}_{ij}$, a 
relation which can
be proven using the explicit definition of the couplings in terms of the
mixing matrices given in Appendix~\ref{sec:appendixpot} and their diagonalization properties.  
We can therefore drop the second line in (\ref{eq:potfiga}) together 
with the pseudo-Goldstone contribution to the potential.

The term proportional to the mass
differences in the first line of the potential~(\ref{eq:potfiga})  
typically yields a very small contribution, since in the EFT
we treat only those species for which
$\delta m\sim {\cal O}(m_{\rm LSP} v^2)$. For a very heavy
LSP, where we could nevertheless have $m_{\rm LSP} v^2\gg M_Z$, 
the size of this term
would still be at most of ${\cal O}(1)$ due to suppressed vector couplings.
To see this, consider the decoupling limit $m_{\rm LSP}\to\infty$. If the mass difference $\delta m_{ij}$
refers to particles within the same electroweak multiplet then $\delta m_{ij}\sim M_{\rm EW}^2/m_{\rm LSP}$
and the mass-difference terms in (\ref{eq:potfiga}) are suppressed as $\delta m_{ij}/M_Z \sim M_{\rm EW}/m_{\rm LSP}$.
If particles $i$ and $j$ belong to different multiplets, 
$\delta m_{ij}/M_Z$ can be large, but is multiplied 
by $v^Z_{ij} \sim M_{\rm EW}/\delta m_{ij}$, as follows 
(\ref{eq:gaugecancel}). 
Since the axial couplings go to zero in the decoupling limit, it 
follows that the gauge-independent, off-diagonal terms in square brackets in 
the first line of (\ref{eq:potfiga}) are given in this limit by the 
pseudo-Goldstone 
couplings $-s^{G^0}_{e_4 e_1} s^{G^0}_{e_3 e_2}$.

The corresponding results for diagrams $b$ and $c$ in Fig.~\ref{fig:pot_matching} can then be easily obtained from 
(\ref{eq:potfiga}): for every line where the fermion flow has been  reversed 
we need to interchange the labels in the vertex couplings (for instance,
$X_{e_3 e_2}\to X_{e_2 e_3}$to go from $a$ to $b$), change the sign of the scalar, pseudo-scalar and axial-vector couplings
(which arises when writing the antiparticle spinors as the charge-conjugate of particle ones), and
add a global sign due to Wick ordering. For the sake of clarity, let us write the results for $Z$-
and neutral Higgs-boson exchange for diagram $b$: 
\begin{align}
({\rm Fig. 3b})^Z  \, = \ &  
\frac{ig_2^2}{\vec{q}^{\,2}+M_Z^2} \, 
\left[ \, \Big( 1+ \frac{\delta m_{e_4 e_1}\,\delta m_{e_3 e_2}}{M_Z^2} \Big)
      v^Z_{e_4 e_1} v^Z_{e_2 e_3} \, \mathbf{1} \otimes  \mathbf{1} 
     +a^Z_{e_4 e_1} a^Z_{e_2 e_3} \, \sigma^i \otimes  \sigma^i  \right]
\nonumber\\
& - \frac{ig_2^2}{\vec{q}^{\,2}+\xi M_Z^2} \, \frac{\delta m_{e_4 e_1}\,\delta m_{e_3 e_2}}{M_Z^2} 
\, v^Z_{e_4 e_1} v^Z_{e_2 e_3} \, \mathbf{1} \otimes  \mathbf{1} 
\; ,
\nonumber\\
({\rm Fig. 3b})^\phi \, = \ &
\frac{ig_2^2}{\vec{q}^{\,2}+m_\phi^2} \, 
      s^\phi_{e_4 e_1} s^\phi_{e_2 e_3} \, \mathbf{1} \otimes  \mathbf{1} 
\; .
\label{eq:potfigb}
\end{align}

The potentials generated by photon exchange can be obtained from the result for the $Z$-boson
case by plugging the corresponding vector coupling and keeping only the $g_{\mu\nu}$ part
of (\ref{eq:Zprop}) with $M_Z\to 0$ for the photon propagator. $W$ and charged Higgs-boson
exchange are relevant for the processes $\chi^0\chi^0\to \chi^\pm\chi^\mp$ and 
$\chi^0\chi^\pm\to \chi^\pm\chi^0$. For these amplitudes, we have
to hermitian conjugate the coupling matrices at vertices involving either an initial $\chi^+$ or a final $\chi^-$.
The potential
from charged boson exchange can be obtained from the neutral boson amplitudes
with trivial replacements in the couplings and the boson mass. 

The amplitudes above provide the potentials in momentum space. 
However, the Schr\"odinger equation which sums the all-order exchange of 
bosons among the chargino and neutralino pairs acquires a simpler form in 
coordinate space. The coordinate-space potentials are obtained by taking 
the Fourier transform 
\begin{align}
V^{\chi\chi \to \chi\chi}_{ \lbrace e_1 e_2\rbrace 
\lbrace e_4 e_3\rbrace }(r) = 
\int \frac{d^3\vec{q}}{(2\pi)^3} \, e^{i\vec{q}\cdot \vec{x}} \,
i\,T^{\chi\chi \to \chi\chi}_{e_1 e_2 e_4 e_3}(\vec{q}^{\,2})
\,,
\label{eq:fourier}
\end{align}
where $r\equiv|\vec{x}\,|$, and $T^{\chi\chi \to \chi\chi}_{e_1 e_2 e_4 e_3}$ 
stands for the momentum-space amplitude as given above 
in~(\ref{eq:potfiga},\ref{eq:potfigaphi},\ref{eq:potfigb}). From the identity 
\begin{align}
\int \frac{d^3\vec{q}}{(2\pi)^3} \, e^{i\vec{q}\cdot \vec{x}} \,\frac{1}{\vec{q}^{\,2}+m^2} = \frac{e^{-m r}}{4\pi r}
\,,
\label{eq:yukawa}
\end{align}
we immediately obtain the well-known Yukawa-like potential for amplitudes with exchange of a force carrier of mass $m$. 
Applied to the massless case, (\ref{eq:yukawa}) gives the Coulomb
potential.\footnote{The $+i\epsilon$ prescription
in the full-theory gauge boson propagators provides the necessary regularization for the $m=0$ case.} 
Before we write the result for the potentials in coordinate space,
let us write the spin operator $\sigma^i \otimes  \sigma^i$ 
in terms of the total spin operator as 
\begin{align}
\sigma^i \otimes  \sigma^i  = 2 ( \vec{S}^{\,2} - \vec{s}^{\,2}_1 - \vec{s}^{\,2}_2 )
= 2 \vec{S}^2 - 3\,(\mathbf{1} \otimes \mathbf{1})
\; ,
\label{eq:totspin}
\end{align}
where $\vec{S}=\vec{s}_1+\vec{s}_2 \equiv 1/2 \, ( \vec{\sigma} \otimes \mathbf{1} + \mathbf{1}  \otimes \vec{\sigma} )$,
with $\vec{s}_{1,2}$  the spin operators
acting on the particles 
at the upper and lower vertices in the scattering diagram, and we have replaced $\vec{s}^{\,2}_{1,2}$ by
$s(s+1)\,(\mathbf{1} \otimes \mathbf{1})=3/4\,(\mathbf{1} \otimes \mathbf{1})$ 
for 
charginos and neutralinos. Working in the basis of eigenstates of total spin for the neutralino and chargino pairs,
we can as well replace $\vec{S}^2$ by $S(S+1)\,(\mathbf{1} \otimes \mathbf{1})=2S\,(\mathbf{1} \otimes \mathbf{1})$
for $S=0,1$ in the potentials. Since the operator $\mathbf{1} \otimes  \mathbf{1}$ is the identity operator, 
we find that  a leading-order potential contribution in the basis of total spin
and in coordinate space thus has the form
\begin{align}
V^{\chi\chi \to \chi\chi}_{ \lbrace e_1 e_2\rbrace \lbrace e_4 e_3\rbrace }(r) = 
  \left( a_{e_1 e_2 e_4 e_3}
  - (3-4S) \, b_{e_1 e_2 e_4 e_3}  \right)
  \frac{e^{-m_X r}}{r}\,,
\label{eq:potstructure}
\end{align}
for the case of the exchange of a vector boson with mass $m_X$, among the incoming and outgoing $\chi\chi$ pairs. 
For leading-order scalar boson and photon exchange, the coefficient $b_{e_1 e_2 e_4 e_3}$ vanishes. 
For $Z$- and $W$-boson exchange the spin-dependent part of the potential arises from the axial-vector coupling, 
see (\ref{eq:potfiga}), (\ref{eq:potfigb}).

We show in Sec.~\ref{sec:Sommerfeld} by analysing the perturbative expansion 
of the annihilation amplitude $\chi_i\chi_j\to X_A X_B$ with ladder-like 
exchanges in the non-relativistic limit,  that the amplitude can be described 
as a product of potential interactions like (\ref{eq:potstructure}) 
times the two-particle propagator of the internal pairs $\chi_m\chi_n$ in the diagrams, integrated over 
corresponding loop momenta, and finally multiplied by the short-distance coefficient $\hat{f}$ with
the appropriate quantum numbers. No additional combinatoric factors arise as long as for the internal states 
we consider as different the states $\chi_{e_1}\chi_{e_2}$ and  $\chi_{e_2}\chi_{e_1}$ and sum over all of them.
Diagrammatically this means for instance
that we have to consider a graph like Fig.~\ref{fig:potential} and the one where particles 
$\chi_{e_4},\chi_{e_3}$ are interchanged. 
At this point, it is convenient to 
switch from the two indices $\{e_1,e_2\}$ that denote a two-particle state $\chi_{e_1}\chi_{e_2}$ 
to a single label $m=1,\dots N_{|Q|}$, where $N_{|Q|}$ 
is the total number of channels for each total charge sector, $|Q|=0,1,2$. 
For
the case of $n_0$ neutralinos and $n_+$ charginos, we have $n_0^2$ $\chi^0\chi^0$ 
 and $2 n_+^2$ $\chi^{\pm}\chi^{\mp}$ neutral states, $2 n_0 n_+$ different states with $Q=\pm 1$ 
({\it i.e.} of the type $\chi^0\chi^\pm$ or $\chi^\pm\chi^0$),
and $n_+^2$ $\chi^{\pm}\chi^{\pm}$ states of charge $Q=\pm2$. Therefore, the total number of neutral states
is  $N_0=n_0^2 + 2 n_+^2$, whereas $N_1=2 n_0 n_+$ and $N_2=n_+^2$. If all four neutralinos and the two charginos
are relevant for the long-distance part of the annihilation ($n_0=4$ and  $n_+=2$), then we have to consider 
the interactions among $N_0=24$ states in the neutral sector, as well as those among
$N_1=16$ and $N_2=4$ states in the singly-charged and doubly-charged sectors, respectively.
For instance,
in the charge-0 sector with all four neutralinos and two charginos, 
the single label runs over states 
\begin{equation}
\chi^0_1 \chi^0_1,\,\chi^0_1 \chi^0_2,\,\chi^0_2 \chi^0_1,\dots,\,\chi^0_4 \chi^0_3,\,\chi^0_4 \chi^0_4,\,
\chi^\pm_1\chi^\mp_1,\,\chi^\pm_1\chi^\mp_2,\,\chi^\pm_2\chi^\mp_1,\,\chi^\pm_2\chi^\mp_2 \quad, 
\label{eq:neutralstates}
\end{equation}
with 24 different states in total, whereas in the charge-1 sector we have 16 channels,
\begin{equation}
\chi^0_1 \chi^\pm_1,\,\chi^\pm_1\chi^0_1,\,\chi^0_1 \chi^\pm_2,\dots,\,\chi^0_4\chi^\pm_2,\,\chi^\pm_2\chi^0_4 \quad, 
\label{eq:singlychargedstates}
\end{equation}
and just 4 in the charge-2 sector,
\begin{equation}
 \chi^\pm_1\chi^\pm_1,\,\chi^\pm_1\chi^\pm_2,\,\chi^\pm_2 \chi^\pm_1,\,\chi^\pm_2\chi^\pm_2 \quad. 
\label{eq:doublychargedstates}
\end{equation}
The ordering of the states in each sector is of course a matter of convention.

An alternative basis of two-particle states with a smaller number of states 
can be used, which takes into account the fact that states 
$\chi_{e_1}\chi_{e_2}$ and $\chi_{e_2}\chi_{e_1}$ have identical particle content. The new basis, that we shall refer
to as ``method-2 basis'' in order to distinguish it from the basis just described above,
is built only by states $\chi_{e_1}\chi_{e_2}$ with  $e_1\le e_2$,
provided we identify
$\chi_{i}=\chi_i^0$ for $i=1,\dots4$ and $\chi_{5}=\chi_1^+$,  $\chi_{6}=\chi_2^+$
$\chi_{7}=\chi_1^-$,  $\chi_{8}=\chi_2^-$.
The two potential contributions of method-1 corresponding to  
$\chi_{e_1}\chi_{e_2}\to \chi_{e_4}\chi_{e_3}$ and $\chi_{e_1}\chi_{e_2}\to \chi_{e_3}\chi_{e_4}$ 
(the so-called ``crossed'' contribution, different from the former if $\chi_{e_3}\ne \chi_{e_4}$) 
are accounted for in method-2 by a single potential entry describing the scattering of state
$(\chi\chi)_{e_1e_2}$ into $(\chi\chi)_{e_4 e_3}$, where we have 
introduced the notation $(\chi\chi)_{\dots}$ to denote a state in the new
basis, and assumed that $e_1 \le e_2$ and $e_4 \le e_3$. The potential 
entries for method-2,
$V^{(\chi\chi) \to (\chi\chi)}_{ \lbrace e_1 e_2\rbrace \lbrace e_4 e_3\rbrace}$,
are obtained from those of method-1 as
\begin{align}
{e_1\ne e_2\;{\rm and}\; e_4\ne e_3 } \,: &\quad  V^{(\chi\chi) \to (\chi\chi)}_{ \lbrace e_1 e_2\rbrace \lbrace e_4 e_3\rbrace}  \, = \  
V^{\chi\chi \to \chi\chi}_{ \lbrace e_1 e_2\rbrace \lbrace e_4 e_3\rbrace} 
+ (-1)^{L+S} \, V^{\chi\chi \to \chi\chi}_{ \lbrace e_1 e_2\rbrace \lbrace e_3 e_4\rbrace} 
\;,\nonumber\\
{ e_1\ne e_2\;{\rm and}\; e_4 =  e_3 } \,: &\quad V^{(\chi\chi) \to (\chi\chi)}_{ \lbrace e_1 e_2\rbrace \lbrace e_4 e_4\rbrace} \, = \ 
\sqrt{2} \, V^{\chi\chi \to \chi\chi}_{ \lbrace e_1 e_2\rbrace \lbrace e_4 e_4\rbrace} \;,
\nonumber\\
{e_1= e_2\;{\rm and}\; e_4\ne e_3} \,: &\quad V^{(\chi\chi) \to (\chi\chi)}_{ \lbrace e_1 e_1\rbrace \lbrace e_4 e_3\rbrace} \, = \ 
\sqrt{2} \, V^{\chi\chi \to \chi\chi}_{ \lbrace e_1 e_1\rbrace \lbrace e_4 e_3\rbrace} \;,
\nonumber\\
{e_1= e_2\;{\rm and}\; e_4 = e_3} \,:&\quad V^{(\chi\chi) \to (\chi\chi)}_{ \lbrace e_1 e_1\rbrace \lbrace e_4 e_4\rbrace} \, = \ 
V^{\chi\chi \to \chi\chi}_{ \lbrace e_1 e_1\rbrace \lbrace e_4 e_4\rbrace} \;.
\label{eq:potmethods}
\end{align}
The $(-1)^{L+S}$ in front of
the crossed diagram amplitude in the first line of (\ref{eq:potmethods}) 
arises from the product $(-1)\times(-1)^L\times(-1)^{S+1}$,
where the $(-1)$ comes from Wick ordering, and the exchange $e_3 \leftrightarrow e_4$ 
produces a factor $(-1)^{S+1}$ in the spin wave function and a change of sign 
in the relative momentum which translates into the factor $(-1)^L$.
An additional rule has to be 
accounted for when building the potential matrix for method-2: since identical spin-1/2
particles cannot form a 2-particle state with odd $L+S$ ({\it i.e.} 
with quantum numbers ${}^3S_1$ and ${}^1P_1$ in the case at hand), 
the entries involving such states for the potential when $L+S$ is odd have to be set to zero.
This prevents that method-2 yields a non-zero annihilation amplitude for a 
forbidden state of two identical neutralinos or charginos 
through an intermediate transition to an allowed state, such as
$(\chi^0\chi^0)_{11} \to (\chi^+\chi^-)_{11} \to \mbox{light particles}$. 
In method-1 this transition is automatically zero because
there are two annihilation terms,  $\chi^+_1\chi^-_1 \to \mbox{light}$ 
and $\chi^-_1\chi^+_1 \to \mbox{light}$, which cancel each other 
by virtue of the symmetry properties of the 
Wilson coefficients $\hat{f}$ under exchange of the particle labels, see
(8) and (6) in~\cite{Beneke:2012tg} and~\cite{Hellmann:2013jxa}, respectively.
We note that the the relations (\ref{eq:potmethods}) and the selection rule just mentioned imply 
that the potential in method-2 depends on the orbital angular momentum $L$ even
at the leading-order in the non-relativistic expansion.

Likewise, the annihilation matrices built from the Wilson coefficients of the 
four-fermion operators  must be supplemented with
additional factors in the method-2 basis.
Namely, the absorptive part of the one-loop annihilation amplitude 
$(\chi\chi)_{e_1 e_2} \to (\chi\chi)_{e_4 e_3}$
involving method-2 states in a ${}^{2S+1}L_J$ 
partial-wave configuration is given by the Wilson coefficient 
$\hat f^{\chi\chi \to \chi \chi}_{\lbrace {e_1 e_2} \rbrace \lbrace {e_4 e_3} \rbrace}(^{2S+1}L_J)$
multiplied by $(1/\sqrt{2})^{n_{id}}$, where $n_{id}=1,2$ if the two-particle 
states $(\chi\chi)_{e_1 e_2}$ or/and $(\chi\chi)_{e_4 e_3}$ are formed of 
identical particles, and $n_{id}=0$ otherwise. 

The basis of states in method-2 for the case of $n_0$ neutralinos and 
$n_+$ charginos, contains $\tilde{N}_0=n_0(n_0+1)/2 +  n_+^2$
neutral states, $\tilde{N}_1=n_0 n_+$ states with $Q=\pm 1$,
and $\tilde{N}_2=n_+(n_+ +1)/2$ states with charge $Q=\pm2$.
The reduction in the number of states in method-2 as compared to method-1
is more significant as more particle species are considered in the 
non-relativistic EFT framework. For instance,
if all neutralino and chargino species are considered then
$(\tilde{N}_0,\tilde{N}_1,\tilde{N}_2)=(14,8,3)$, 
whereas for method-1 $(N_0,N_1,N_2)=(24,16,4)$.
The list of method-2 states can be read off
from (\ref{eq:neutralstates},\ref{eq:singlychargedstates},\ref{eq:doublychargedstates})
by dropping states that differ only in $\chi_{i}\chi_{j}$ where $i>j$.
In what follows, quantities with indices referring to two-particle states, such as the 
Wilson coefficients that built the annihilation matrices and the potentials, 
will be written using the single-label notation with lower-case
letters, 
{\it i.e.} $\hat{f}_{mn}({}^{2S+1}L_J)$ and $V_{mn}(r)$, 
with $m,n=1,\dots N_{|Q|}$ in the case of method-1, and
$m,n=1,\dots \tilde{N}_{|Q|}$ for method-2.
We shall also adopt the convention that the potential matrix element $V_{mn}$ 
describes the scattering $m\to n$, whereas
for the annihilation matrices, the entry $\hat{f}_{mn}({}^{2S+1}L_J)$ is the  
absorptive part of the one-loop amplitude for $m\to n$.
The formula for the  Sommerfeld enhancement factor (\ref{eq:SFdef}) using this notation reads
\begin{eqnarray}
S_{a}[\hat f(^{2S+1}L_J) ] = 
\frac{ 
 \left[ \psi^{(L,S)}_{ba}\right]^*
     \hat f_{cb}(^{2S+1}L_J)
    \, \psi^{(L,S)}_{ca}}
{ \hat f_{aa}(^{2S+1}L_J)|_{\rm LO}}
\ , 
\end{eqnarray}
for an incoming two-particle state $a$, and the same equation holds in 
both methods.
In Appendix~\ref{sec:appendixmethodIvsII} we demonstrate explicitly for a
simplified example that method-1 and method-2 yield the same Sommerfeld factors.

To provide an example, let us give both the potential and the 
annihilation matrices in the well-known wino limit of the MSSM. The relevant 
particles in this limit are the pure-wino lightest neutralino
$\chi_1^0$, with mass $m_\chi$, and its mass-degenerate chargino partners 
$\chi^\pm_1$. In the neutral sector,
the potential matrix of method-1 for both $S=0,1$ 
derives from the formulae given in Appendix~\ref{sec:appendixpot}
after plugging in the values of the  neutralino and chargino mixing matrix entries relevant to the calculation in
the pure-wino, {\it i.e.} $Z_{N\, i1} \ = \ \delta_{i2}$ and $Z_{\pm\, i1} \ = \ \delta_{i1}$:
\begin{align}
V^{(1)}_{Q=0}(r)  \ &= \
  \left(\begin{array}{ccc}
  0 &  \quad - \alpha_2 \,\frac{e^{-M_W r}}{r} &  \quad - \alpha_2 \,\frac{e^{-M_W r}}{r}  \\
  - \alpha_2 \,\frac{e^{-M_W r}}{r} & \quad  -\frac{\alpha}{r} - \alpha_2 \,c_W^2 \,\frac{e^{-M_Z r}}{r} & 0 \\
  - \alpha_2 \,\frac{e^{-M_W r}}{r} & \quad 0 & 
      \quad  -\frac{\alpha}{r} - \alpha_2 \,c_W^2 \,\frac{e^{-M_Z}r}{r} 
  \end{array}\right) \ ,
\label{eq:winopotI}
\end{align}
where the matrix indices ($m,n=1,2,3$)
correspond to channels $\chi_1^0\chi_1^0,\, \chi_1^+\chi_1^-,\, \chi_1^-\chi_1^+$.
The potential for method-2 then follows from the rules given in (\ref{eq:potmethods}) together
with the rule that sets to zero entries involving a forbidden ${}^3S_1$ or ${}^1P_1$ $\chi^0_1\chi^0_1$ state: 
\begin{align}
V^{(2)}_{Q=0, {\rm even} \,L+S}(r)  \ &= \
  \left(\begin{array}{cc}
  0 &  \quad - \sqrt{2}\,\alpha_2 \,\frac{e^{-M_W r}}{r}  \\
  - \sqrt{2}\,\alpha_2 \,\frac{e^{-M_Wr}}{r} & \quad  -\frac{\alpha}{r} - \alpha_2 \,c_W^2 \,\frac{e^{-M_Z r}}{r} 
  \end{array}\right) \ ,
 \label{eq:winopotIIeven}
  \\
V^{(2)}_{Q=0,{\rm odd}\,L+S }(r)  \ &= \
  \left(\begin{array}{cc}
  0 &  0  \\
 0 & \quad  -\frac{\alpha}{r} - \alpha_2 \,c_W^2 \,\frac{e^{-M_Z r}}{r} 
  \end{array}\right) \ ,  
\label{eq:winopotIIodd}
\end{align}
where the matrix indices $m,n=1,2$ refer now to channels
$\chi_1^0\chi_1^0,\,\chi_1^+\chi_1^-$. The annihilation matrices for any 
MSSM model to ${\cal O}(v_{\rm rel}^2)$
can be obtained from the analytic formulae given in~\cite{Beneke:2012tg,Hellmann:2013jxa}. 
In the pure-wino limit, the explicit expressions for the absorptive parts of the Wilson coefficients
have been worked out in detail in Appendix~C of~\cite{Hellmann:2013jxa}, and can be read off from (147) and Tables~5-7
therein. In the neutral sector, the leading order $S$-wave annihilation matrices read
\begin{align}
[\hat f(^{1}S_0)]^{(1)}    \; &= \; \frac{\pi \alpha_2^2}{m_\chi^2}
  \left(\begin{array}{rrr}
  2 & 1 &  1  \\
  1 & \frac{3}{2}& \frac{3}{2} \\
  1 & \frac{3}{2} &  \frac{3}{2} 
  \end{array}\right) \quad , \quad
[\hat f(^{3}S_1)]^{(1)}    \; = \; \frac{25}{24}\,\frac{\pi \alpha_2^2}{m_\chi^2}
  \left(\begin{array}{rrr}
  0 &  0 &  0  \\
  0 &  1 & -1 \\
  0 &  -1 &  1
  \end{array}\right) \ , 
\label{eq:winoGammaSwaveI}
\end{align}
in method-1, and 
\begin{align}
[\hat f(^{1}S_0)]^{(2)}    \; &= \; \frac{\pi \alpha_2^2}{m_\chi^2}
  \left(\begin{array}{cc}
  1 &  \frac{1}{\sqrt{2}}   \\
  \frac{1}{\sqrt{2}} & \frac{3}{2} 
  \end{array}\right) \quad , \quad
[\hat f(^{3}S_1)]^{(2)}    \; = \;  \frac{25}{24} \,\frac{\pi \alpha_2^2}{m_\chi^2}
  \left(\begin{array}{rr}
  0 &  0  \\
  0 &  1
  \end{array}\right) \ , 
\label{eq:winoGammaSwaveII}
\end{align}
in method-2. The results (\ref{eq:winoGammaSwaveII}) have been given before in \cite{Hisano:2006nn}.
Let us also write the $P$-wave annihilation matrices in both methods:
\begin{align}
\Big[\frac{\hat f(^{1}P_1)}{M^2}\Big]^{(1)}   \; &= \; 
\frac{1}{6}\,\frac{\pi \alpha_2^2}{m_\chi^4}
  \left(\begin{array}{rrr}
  0 & 0 &  0  \\
  0 & 1& -1 \\
  0 & -1 &  1
  \end{array}\right) \quad , \quad
\Big[\frac{\hat f(^{3}P_{\cal J})}{M^2}\Big]^{(1)}    \; = \; 
\frac{7}{3}\,\frac{\pi \alpha_2^2}{m_\chi^4}
  \left(\begin{array}{rrr}
  2 &  1 &  1  \\
  1 &  \frac{3}{2} & \frac{3}{2} \\
  1 &  \frac{3}{2} &  \frac{3}{2}
  \end{array}\right) \ , 
\label{eq:winoGammaPwaveI}
\end{align}
\begin{align}
\Big[\frac{\hat f(^{1}P_1)}{M^2}\Big]^{(2)}    \; &= \; 
\frac{1}{6}\,\frac{\pi \alpha_2^2}{m_\chi^2}
  \left(\begin{array}{rr}
  0 & 0   \\
  0 & 1
  \end{array}\right) \quad , \quad
\Big[\frac{\hat f(^{3}P_{\cal J})}{M^2}\Big]^{(2)}     \; = \; 
\frac{7}{3}\,\frac{\pi \alpha_2^2}{2 m_\chi^2}
  \left(\begin{array}{rr}
  1 &  \frac{1}{\sqrt{2}}  \\
  \frac{1}{\sqrt{2}} &  \frac{3}{2}
  \end{array}\right) \ .
\label{eq:winoGammaPwaveII}
\end{align}
Recall that the entries in the annihilation matrices of method-1 
which differ only in the replacement $\chi_1^+\chi_1^- \leftrightarrow 
\chi_1^-\chi_1^+$ as in or out state  
are equal up to a factor $(-1)^{L+S}$, which arises as a consequence of the
exchange of labels in the Wilson coefficients (see Eqs.~(2.7) and~(2.4) 
of~\cite{Beneke:2012tg,Hellmann:2013jxa}, respectively).
The potential and annihilation matrices for the other charge
sectors can be obtained similarly.


\section{Sommerfeld enhancement}
\label{sec:Sommerfeld}

In this section we discuss the computation of the Sommerfeld factors 
$S_{ij}[\hat f(^{2S+1}L_J)]$ defined in (\ref{eq:SFdef}). Within the 
non-relativistic MSSM the Sommerfeld enhancement arises from the 
matrix elements of annihilation operators such as
$\mathcal O^{\chi \chi \to \chi \chi }_{ \lbrace e_4 e_3\rbrace 
\lbrace e_2 e_1\rbrace }\!\left( {}^{2S+1}L_J \right)$. They receive 
large quantum corrections, which have to be summed to all orders. 
Generalizing (\ref{eq:wave}), we determine $\psi^{(L,S)}_{e_1 e_2, \,ij}$ by 
computing the left-hand side of 
\begin{equation}
\langle 0| \chi^{c \dagger}_{e_2}\Gamma 
K\big[-\frac{i}{2}\overleftrightarrow{\bff{\partial}}\big]\chi^{}_{e_1}
| \chi_i \chi_j \rangle = 
\,\langle \xi^{c \dagger}_{j} \Gamma \xi_{i} \rangle \, 
K[\vec{p}\,]\,
\Big( \psi^{(L,S)}_{e_1 e_2, \,ij} + (-1)^{L+S}\, \psi^{(L,S)}_{e_2 e_1, \,ij} \Big)
\, . 
\label{eq:medef}
\end{equation}
Here $\Gamma = 1_{2 \times 2}$, $\vec{\sigma}$ for $S=0,1$, respectively, 
and $K$ is a polynomial in relative momentum (derivatives) corresponding 
to a given angular momentum $L$. $\vec{p}\,$ denotes the relative 
momentum of the $\chi_i$ and $\chi_j$ particle as defined below 
in (\ref{eq:defE1}). With this definition $\psi^{(L,S)}_{e_1 e_2, \,ij}$ 
is normalized to one, when the Sommerfeld effect is neglected, 
and the matrix element is evaluated in the tree approximation.

In the first part of this section, we establish the relation between 
the non-relativistic (NR) matrix element above, diagrammatic resummation 
and the solution of a multi-channel Schr\"odinger equation. We then 
describe the actual solution of the Schr\"odinger equation. We shall 
find that the standard method fails to provide an accurate result 
in many relevant cases with kinematically closed channels and derive 
an alternative method that solves this problem. This is our main 
result, since it allows to compute the Sommerfeld factors 
for general MSSM parameter points. In practice, the solution with 
the full number of channels is time-consuming and often not relevant, 
since the heavier channels have only a small effect on the final 
result. In the final part of this section we therefore describe an 
approximation to the treatment of heavier channels, which is accurate 
and fast for most practical purposes.

\subsection{NR matrix elements and the Schr\"odinger equation}

%
\begin{figure}[t]
\begin{center}
\includegraphics[width=0.6\textwidth]{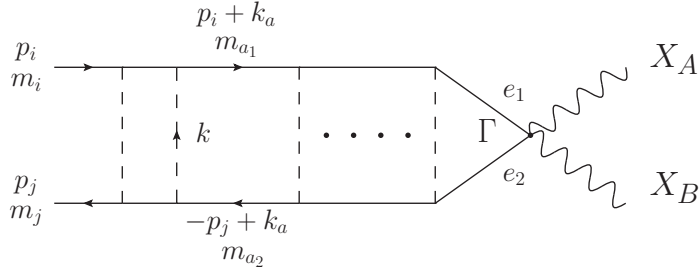}
\caption{Diagrammatic representation of the leading (ladder) 
contributions to the non-relativistic matrix element (\ref{eq:medef}).}
\label{fig:coulombladder}
\end{center}
\end{figure}
%

The enhancement of non-relativistic scattering originates from 
the potential loop momentum region, when the momentum $k$ of the 
exchanged virtual particle satisfies $k^0\ll |\vec{k}\,|\ll \mu$, where 
$\mu$ is the reduced mass of the two non-relativistic particles. It can 
be shown by power-counting arguments, see for 
instance \cite{Beneke:2013jia}, that no other region requires 
resummation, and that the enhancement from the potential region 
is only present in ladder diagrams. 
Relating the sum of ladder diagrams shown in Fig.~\ref{fig:coulombladder} 
to the solution of a Schr\"odinger equation involves a number of 
steps, which we now sketch. 

The first consists of kinematic simplifications of the propagators 
in the potential region. We already mentioned that the $(k^0)^2$ term 
can be dropped in the propagator of the exchanged particle, so that 
the interaction corresponds to a potential between the heavy particles. 
The pair of heavy-particle propagators in each ladder rung 
can also be simplified. Let
\begin{equation}
\vec{p}_i = \frac{\mu_{ij}}{m_j} \,\vec{P} + \vec{p},
\qquad 
\vec{p}_j = \frac{\mu_{ij}}{m_i} \,\vec{P} - \vec{p},
\end{equation}
be the momenta of the on-shell external state $\chi_i\chi_j$, 
and $\mu_{ij}$ the reduced mass. Since 
$|\vec{P}\,|, |\vec{p}\,|\ll \mu_{ij}$, the center-of-mass energy of 
the annihilation process is 
\begin{equation}
\sqrt{s} = m_i+m_j+{\cal E} \equiv 2 m_{\rm LSP} + E,
\qquad {\cal E}\equiv \frac{\vec{p}^{\,2}}{2\mu_{ij}}\,,
\label{eq:defE1}
\end{equation}
up to higher-order terms in the non-relativistic expansion. For later 
purposes it proves convenient to introduce the variable $E$, which 
measures energy from the common reference value $2m_{\rm LSP}$. 
Since mass differences $m_i+m_j - 2 m_{\rm LSP}$ scale as 
$m_{\rm LSP} v^2$ by assumption, both energy variables have 
the non-relativistic scaling $E\sim {\cal E}\sim m_{\rm LSP} v^2$. 
The masses of the two-particle state $a$ in a given ladder rung 
are denoted by $m_{a_1}$, $m_{a_2}$, and $M_a \equiv m_{a_1}+m_{a_2}$. 
The loop momentum energy-component integrals in each ladder rung 
can then be performed employing
\begin{eqnarray}
&& \int \frac{dk_a^0}{2\pi}\, 
\frac{2 m_{a_1}}{(p_i+k_a)^2-m_{a_1}^2+i\epsilon}\,
\frac{2 m_{a_2}}{(-p_j+k_a)^2-m_{a_2}^2+i\epsilon}
\nonumber\\
&& 
= \,\frac{-i}{E-[M_a-2 m_{\rm LSP}]-
\frac{(\vec{p}+\vec{k}_a)^2}{2 \mu_{a}}}\,.
\end{eqnarray}
To arrive at this result we systematically neglected higher-order 
terms in the non-relativistic expansion. As discussed before, we 
must require that the mass differences along a given heavy 
particle line are small, that is $m_{a_1}-m_i$ and $m_{a_2}-m_j$ should 
be of order $m_{\rm LSP} v^2$, 
but we do not have to assume that $m_j-m_i$ and $m_{a_2}-m_{a_1}$ 
are small. Note than within these approximations the denominator 
of the kinetic energy term, $\mu_a$, could be equivalently substituted 
by $\mu_{ij}$ or $m_{\rm LSP}/2$, or any other two-particle state 
reduced mass.

With these simplifications to the ladder-diagram sum, 
the annihilation matrix element, including the 
tree diagram with no exchange,  
can be written as 
\begin{eqnarray}
&& \langle 0| \chi^{c \dagger}_{e_2}\Gamma 
K\big[-\frac{i}{2}\overleftrightarrow{\bff{\partial}}\big]\chi^{}_{e_1}
| \chi_i \chi_j \rangle = 
\langle \xi^{c \dagger}_{j} \Gamma \xi_{i} \rangle 
\nonumber\\
&&\hspace*{0.cm}
\times\, 
\lim_{\hat E\to E} (-1)\left(\hat E-\frac{\vec{p}^{\,2}}{2\mu_{ij}}\right)
\,\int\frac{d^3\vec{q}}{(2\pi)^3} \,K[\vec{q}\,]\,
\Big( \tilde{G}^{ie}(\vec{p},\vec{q};\hat E) + (-1)^{L+S} \, \tilde{G}^{i\bar{e}}(\vec{p},\vec{q};\hat E) \Big)
\,,\qquad
\label{eq:meladdersum}
\end{eqnarray}
where $i$ refers to the initial two-particle state $ij$ and 
$e$ ($\bar{e}$) to the state $e_1 e_2$ ($e_2e_1$) which annihilates into
light particles. Below 
we employ this compact notation to label two-particle states 
by compound indices $a,b,\dots$.
The function $\tilde G$ is defined through
\begin{eqnarray}
\tilde{G}^{ab}(\vec{p},\vec{q}; E) &=& 
-\frac{\delta^{ab}}{E-[M_a-2 m_{\rm LSP}]- \frac{\vec{p}^{\,2}}{2\mu_{a}}}\,
(2\pi)^3\delta^{(3)}(\vec{p}-\vec{q}\,)
\nonumber\\
&&\hspace*{-1.5cm}+\, 
\frac{1}{E-[M_a-2 m_{\rm LSP}]- \frac{\vec{p}^{\,2}}{2\mu_{a}}}
\,i H^{ab}(\vec{p},\vec{q}; E)\,
\frac{1}{E-[M_b-2 m_{\rm LSP}]-\frac{\vec{q}^{\,2}}{2\mu_{b}}}
\qquad
\label{eq:greenf}
\end{eqnarray}
and 
\begin{eqnarray}
H^{ab}(\vec{p},\vec{q}; E) &=& i \sum_{n=0}^\infty 
\int\left[\prod_{i=1}^n\frac{d^3\vec{k_i}}{(2\pi)^3}\right]
\hat{V}^{a a_1}(\vec{k}_1) 
\,\frac{1}{E-[M_{a_1}-2 m_{\rm LSP}]-
\frac{(\vec{p}+\vec{k}_1)^{2}}{2\mu_{a_1}}}
\nonumber\\
&& \hspace*{-1.5cm}
\times \,\hat{V}^{a_1 a_2}(\vec{k}_2-\vec{k}_1) \ldots 
\frac{1}{E-[M_{a_n}-2 m_{\rm LSP}]-\frac{(\vec{p}+\vec{k}_n)^{2}}{2\mu_{a_n}}}
\,\hat{V}^{a_n b}(\vec{q}-\vec{p}-\vec{k}_n)\,. 
\qquad
\label{eq:hfn}
\end{eqnarray}
(For the term $n=0$ we set $a_0=a$ and $\vec{k}_0=0$.)
The first term on the right-hand side of (\ref{eq:greenf}) accounts 
for the tree diagram, a term with given $n$ in (\ref{eq:hfn}) for 
the $(n+1)$-loop ladder diagram with $n+1$ exchanges.\footnote{The 
loop integration of the last loop before the annihilation vertex 
is the $\vec{q}$ integral in (\ref{eq:meladdersum}).} The factors  
$\hat{V}^{ab}(k)$ contain the propagator of the exchanged particle and 
coupling factors from the two vertices. These are the momentum-space 
potentials discussed in Sec.~\ref{sec:potMSSM}. Note that using 
the on-shell condition for the external particles and non-relativistic 
approximations, the Dirac matrices from the numerator can be 
reduced to the tree structure $\xi^{c \dagger}_{j} \Gamma \xi_{i}$. 
Depending on whether $\Gamma = 1_{2 \times 2}$ or 
$\vec{\sigma}$, the potentials for $S=0$ or $S=1$ must be used in 
the above equations, which causes the dependence of 
$\psi^{(L,S)}_{e_1 e_2, \,ij}$ on the total spin $S$. 
The angular-momentum dependence 
arises through $K[\vec{q}\,]$ in (\ref{eq:meladdersum}), 
which reads $K=1$ for $L=0$ and $K=\vec{q}\,$ for $L=1$. 
The limiting procedure in (\ref{eq:meladdersum}) is required 
since the factor $[\hat E-\vec{p}^{\,2}/(2\mu_{ij})]$ 
vanishes for $\hat E=E$, while $\tilde{G}^{ab}(\vec{p},\vec{q}; E)$ 
develops a singularity that corresponds to scattering 
states with relative-momentum kinetic-energy $E$.

The second step consists of verifying that 
$\tilde{G}^{ab}(\vec{p},\vec{q}; E)$ 
is in fact the 
momentum-space Green function of a certain Schr\"odinger operator. 
Indeed, one easily checks that it satisfies the Lippmann-Schwinger 
equation
\begin{eqnarray}
&&\left(\frac{\vec{p}^{\,2}}{2\mu_{a}} - [E-M_a+2 m_{\rm LSP}]\right)
\tilde{G}^{ab}(\vec{p},\vec{q}; E)
\nonumber\\
&&\hspace*{1cm}
+\,\int \frac{d^3\vec{k}}{(2\pi)^3}\,\hat{V}^{ac}(\vec{k}\,)\,
\tilde{G}^{cb}(\vec{p}-\vec{k},\vec{q}; E)
= \delta^{ab}\,(2\pi)^3\,\delta^{(3)}(\vec{p}-\vec{q}\,),
\end{eqnarray}
hence the Fourier-transform $G^{ab}(\vec{r},\vec{r}^{\,\prime}; E)$ 
is the Green function for the Schr\"odinger equation
\begin{eqnarray}
\left(-\frac{\vec{\nabla}^{\,2}}{2\mu_{a}} - E\right)
G^{ab}(\vec{r},\vec{r}^{\,\prime}; E)
+ V^{ac}(r)
G^{cb}(\vec{r},\vec{r}^{\,\prime}; E)
= \delta^{ab}\,\delta^{(3)}(\vec{r}-\vec{r}^{\,\prime})
\end{eqnarray}
with the coordinate-space potential
\begin{equation}
V^{ac}(r) = \hat{V}^{ac}(r) +\delta^{ac} \,\big[M_a-2 m_{\rm LSP}\big]\,.
\label{eq:Vhatpot1}
\end{equation}
Note that this result could be obtained directly as an equation 
of motion from the 
effective Lagrangian given in Sec.~\ref{sec:eft} by including the 
potential in the unperturbed Lagrangian.

The third and final step uses elementary scattering theory to 
express (\ref{eq:meladdersum}) in terms of eigenfunctions with 
energy $E$ rather than the full Green function. Let us define 
the Green operators
\begin{equation}
G(\hat E) = \frac{1}{H-\hat E-i\epsilon}, \qquad
G_0(\hat E) = \frac{1}{H_0-\hat E-i\epsilon}
\end{equation}
for the interacting and free Hamiltonians of the Schr\"odinger 
problem and the corresponding momentum eigenstates
\begin{equation}
H|\vec{p}+, a\rangle = \frac{\vec{p}^{\,2}}{2\mu_{a}}|\vec{p}+, a\rangle,
\qquad 
H_0|\vec{p}\,,a \rangle = \frac{\vec{p}^{\,2}}{2\mu_{a}}|\vec{p}\,,a\rangle\,.
\end{equation}
Note that $G$, $G_0$ are matrix-valued operators with dimensionality 
related to the number of two-particle states $(\chi\chi)_a = \chi_{a_1}
\chi_{a_2}$. Accordingly, the 
eigenstates carry a compound index $a$ that refers to the 
component of the wave-function proportional to the two-particle state $a$. 
The states $|\vec{p}+,a\rangle$ are the exact stationary scattering states 
of $H$ with eigenvalue $E=\vec{p}^{\,2}/(2\mu_{a})$, 
while $|\vec{p}\,,a\rangle$ are plane waves. We may represent the 
former in the plane-wave basis by $[\tilde \psi_{E}(\vec{q}\,)]_{ab} = 
\langle\vec{q}\,,a|\vec{p}+,b\rangle$. The Lippmann-Schwinger equation 
can be used to derive $|\vec{p}+,a\rangle = G G_0^{-1}|\vec{p}\,,a\rangle$. 
With these preparations we use (\ref{eq:medef}), (\ref{eq:meladdersum}) 
to compute
\begin{eqnarray}
&&  K[\vec{p}\,]\, \psi^{(L,S)}_{e_1 e_2, \,ij} = 
\lim_{\hat E\to E} (-1)\left(\hat E-\frac{\vec{p}^{\,2}}{2\mu_{ij}}\right)
\,\int\frac{d^3\vec{q}}{(2\pi)^3} \,K[\vec{q}\,]\,
\underbrace{\tilde{G}^{ie}(\vec{p},\vec{q};\hat E)}_{\langle 
\vec{p},i|G(\hat E)|\vec{q}\,,e\rangle}
\nonumber\\
&&\hspace*{0.5cm}
=\, \lim_{\hat E\to E} 
\,\int\frac{d^3\vec{q}}{(2\pi)^3} \,K[\vec{q}\,]\,
\underbrace{ \langle \vec{p},i| 
\left(\frac{\vec{p}^{\,2}}{2\mu_{ij}}-\hat E\right)}_{\langle \vec{p},i | 
G_0^{-1}(\hat E)}\, G(\hat E)
|\vec{q}\,,e\rangle
= \int\frac{d^3\vec{q}}{(2\pi)^3} \,K[\vec{q}\,]\,
[\tilde\psi_{E}(\vec{q}\,)]_{ei}^*\,.\,
\qquad
\label{eq:sfpsi}
\end{eqnarray}
The same result can also be derived from the spectral representation
\begin{equation}
G(\hat E) = \sum_a \int \frac{d^3\vec{p}}{(2\pi)^3}\,
\frac{|\vec{p}+,a\rangle \langle \vec{p}+,a |}
{\frac{\vec{p}^{\,2}}{2\mu_{a}}-\hat E - i\epsilon}
\end{equation}
of the Green operator (ignoring the bound states not relevant to our 
discussion). 
Finally, since $K=1$ for $L=0$ and $K[\vec{q}\,]=\vec{q}\,$ for $L=1$, 
we obtain 
\begin{eqnarray}
\psi^{(0,S)}_{e_1 e_2, \,ij} = [\psi_{E}(0)]_{e_1 e_2, \,ij}^*
\qquad\mbox{and}\qquad 
\vec{p}\,\psi^{(1,S)}_{e_1 e_2, \,ij} = -i\,
[\vec{\nabla}\psi_{E}(0)]_{e_1 e_2, \,ij}^*
\label{eq:psiwf0}
\end{eqnarray}
in terms of coordinate-space scattering wave-functions at the 
origin. Note that the result carries two compound indices referring 
to two-particle states. The second, $ij$, refers to the incoming 
two-particle state $|\vec{p}\,,i\rangle$ with kinetic energy 
$E-(M_i-2m_{\rm LSP})$ in 
the cms frame of the annihilation. The first, $e={e_1 e_2}$, specifies 
that only the component proportional to the two-particle state $e$ of 
the wave-function at the origin for this incoming state is picked out 
by the annihilation operator $\chi^{c \dagger}_{e_2}\chi^{}_{e_1}$ 
that defines $\psi^{(L,S)}_{e_1 e_2, \,ij}$. The wave-function 
$[\psi_E(\vec{r}\,)]_{a,ij}$ can 
be obtained directly from the matrix-Schr\"odinger equation
\begin{equation}
\left(\left[-\frac{\vec{\nabla}^{\,2}}{2\mu_{a}} - E\right] \delta^{ab}
+ V^{ab}(r)\right) [\psi_E(\vec{r}\,)]_{b,ij} =0
\label{eq:schroedinger1}
\end{equation}
with the potential (\ref{eq:Vhatpot1}). 
The label $ij$ refers to the fact 
that this equation should be solved with the initial condition corresponding 
to a particular incoming two-particle state $ij$.
Let us finally mention that the complex conjugated scattering
wave-function appears in (\ref{eq:sfpsi}) because of the convention used 
for the left and right states in the definition 
of the Green function (\ref{eq:greenf}). Using the opposite convention
we would end up with $[\psi_E(\vec{q}\,)]_{ei}$ in (\ref{eq:sfpsi}) 
but with $V^{ba}$ instead
of $V^{ab}$ in the Schr\"odinger 
equation~(\ref{eq:schroedinger1}).\footnote{For a symmetric 
potential the solutions for $\psi_E(\vec{r}\,)$ and
$[\psi_E(\vec{r}\,)]^*$ are identical and the correct result would be 
obtained even if the conventions for the Green function
and the potential were not consistently taken care of. However, in 
the MSSM the potential matrix is complex-hermitian in general.}

\subsection{Solution of the multi-channel Schr\"odinger equation}
\label{sec:schrsolution}

We first solve the matrix-Schr\"odinger equation for the Sommerfeld factors
by following closely the method described in \cite{Slatyer:2009vg}. 
Although, eventually, we will not use this method, we discuss it 
to set up the framework for the subsequent improvement. Under our 
assumption for the mass splittings, we can approximate 
$\mu_{a} \approx m_{\rm LSP}/2$ in the kinetic term, since the 
difference is a $v^2$ correction, which we consistently neglect in the 
long-distance part. The Schr\"odinger equation now reads 
\begin{equation}
\left(\left[-\frac{\vec{\nabla}^{\,2}}{m_{\rm LSP}} - E\right] \delta^{ab}
+ V^{ab}(r)\right) [\psi_E(\vec{r}\,)]_{bi} =0\,,
\label{eq:schroedinger2}
\end{equation}
and we recall the definitions 
\begin{equation}
\sqrt{s} = 2 m_{\rm LSP}+E \qquad \mbox{and}\qquad
V^{ab}(r) = \hat{V}^{ab}(r) +\delta^{ab} \,\big[M_a-2 m_{\rm LSP}\big]\,.
\label{eq:Vpot2}
\end{equation}
We also use the velocity variable $v$ defined by 
\begin{equation}
E \equiv m_{\rm LSP} v^2\,.
\end{equation}
This implies that the threshold for 
co-annihilation channels occurs at finite velocities 
$v= ((m_{i}+m_{j}-2 m_{\rm LSP})/m_{\rm LSP})^{1/2}$. 
With the above redefinitions the dependence on the initial scattering 
state appears only in the initial condition for the solution as 
indicated by the subscript $i$ of $[\psi_E(\vec{r}\,)]_{bi}$, but not 
in the equation itself. We note 
that the MSSM matrix potential is hermitian, but in general not symmetric, 
since the entries of the potential matrix can be complex numbers. 
The potential is spherically symmetric. For large $r$, the potential 
approaches the diagonal matrix $V_{\rm inf}$ with diagonal entries 
$V_{{\rm inf},a} = M_a-2 m_{\rm LSP}$. We therefore define the 
wave numbers
\begin{equation}
k_a^2 = m_{\rm LSP} (E+i\epsilon-V_{{\rm inf},a}),
\end{equation}
which determine the asymptotic behaviour of the solutions. When 
$E>V_{{\rm inf},a}$, channel $a$ is kinematically open and $k_a$ 
is real. Otherwise the channel is closed and $k_a$ is imaginary.

The scattering solution describing an incoming plane wave propagating 
along the $z$-direction and an outgoing scattered spherical wave 
takes the asymptotic form 
\begin{equation}
[\psi_E(\vec{r}\,)]_{ai} \stackrel{r\to\infty}{=} \delta_{ai} \, e^{ik_a z}
+ f_{ai}(\theta) \, \frac{e^{ik_a r}}{r}.
\label{eq:Psi_asympt}
\end{equation}
where, due to the azimuthal symmetry of the problem $f_{ai}(\theta)$ does 
not depend on the azimuthal angle.\footnote{
\label{ft:Coulombpotential}
The asymptotic form (\ref{eq:Psi_asympt}) applies to radial potentials
vanishing faster than $1/r$ as $r\rightarrow\infty$. This holds for 
Yukawa potentials, arising from massive mediator exchange, which is -- 
besides the Coulomb interaction from photon exchange -- the relevant 
case for $\chi\chi$ pair annihilations. For Coulomb potentials, 
(\ref{eq:Psi_asympt}) does not apply. However, photon-exchange does 
not change the neutralino state, so the $1/r$ potentials arise exclusively 
in the diagonal entries of the potential matrix $V(r)$, when
written in the two-particle mass-eigenstate basis. For large values of $r$, 
$V(r)$ then always approaches a diagonal matrix (the non-diagonal 
Yukawa potentials being exponentially suppressed) 
with entries $V_{{\rm inf},a} + c_a \alpha/r$, containing the Coulomb 
potential contributions as well as constant mass splitting terms. 
Accounting for the presence of long-range Coulomb potentials, the 
$\exp{(ik_a z)}$ factor in (\ref{eq:Psi_asympt}) should be 
replaced by the incoming wave-function in presence of the Coulomb 
potential in the $aa$ component of $V(r)$, and in the outgoing 
scattered wave, one has to replace $\exp{(ik_a r)}$ by
$\exp{\left(i (k_a r + m_{\rm LSP} c_a \alpha/(2k_a) 
\ln(2k_ar))\right)}$. Since it can be shown that 
the subsequent derivation of the 
Sommerfeld factor including Coulomb potentials on the diagonals of $V(r)$
is completely analogous to the short-range potential case, and leads to 
the same result for the enhancement factor, we will for brevity refer 
to the asymptotic behaviour (\ref{eq:Psi_asympt}) in the following.} 
Since the potential is non-diagonal, 
the scattered spherical wave is not proportional to $\delta_{ai}$. 
The asymptotic behaviour (\ref{eq:Psi_asympt}) should be matched 
to the behaviour of the general solution 
\begin{equation}
[\psi_E(\vec{r}\,)]_{ai} = \sum_{L} \frac{[u_L(r)]_{ab}}{r}\,
A_{bi}\, 
P_{L}(\cos \theta),
\label{eq:psigeneral}
\end{equation}
where $P_L(\cos\theta)$ denotes the Legendre polynomials. This expresses 
$[\psi_E(\vec{r}\,)]_{ai}$ as a superposition $A_{bi}$  
of basis solutions $[u_L(r)]_{ab}$
of the radial Schr\"odinger equation 
\begin{equation}
\left(\left[-\frac{d^2}{dr^2}
+\frac{L (L+1)}{r^2} - m_{\rm LSP} E \right] \delta^{ab}
+ m_{\rm LSP} V^{ab}(r)\right) [u_L(r)]_{bi} =0
\label{eq:schroedingerpartial}
\end{equation}
for each partial wave. When $V$ is a $N\times N$ matrix, there exist 
$2 N$ linearly independent solutions to (\ref{eq:schroedingerpartial}) 
corresponding to the $N$ independent initial conditions $i$, each 
of which is an $N$-component vector with index $a$. 
$N$ solutions are
irregular at the origin, hence restricting us to the set of $N$ regular 
ones. The asymptotic behaviour of the $a$th component of the 
regular linear independent solution for initial state $i$ is given by
\begin{equation}
[u_L(r)]_{ai} \stackrel{r \rightarrow \infty}{=}
n_{ai}\, \sin\left(k_a r - \frac{L \pi}{2} + \delta_{ai} \right)
\label{eq:Rl_infty}
\end{equation}
with constant coefficients $n_{ai}$ and scattering phases  
$\delta_{ai}$.\footnote{\label{ft:Rl_Coulomb}
Both depend on $L$, but we do not indicate this explicitly. 
Eq.~(\ref{eq:Rl_infty}) assumes short-range potentials. In the 
presence of Coulomb potentials in the diagonal entries of 
the potential matrix $V(r)$ into account, the
asymptotic behaviour of the regular basis solutions reads
\begin{displaymath}
[u_L(r)]_{ai} \stackrel{r \rightarrow \infty}{=}
n_{ai}\, \sin\left(k_a r - \frac{L \pi}{2} 
+ \frac{m_{\rm LSP} c_a \alpha}{2 k_a}\, \ln(2k_a r)
\delta_{ai} \right)\,.
\end{displaymath}
The modifications match those of the scattering solution, see footnote
\ref{ft:Coulombpotential}.}
Matching the asymptotic behaviours of 
(\ref{eq:Psi_asympt}) and (\ref{eq:psigeneral}), we obtain
\begin{equation}
A_{bi} = i^L (2L+1) \,\frac{[M^{-1}]_{bi}}{k_i}
\qquad\mbox{(no sum over $i$)}
\label{eq:A}
\end{equation}
with 
\begin{equation}
M_{ai}\equiv n_{ai} \,e^{-i\delta_{ai}}\,.
\label{eq:Mdef}
\end{equation}

The basic ingredients (\ref{eq:psiwf0}) for the Sommerfeld factors 
now require the wave function and its derivatives at the origin. 
We suppose that the annihilation operator is constructed such that it 
overlaps only with a single partial wave $L$, the relevant cases 
in practice being $L=0$ ($S$-wave annihilation) and $L=1$ 
($P$-wave annihilation). The leading term in the Taylor expansion of 
$[u_L(r)]_{ai}$ around the origin is given by 
\begin{equation}
[u_L(r)]_{ai} = \frac{1}{(L+1)!}\, [u_l^{(L+1)}(0)]_{ai}\,
r^{L+1} +\ldots
\end{equation}
with $[u_L^{(L+1)}(0)]_{ai}$ denoting the $(L+1)$-th derivative at 
the origin. 
Hence using (\ref{eq:A}) in (\ref{eq:psigeneral}), we have 
\begin{eqnarray}
[\psi_E(\vec{r}\,)]_{ai} \stackrel{r\to 0}{=} 
P_L(\cos\theta) r^L\,\frac{i^L (2L+1)}{(L+1)!}
\,  [u_L^{(L+1)}(0)]_{ab} \,\frac{[M^{-1}]_{bi}}{k_i}\,.
\label{eq:psiE0}
\end{eqnarray}
The quantity 
$\psi^{(L,S)}_{e_1 e_2, \,ij}$ is defined in (\ref{eq:medef}) as the ratio 
of the annihilation matrix element to the same 
matrix element evaluated in the tree approximation, 
which corresponds to replacing $V$ by $V_{\rm inf}$ in the 
Schr\"odinger equation. The free Schr\"odinger equation 
can be solved exactly, and one finds 
\begin{eqnarray}
[\psi_E^{\rm free}(\vec{r}\,)]_{ai} \stackrel{r\to 0}{=} 
P_L(\cos\theta) r^L\,\frac{i^L (2L+1)}{(2L+1)!!} \,k_i^L\,\delta_{ai}\,.
\label{eq:psiE0free}
\end{eqnarray}
Comparison with (\ref{eq:psiE0}) gives 
\begin{eqnarray}
\psi^{(L,S)}_{e_1 e_2, \,ij} = 
\frac{(2L+1)!!}{(L+1)!}\, 
[u_L^{(L+1)}(0)]_{eb}^* \,\frac{[M^{-1}]^*_{bi}}{k_i^{L+1}}
\label{eq:psifactor1}
\end{eqnarray}
for general $L$.

We can avoid the computation of the phase-shift matrix $M_{bi}$ by 
making use of the fact that the Wronskian matrix is $r$-independent. 
Let $[v_L(r)]_{ai}$ be the $N$ linearly independent singular basis 
solutions with asymptotic behaviours
\begin{equation}
[v_L(r)]_{ai} \stackrel{r\to 0}{=} \frac{\delta_{ai}}{r^L}\,,
\qquad
[v_L(r)]_{ai} \stackrel{r\to \infty}{=} [T^\dagger]_{ai} \,e^{-ik_a r}\,,
\label{eq:asymptoticirregular}
\end{equation}
which defines the matrix $T$.
The Wronskian is defined as
\begin{equation}
[W_L]_{ij} = [v_L^\dagger(r)]_{ia} [u_L^\prime(r)]_{aj} - 
[v_L^{\dagger\,\prime}(r)]_{ia} [u_L(r)]_{aj}\,,
\label{eq:Wl}
\end{equation}
where the prime denotes the derivative.\footnote{
\label{ft:Wl_Slatyer}
Eq.~(\ref{eq:Wl}) is a generalization of the expression $W_L$ considered in
\cite{Slatyer:2009vg}. There the potential matrix was assumed to be 
real-symmetric, such that the transpose of the matrix $v_L$  
appeared in the definition of $W_L$, instead of the hermitian conjugate.
Due to the generic hermiticity property of the Schr\"odinger equation, 
the definition of $W_L$ with hermitian conjugates (\ref{eq:Wl}) looks 
more natural even in the case of real-symmetric potentials.} 
Inserting the asymptotic behaviour 
of the regular and singular solutions, we obtain
\begin{eqnarray}
[W_L(r)]_{ij} \stackrel{r\to 0}{=} \frac{2L+1}{(L+1)!}\,
[u_L^{(L+1)}(0)]_{ij}\,,
\qquad
[W_L(r)]_{ij} \stackrel{r\to \infty}{=} i^L \sum_a k_a 
T_{ia} M_{aj}\,.
\end{eqnarray}
Both expressions must be equal due to the constancy of the 
Wronskian, which implies
\begin{equation}
[u_L^{(L+1)}(0)]_{ab} \,[M^{-1}]_{bi} = 
\frac{(L+1)!}{2L+1}\,i^L\,k_i \,T_{ai}
\label{eq:buildT}
\end{equation}
Plugging this into (\ref{eq:psifactor1}) gives
\begin{eqnarray}
\psi^{(L,S)}_{e_1 e_2, \,ij} = 
(2L-1)!!\,i^{-L}\,\frac{T_{ei}^*}{k_i^{L}}\,,
\label{eq:psifactor2}
\end{eqnarray}
which expresses $\psi^{(L,S)}_{e_1 e_2, \,ij}$ in terms of the large-$r$ 
behaviour of the singular solutions. This can be obtained from the 
regular solutions as described below. The Sommerfeld factor 
is defined through the square of the annihilation amplitude. 
The definition (\ref{eq:SFdef}) can now be evaluated in the 
form
\begin{eqnarray}
S_{i}[\hat f(^{2S+1}L_J)] =  
\left(\frac{(2L-1)!!}{k_i^L}\right)^2\,
\frac{[T^\dagger]_{ie}\,
\hat f^{\chi\chi \to \chi \chi}_{e e^\prime}(^{2S+1}L_J) 
\,T_{e^\prime i}}
{\hat  f^{\chi\chi\to \chi\chi}_{ii}(^{2S+1}L_J)|_{\rm LO}}
\,,
\label{eq:SFwithT}
\end{eqnarray}
which is equivalent to the result first derived in \cite{Slatyer:2009vg}. 
Note, in compound-index notation, 
$\hat f^{\chi\chi \to \chi \chi}_{e e^\prime} = 
\hat f^{\chi\chi \to \chi \chi}_{\lbrace e_1 e_2 \rbrace \lbrace e_4 e_3 
\rbrace}$ and $\hat  f^{\chi\chi\to \chi\chi}_{ii} = \hat  
f^{\chi\chi\to \chi\chi}_{\lbrace i j \rbrace \lbrace i j \rbrace}$.

In practice, the matrix-Schr\"odinger equation must be solved 
numerically. We obtain the matrix $T$ from the regular solutions 
by the following three steps:
\begin{itemize}
\item[(1)] Determine the $N$ linearly independent regular solution 
vectors $[u_L(r)]_{ai}$ for $i=1,\ldots, N$ for given $L$ by solving 
the radial Schr\"odinger equation with initial conditions
\begin{equation}
[u_L(r_0)]_{ai} = \frac{1}{2L+1}\,\hat{r}_0^{L+1}\,\delta_{ai}\,,
\qquad
[u_L^\prime(r_0)]_{ai} = \frac{L+1}{2L+1}\,\hat{r}_0^{L}\,\delta_{ai}\,,
\label{eq:initial}
\end{equation}
where $\hat{r}_0$ is close to zero. In practice, we work with the 
dimensionless, scaled variable $\hat r = m_{\rm LSP} v r$ and 
use $\hat{r}_0=10^{-7}$. The normalization is chosen such that 
$[u_L^{(L+1)}(0)]_{ai} = \delta_{ai} (L+1)!/(2 L+1)$, in which case 
the Wronskian equals $[W_{L}(r)]_{ij}=\delta_{ij}$.
\item[(2)] Pick a large value $r_{\infty}$, such that the asymptotic 
behaviour (\ref{eq:asymptoticirregular}) of the irregular solution 
applies, and the Wronskian evaluates to 
\begin{equation}
[W_L]_{ij} = T_{ia} U_{aj} \stackrel{!}{=}\delta_{ij}
\end{equation}
with
\begin{equation}
U_{aj}(r_\infty) = e^{i k_a r_\infty}\,\left(
[u_L^\prime(r_\infty)]_{aj} - i k_a [u_L(r_\infty)]_{aj}\right)\,,
\label{eq:Umatrix}
\end{equation}
and $[u_L(r)]_{aj}$ known from step 1.
Hence the matrix $T$ appearing in (\ref{eq:SFwithT}) follows from 
matrix inversion, 
\begin{equation}
T = U^{-1}\,.
\label{eq:Uinversion}
\end{equation}
\item[(3)] Since in practice the Schr\"odinger equation can only 
be solved up to some finite value of $r$, check the stability of the 
result by varying (and increasing) $r_\infty$ until $T$ is 
independent of  $r_\infty$ within a certain target accuracy.
\end{itemize}

The procedure described here has been implemented in {\sc Mathematica} 
and works well, when {\em all$\,$} $N$ states included in the 
multi-channel Schr\"odinger equation are degenerate to a high degree. 
This is the case in MSSM parameter regions where the Sommerfeld 
enhancement is most effective, such as the wino or Higgsino 
limit for the neutralino, and when the other states not related to 
the wino or Higgsino electroweak multiplet are decoupled and ignored. 
However, our aim is to compute the Sommerfeld-enhanced 
radiative corrections in a larger part of the MSSM parameter space, 
when the mass splittings become larger than in the wino or Higgsino 
limit. In this case the method outlined above encounters severe 
numerical problems, as we now describe, and fails to provide the 
desired solution.

%
\begin{figure}[t]
\begin{center}
\includegraphics[width=0.6\textwidth]{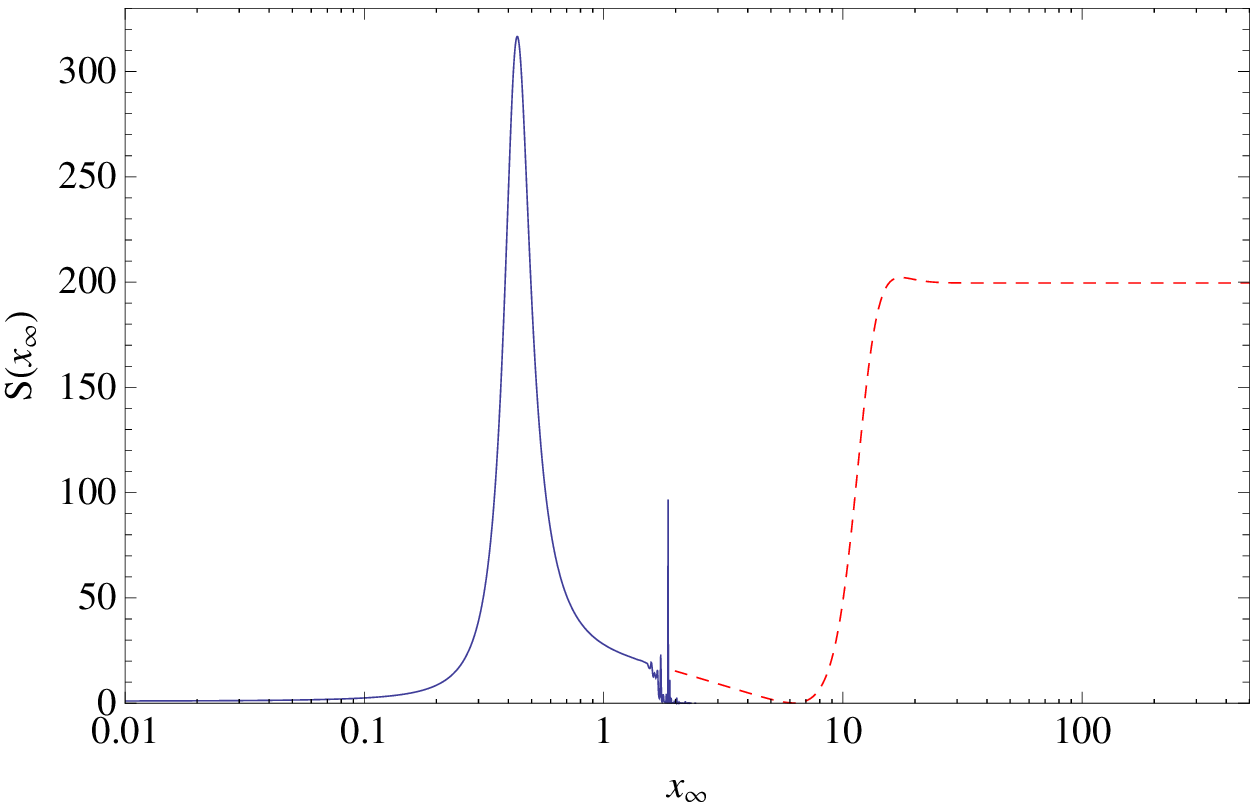}
\includegraphics[width=0.6\textwidth]{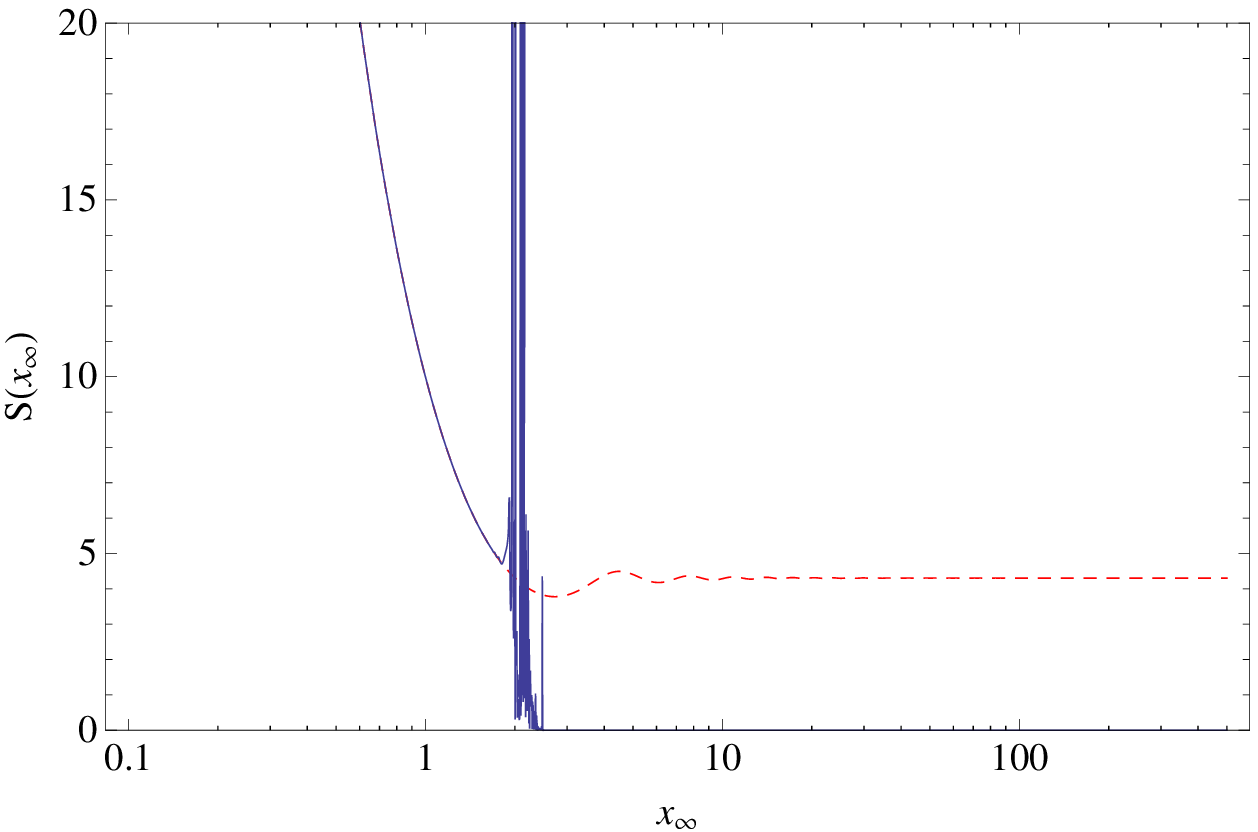}
\caption{$S$- and $P$-wave Sommerfeld factors 
$S\equiv S_{\chi_1^0\chi_1^0}[\hat f(^{1}S_0)]$ (upper plot) 
and $S_{\chi_1^0\chi_1^0}[\hat f(^{3}P_J)]$ (lower plot)
for $v=0.012$ in a wino-like model as described in the text. 
The light-grey (red) dashed curve shows $S(x_\infty)$, when only 
the $\chi_1^0\chi_1^0$ and $\chi_1^+\chi_1^-$ two-particle 
states in the Schr\"odinger equation and the annihilation 
process are kept and the asymptotic regime is reached for 
$x_\infty>50$. The dark-grey (blue) solid curve shows the result 
when the $\chi_1^0\chi_2^0$ state is included. In this case, the 
evaluation fails for $x_\infty>2$ and no reliable result is 
obtained.}
\label{fig:Sinfoldmethod}
\end{center}
\end{figure}
%

For the purpose of illustration we consider a MSSM parameter point 
where the LSP is wino-like. The relevant masses are $m_{\rm LSP} = 
m_{\chi_1^0} = 2749.4\,$GeV, $m_{\chi_1^+} = 2749.61\,$GeV 
and $m_{\chi_2^0} = 2950.25\,$GeV. Further details of the models 
are not relevant for the following discussion. We first compute 
the Sommerfeld factor $S\equiv S_{\chi_1^0\chi_1^0}[\hat f(^{1}S_0)]$ 
as function of $x_\infty = m_{\rm LSP} v r_\infty$ by keeping 
only the $\chi_1^0\chi_1^0$ and $\chi_1^+\chi_1^-$ two-particle 
states in the Schr\"odinger equation and the annihilation 
process. The velocity is chosen $v=0.012$, slightly below the 
threshold for the $\chi_1^+\chi_1^-$ state. The result $S(x_\infty)$ 
is shown as the light-grey (red), dashed curve 
in the upper panel of Fig.~\ref{fig:Sinfoldmethod}. 
After a rapid variation with a peak structure, the Sommerfeld 
factor reaches a plateau and for $x_\infty > 50$ stays at the 
constant value $S(\infty)\approx 199.59$. Next, we add the 
$\chi_1^0\chi_2^0$ state to the problem, so that the Schr\"odinger 
system is now for a $3\times 3$ matrix. Since the new state 
is $200\,$GeV heavier and moreover rather weakly coupled to 
the two lowest, nearly degenerate wino states, we expect that it 
should have little effect on the value of the Sommerfeld factor. 
However, now the numerical solution fails when $x_\infty$ is slightly 
large than 1, as seen from the dark-grey (blue) curve in 
Fig.~\ref{fig:Sinfoldmethod}, which drops to 0 after a few spikes. 
It is not possible to reach the plateau, where $S(x_\infty)$ 
stabilizes. A similar behaviour impedes the calculation of the 
$P$-wave Sommerfeld factor, as shown in the lower panel of 
Fig.~\ref{fig:Sinfoldmethod}. 

The numerical instability originates from the presence of kinematically 
closed two-particle state channels, here the $\chi_1^0\chi_2^0$ state. 
When the mass splittings become 
larger or $v$ small, the situation $2 m_{\rm LSP} + m_{\rm LSP} v^2
< M_b$ is rather generic. To quantify the issue suppose that 
$b$ in the Schr\"odinger equation (\ref{eq:schroedingerpartial}) 
is a closed channel, while $a$ refers to 
the two-LSP $\chi_1^0\chi_1^0$ channel. The solution $[u_l(x)]_{bi}$ for the 
closed channel involves an exponentially growing component 
proportional to $e^{\kappa_b r}$ where $\kappa_b^2 = m_{\rm LSP} 
(M_b - [2 m_{\rm LSP} + m_{\rm LSP} v^2])$. In the absence of channel 
mixing, the solutions for the kinematically open channels would be 
oscillating. The off-diagonal potentials $V^{ab}(r)$ that couple the 
different channels decrease as $e^{-M_{\rm EW} r}/r$, 
where $M_{\rm EW}$ is the mass scale set by the electroweak gauge 
boson exchange that mediates the channel mixing. Hence, the 
open-channel solutions  $[u_l(x)]_{ai}$ 
inherit the exponential growth from the closed channels. For the 
two-LSP $\chi_1^0\chi_1^0$ channel, exponential growth occurs 
when at least one of the included kinematically closed channels $b$
satisfies
\begin{equation}  
M_b - [2 m_{\rm LSP} + m_{\rm LSP} v^2] > \frac{M_{\rm EW}^2}{m_{\rm LSP}}
\,.
\end{equation}
Since typically $m_{\rm LSP}\gg M_{\rm EW}$ for the dark matter scenarios 
of interest, this condition is easily satisfied unless all two-particle 
states included in the computation are very degenerate within a 
few GeV or less. In consequence the formally linearly independent 
solutions $[u_l]_{ai}$ degenerate and the matrix $U_{ai}$ 
becomes ill-conditioned with exponentially growing entries in the 
rows corresponding to open channels $a$. The matrix inversion 
(\ref{eq:Uinversion}) can no longer be done in practice for $r_\infty$ 
large enough such that the asymptotic regime is reached, 
which causes the instability seen in Fig.~\ref{fig:Sinfoldmethod}.

\subsection{Improved method}
\label{sec:schrsolutionimp}

To avoid the matrix inversion (\ref{eq:Uinversion}) we seek a method 
where the inverse of $U$ defined in (\ref{eq:Umatrix}) follows 
directly from the solution of a differential equation system. This 
can indeed be found by an adaptation of the reformulation of the 
Schr\"odinger problem described in \cite{Ershov:2011zz}.

We work with the dimensionless radial variable $x =  m_{\rm LSP} v r$ 
and dimensionless wave numbers $\hat{k}_a = k_a/(m_{\rm LSP} v) = 
[1+i\epsilon-(M_a-2 m_{\rm LSP})/E]^{1/2}$. 
The radial equation (\ref{eq:schroedingerpartial}) reads 
\begin{equation}
[u_L^{\prime\prime}(x)]_{ai} + 
\left[\left(1-\frac{L (L+1)}{x^2}\right) \delta^{ab}
- \frac{V^{ab}(x)}{E}\right] [u_L(x)]_{bi} =0\,.
\label{eq:schroedingerpartialx}
\end{equation}
We separate the asymptotic value of the potential by defining 
$V(x) = V_{\rm inf} + \hat{V}(x)$, such that $V_{\rm inf}$ contains 
the mass splittings, and is diagonal, while 
$\hat{V}(x)\stackrel{x\to\infty}{=}0$. The method proposed 
in~\cite{Ershov:2011zz} starts with the variable phase 
ansatz\footnote{We do not indicate the dependence on $L$ of the 
functions of the right-hand side of the ansatz. Furthermore, in 
subsequent equations we 
do not write explicitly when an index appearing twice is 
{\em not} summed, since this should be clear from the index structure 
of the equation. For instance, in the following equation 
as well as later ones, such as (\ref{eq:schroedingerfreex}), the 
left-hand side has indices $ai$, so $a$ should not be summed.}
\begin{equation}
[u_L(x)]_{ai} = f_a(x)\alpha_{ai}(x)-g_a(x)\beta_{ai}(x)
\qquad \mbox{(no sum over $a$).}
\end{equation}
Here 
\begin{equation}
f_a(x) = \sqrt{\frac{\pi x}{2}}\,J_{L+\frac{1}{2}}(\hat{k}_a x)
\qquad
g_a(x) = -\sqrt{\frac{\pi x}{2}}\,
\left[Y_{L+\frac{1}{2}}(\hat{k}_a x)
-i J_{L+\frac{1}{2}}(\hat{k}_a x)\right]
\label{eq:freesol}
\end{equation}
are linearly independent Bessel function solutions of the 
free Schr\"odinger equations
\begin{equation}
[u_L^{\prime\prime}(x)]_{ai} + 
\left[\hat{k}_a^2-\frac{L (L+1)}{x^2}\right] [u_L(x)]_{ai} =0
\label{eq:schroedingerfreex}
\end{equation}
The Wronskian of the free solutions is 
\begin{equation}
f^\prime_a(x) g_a(x) - f_a(x) g_a^\prime(x) = 1. 
\end{equation}
The ansatz doubles the set of unknown functions by introducing 
$\alpha_{ai}(x)$ and $\beta_{ai}(x)$ and 
the additional freedom can be eliminated by imposing 
the condition
\begin{equation}
f_a(x) \alpha_{ai}^\prime(x) - g_a(x) \beta_{ai}^\prime(x) = 0
\end{equation}
for every $ai$~\cite{Ershov:2011zz}. The condition reduces
the $N$ second-order
differential equations~(\ref{eq:schroedingerpartialx})
to a system of $2N$ coupled differential equations of first order
for $\alpha_{ai}(x)$ and $\beta_{ai}(x)$.

The important quantities in the present approach are the 
matrix-functions $N$, $\tilde\alpha$, defined by
\begin{equation}
N_{ab} = f_a g_a \delta_{ab} - g_a O_{ab} g_b
\end{equation}
with $O_{ab}$ defined through $\beta_{ai} = O_{ab}\,\alpha_{bi}$, 
and 
\begin{equation}
\tilde{\alpha}_{ai} = \frac{\alpha_{ai}}{g_a}\,.
\end{equation} 
From the original Schr\"odinger equation for $[u_L]_{ai}$ and 
the above definitions, one derives that $N$ and $\tilde\alpha$ 
satisfy the first-order differential equations 
\begin{eqnarray}
N^\prime_{ab} &=& \delta_{ab} + \left(\frac{g_a^\prime}{g_a}
+\frac{g_b^\prime}{g_b} \right) N_{ab} - 
N_{ac}\,\frac{\hat{V}_{cd}}{E}\,N_{db}\,,
\label{eq:diffN}
\\
\tilde{\alpha}^\prime_{ai} &=& 
Z_{ab} \tilde{\alpha}_{bi} 
\qquad\mbox{with}\qquad 
Z_{ab}\equiv -\frac{g_a^\prime}{g_a}\,\delta_{ab}+ 
\frac{\hat{V}_{ac}}{E}\,N_{cb}\,.
\label{eq:diffalphatilde}
\end{eqnarray}
The initial conditions (\ref{eq:initial}) translate into
\begin{equation}
N_{ab}(x_0) = \frac{x_0}{2 L+1}\,\delta_{ab},
\qquad
\tilde{\alpha}_{ai}(x_0) =  x_0^L\,\delta_{ai}\,. 
\label{eq:initial2}
\end{equation}
where $x_0$ is a number close to zero, which we set to $10^{-7}$. 
Up to this point we essentially reviewed the modification of the 
variable phase method proposed in \cite{Ershov:2011zz}.

The crucial observation is now that
\begin{equation} 
[u_L^\prime]_{ai} = \tilde{\alpha}_{ai} +\frac{g_a^\prime}{g_a}\,  
[u_L]_{ai}\,,
\end{equation}
which implies that the matrix $U_{ai}$ from (\ref{eq:Umatrix}) 
is asymptotically trivially related to $\tilde{\alpha}_{ai}$, since  
\begin{eqnarray} 
U_{ai} &=& e^{i \hat{k}_a x}\,\left(
[u_L^\prime(x)]_{ai} - i \hat{k}_a [u_L(x)]_{ai}\right)
\nonumber\\
&=& e^{i \hat{k}_a x}\,\tilde{\alpha}_{ai} + 
e^{i \hat{k}_a x}\,\left(
\frac{g_a^\prime}{g_a} - i \hat{k}_a \right)[u_L(x)]_{ai}
\stackrel{\;x\to\infty\;}{=}
e^{i \hat{k}_a x}\,\tilde{\alpha}_{ai}
\label{eq:Ualpharelation}
\end{eqnarray} 
The last equality follows, because $g_a^\prime/g_a - i \hat{k}_a$ 
vanishes for large $x$. Specifically, for the relevant cases 
$L=0,1$, 
\begin{equation}
\frac{g_a^\prime}{g_a} - i \hat{k}_a = 
\left\{
\begin{array}{lcl} 
\quad & 0 & \qquad L=0\\[0.2cm]
&\displaystyle
-\frac{1}{x (1-i \hat{k}_a x)} & \qquad L=1
\end{array}
\right.
\end{equation} 

We recall that our task is to compute the inverse $U^{-1}$ for 
large enough $x$, such that the result is practically $x$-independent. 
This is now easily accomplished, since using 
$d(\tilde{\alpha}^{-1}\tilde{\alpha})/dx=0$,
(\ref{eq:diffalphatilde}) 
can be turned into the differential equation 
\begin{equation}
\tilde{\alpha}^{-1\,\prime}_{ia} = -\tilde{\alpha}^{-1}_{ib} Z_{ba} ,
\qquad
\tilde{\alpha}^{-1}_{ia}(x_0) =  x_0^{-L}\,\delta_{ia}\,. 
\label{eq:alphainv}
\end{equation}
with $Z$ given in (\ref{eq:diffalphatilde}),
and then 
\begin{equation}
T_{ia}(x_\infty) = [U^{-1}]_{ia}(x_\infty) = 
e^{-i \hat{k}_a x_\infty}\,\tilde{\alpha}^{-1}_{ia}(x_\infty)
\label{eq:T2}
\end{equation}
from the solution to (\ref{eq:alphainv}) 
for sufficiently large $x_\infty$. 

To summarize, the matrix $T$ and hence the Sommerfeld factors 
(\ref{eq:SFwithT}) are computed by the following three steps:
\begin{itemize}
\item[(1)] Solve for $N_{ab}$ the first-order differential equations 
(\ref{eq:diffN}) with initial conditions (\ref{eq:initial2}) for 
every $b=1,\ldots, N$.
\item[(2)] Solve (\ref{eq:alphainv}) with $Z$ 
given by (\ref{eq:diffalphatilde}).
\item[(3)] Evaluate (\ref{eq:T2}) for several $x_\infty$ and 
check (by varying and increasing $x_\infty$) that $T$ is 
independent of  $x_\infty$ within a certain target accuracy.
\end{itemize}

\begin{figure}[t]
\begin{center}
\includegraphics[width=0.6\textwidth]{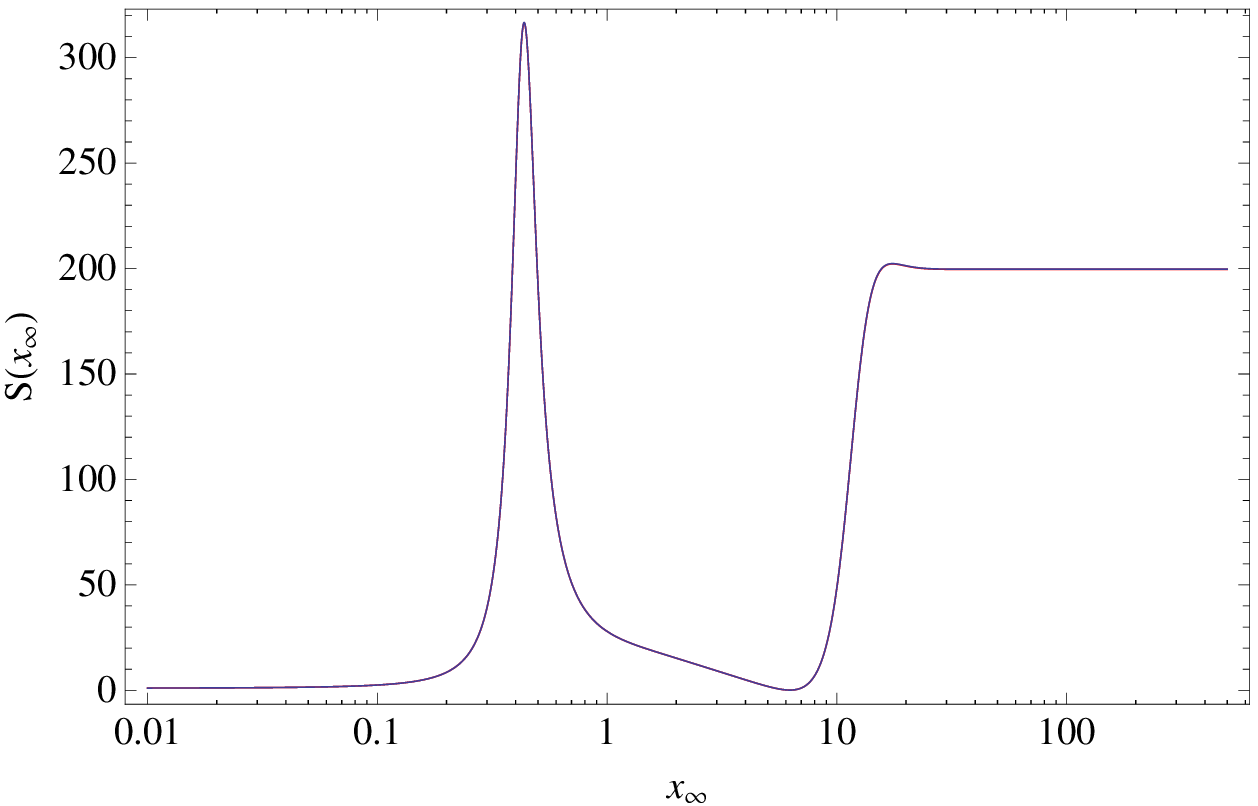}
\includegraphics[width=0.6\textwidth]{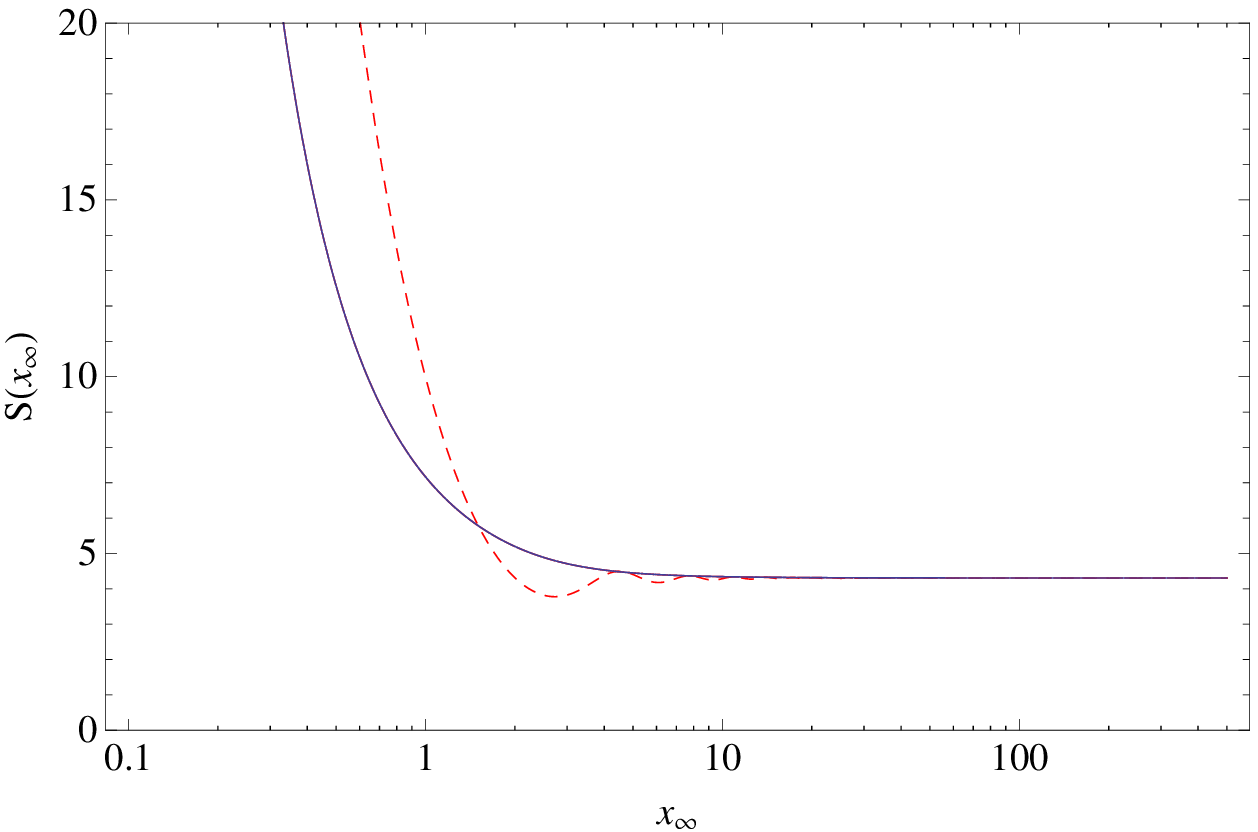}
\caption{$S$- and $P$-wave Sommerfeld factors 
$S\equiv S_{\chi_1^0\chi_1^0}[\hat f(^{1}S_0)]$ (upper plot) 
and $S_{\chi_1^0\chi_1^0}[\hat f(^{3}P_J)]$ (lower plot)
for $v=0.012$ in a wino-like model as described at the end of  
Sec.~\ref{sec:schrsolution} using the improved solution method.
The light-grey (red) curve shows $S(x_\infty)$, when only 
the $\chi_1^0\chi_1^0$ and $\chi_1^+\chi_1^-$ two-particle 
states in the Schr\"odinger equation and the annihilation 
process are kept. The dark-grey (blue) curve shows the result 
when the $\chi_1^0\chi_2^0$ state is included. In both plots 
the two cases are now indistinguishable. The lower plot 
shows in addition (light-grey (red) dashed) the $P$-wave 
Sommerfeld factor with the ``old'' 
method for the two-state problem.}
\label{fig:Sinfnewmethod}
\end{center}
\end{figure}

To illustrate the improved performance of the method developed in 
this section, we show 
in the upper plot of Fig.~\ref{fig:Sinfnewmethod} the same result 
as in Fig.~\ref{fig:Sinfoldmethod}, that is, the $S$-wave 
Sommerfeld factor $S_{\chi_1^0\chi_1^0}[\hat f(^{1}S_0)]$ in 
the wino-LSP model described at the end of  
Sec.~\ref{sec:schrsolution}, but now using the new method. 
The solution for the three-state case that was impossible to obtain with 
the standard method can now be evolved to sufficiently large 
$x_\infty$, such that the Sommerfeld factor can be extracted 
without difficulty. In fact, the two curves referring 
to the two-state solution and the one including the heavier 
$\chi_1^0\chi_2^0$ state cannot be distinguished on the scale 
of the plot. As expected, the heavier state has no effect on 
the Sommerfeld enhancement in this particular model -- the enhancement 
factor changes only by a tiny amount 
from $S(\infty)\approx 199.59$ to $S(\infty)\approx 199.72$. 

%
%

The lower plot of Fig.~\ref{fig:Sinfnewmethod} shows the 
corresponding results for a $P$-wave Sommerfeld factor, here for 
$S_{\chi_1^0\chi_1^0}[\hat f(^{3}P_J)]$. Once again the 
two-state solution (light-grey/red curve [invisible]) and the one 
including the heavier $\chi_1^0\chi_2^0$ state (dark grey/blue curve) 
cannot be distinguished on the scale of the plot. The plateau 
is reached for $x_\infty > 50$ with $S(\infty)\approx 4.31$. 
We also show the two-state solution with the standard method 
(light-grey/red dashed curve). All results agree for large 
enough $x_\infty$, but contrary to the $S$-wave case there is a 
visible difference for $x_\infty <10$. The difference  arises 
from the approximation in the  
second line of (\ref{eq:Ualpharelation}), which implies that 
the $U$ matrix in the standard and new method agree only 
for large $x_\infty$ in general. The ``approximation'' is exact 
only for $L=0$. For the $P$-wave case the difference vanishes 
as $1/x_\infty^2$. Since there is no numerical restriction on 
$x_\infty$ with the new method, the difference can always be 
made sufficiently small as seen in the lower plot. Note that 
it would not be possible to compute the $P$-wave Sommerfeld 
factor with the standard method for the three-state problem, 
since the matrix inversion becomes unstable at $x_\infty \approx 2$, 
as in the $S$-wave case.

\subsection{Second-derivative operators}
\label{sec:derivops}

The above method allows us to calculate the matrix elements of the 
$S$- and $P$-wave operators 
${\cal O}^{\chi \chi \to \chi \chi }_{ \lbrace e_4 e_3\rbrace 
\lbrace e_2 e_1\rbrace }$ and the associated 
operators ${\cal Q}^{\chi \chi \to \chi \chi }_{ \lbrace e_4 e_3\rbrace 
\lbrace e_2 e_1\rbrace }$ proportional to mass differences. In order 
to fully describe the Sommerfeld enhancement of the short-distance 
annihilation cross section at order $v^2$, we further require the 
matrix element of the second-derivative operators 
${\cal P}^{\chi \chi \to \chi \chi }_{ \lbrace e_4 e_3\rbrace 
\lbrace e_2 e_1\rbrace }$. For spin-0 it is defined 
as~\cite{Hellmann:2013jxa} 
\begin{equation}
{\cal P}^{\chi \chi \to \chi \chi }_{ \lbrace e_4 e_3\rbrace 
\lbrace e_2 e_1\rbrace }(^{1}S_0) = 
\frac{1}{2}\left[ \chi^\dagger_{e_4} \chi^c_{e_3} \  
\chi^{c \dagger}_{e_2} \big(-\frac{i}{2}
\overleftrightarrow{\bff{\partial}}\big)^2 \chi^{}_{e_1}
+  \chi^\dagger_{e_4} \big(-\frac{i}{2}
\overleftrightarrow{\bff{\partial}}\big)^2 \chi^c_{e_3}
\ \chi^{c \dagger}_{e_2} \chi^{}_{e_1} \right],
\end{equation}
with obvious generalization to spin-1. We show in this section 
that the matrix element of 
${\cal P}^{\chi \chi \to \chi \chi }_{ \lbrace e_4 e_3\rbrace 
\lbrace e_2 e_1\rbrace }$ can be obtained from the non-derivative 
$S$-wave operators by the relations (\ref{eq:p2Swaveeom}), 
(\ref{eq:kappa}). To this end, we show that 
\begin{equation}
\langle 0| \chi^{c \dagger}_{e_2}\Gamma
 \big(-\frac{i}{2}
\overleftrightarrow{\bff{\partial}}\big)^2\chi^{}_{e_1} |\chi_i\chi_j\rangle
= \kappa_{e e^\prime}^* \,
\langle 0| \chi^{c \dagger}_{e_2^\prime}\Gamma\chi^{}_{e_1^\prime} |
\chi_i\chi_j\rangle\,, 
\label{eq:merel}
\end{equation}
where in compound index notation (\ref{eq:kappa}) reads 
\begin{eqnarray}
\kappa_{\, e e^\prime} = \vec{p}^{\,2}_{e}\, \delta_{e e^\prime}
+ 2 \, \mu_{e}\alpha_2\, \sum_a    m_{\phi_a}  
c^{(a)}_{e e^\prime}\,. 
\label{eq:kappa2}
\end{eqnarray}

We start from (\ref{eq:meladdersum}) and (\ref{eq:sfpsi}), which for the 
specific case at hand imply
\begin{eqnarray}
\langle 0| \chi^{c \dagger}_{e_2}\Gamma 
 \big(-\frac{i}{2}
\overleftrightarrow{\bff{\partial}}\big)^2\chi^{}_{e_1}
| \chi_i \chi_j \rangle = 
\langle \xi^{c \dagger}_{j} \Gamma \xi_{i} \rangle 
\int\frac{d^3\vec{q}}{(2\pi)^3} \,\vec{q}^{\;2}
\Big( \,[\tilde{\psi}_E]^*_{ei}(\vec{q}\,) + 
(-1)^S\, [\tilde{\psi}_E]^*_{\bar{e}i}(\vec{q}\,)\Big)\,.\qquad
\label{eq:meladdersumd2}
\end{eqnarray}
The momentum-space Schr\"odinger equation~(\ref{eq:schroedinger1}) 
leads to 
\begin{eqnarray}
\int\frac{d^3\vec{q}}{(2\pi)^3} \,\vec{q}^{\;2}\,
[\tilde{\psi}_E]^*_{ei}(\vec{q}\,) &=&
2 \mu_e (2 m_{\rm LSP}-M_e+E)\int\frac{d^3\vec{q}}{(2\pi)^3} \,
[\tilde{\psi}_E]^*_{ei}(\vec{q}\,)
\nonumber\\   
&& \hspace*{-2cm}
-\, 2 \mu_e  \int\frac{d^3\vec{q}}{(2\pi)^3}\frac{d^3\vec{k}}{(2\pi)^3}
\,[\hat{V}^{e e^\prime}(\vec{k}\,)]^*\,[\tilde{\psi}_E]^*_{e^\prime i}(\vec{q}
-\vec{k}) 
\label{eq:lippmannpsi}
\end{eqnarray}
The momentum-space potential is a sum of terms from the exchange of 
gauge and Higgs bosons, 
\begin{equation}
\hat{V}^{ee^\prime}(\vec{k}\,) = 4\pi \alpha_2 \sum_a 
\frac{c^{(a)}_{ee^\prime}}{\vec{k}^{2}+m_{\phi_a}^2}\,,
\label{eq:yukawasum}
\end{equation}
where $m_{\phi}$ is the mass of the exchanged gauge boson, 
$\alpha_2=g_2^2/(4\pi)$ the SU(2) gauge coupling and $c^{(a)}_{ee^\prime}$ 
the coefficients of the potentials given
in Tab.~\ref{tab:potentials} of Appendix~\ref{sec:appendixpot}. 
Shifting $\vec{q}\to \vec{q}+\vec{k}$ in (\ref{eq:lippmannpsi}) 
factorizes the two integrations. Employing dimensional regularization 
gives the finite result
\begin{equation}
\int \frac{d^{d-1}\vec{k}}{(2\pi)^{d-1}} \frac{1}{\vec{k}^{2}+m_{\phi}^2} 
\stackrel{d\to 4}{=} -\frac{m_\phi}{4\pi}
\label{eq:intlinear}
\end{equation}
for the linearly divergent integral. Inserting this into 
(\ref{eq:meladdersumd2}), we obtain 
\begin{eqnarray}
&& 
\langle 0| \chi^{c \dagger}_{e_2}
 \big(-\frac{i}{2}
\overleftrightarrow{\bff{\partial}}\big)^2\chi^{}_{e_1}
| \chi_i \chi_j \rangle = 
\langle \xi^{c \dagger}_{j} \Gamma \xi_{i} \rangle \, 
\int\frac{d^3\vec{q}}{(2\pi)^3} \,
[\tilde{\psi}_E]^*_{e^\prime i}(\vec{q}\,)
\nonumber\\
&&\hspace*{1cm}
\times\,\left[2 \mu_e (2 m_{\rm LSP}-M_e+E) 
\delta_{e^\prime e}+ 2 \mu_e \alpha_2 
\sum_a m_{\phi_a} c^{(a)*}_{ee^\prime}\right]
+ (-1)^S \, \{ e \to \bar{e} \}
\nonumber\\
&&\hspace*{1cm}
= \langle \xi^{c \dagger}_{j} \Gamma \xi_{i} \rangle \, 
\int\frac{d^3\vec{q}}{(2\pi)^3} \,
[\tilde{\psi}_E]^*_{e^\prime i}(\vec{q}) \, \Big( \kappa_{e e^\prime}^* + (-1)^S\,\kappa_{\bar{e}e^\prime}^* \Big)
\, , 
\label{eq:d2me}
\end{eqnarray}
where the last equality follows from the definition~(\ref{eq:kappa2}) of 
the matrix $\kappa$ and the expression for the relative momentum 
of the two-particle state $e$, $\vec{p}_{e}^{\,2}=
2\mu_e(2 m_{\rm LSP}-M_{e}+E)+{\cal O}(\vec{p}_{e}^{\,4})$.
We can rename the dummy index $e^\prime \to \bar{e}^\prime$ in the 
second term of~(\ref{eq:d2me}),
and use that $\kappa_{\bar{e}\bar{e}^\prime}=\kappa_{e e^\prime}$ (since the potentials for
$\chi_{e_1}\chi_{e_2} \to \chi_{e_1^\prime}\chi_{e_2^\prime}$ and  
$\chi_{e_2}\chi_{e_1} \to \chi_{e_2^\prime}\chi_{e_1^\prime}$ involve the same coupling structure),
to write~(\ref{eq:d2me}) as
\begin{eqnarray}
\langle 0| \chi^{c \dagger}_{e_2}
 \big(-\frac{i}{2}
\overleftrightarrow{\bff{\partial}}\big)^2\chi^{}_{e_1}
| \chi_i \chi_j \rangle =  \kappa_{e e^\prime}^* \,  \langle \xi^{c \dagger}_{j} \Gamma \xi_{i} \rangle \, 
\!\!\int\!\! \frac{d^3\vec{q}}{(2\pi)^3} \,
\Big( [\tilde{\psi}_E]^*_{e^\prime i}(\vec{q}\,) + (-1)^S\,
      [\tilde{\psi}_E]^*_{\bar{e}^\prime i}(\vec{q}\,) \Big)
\,, \quad
\label{eq:d2me2}
\end{eqnarray}
which proves (\ref{eq:merel}). Note that the second term in the 
definition~(\ref{eq:kappa2}) of $\kappa$ is only present for the exchange of 
massive particles, and not for the Coulomb potential.
The factor $\kappa$ can be absorbed into an 
effective annihilation matrix $\hat g_{\kappa}$ as has been 
done in (\ref{eq:kappa}).

In parameter-space regions where the Sommerfeld effect is important 
and the exchange of a particle with mass $m_{\phi} \ll m_{\rm LSP}$ 
leads to an enhancement of the radiative correction, all terms 
in $\kappa$ are parametrically of the 
same order, since $M_e-m_{\rm LSP} \sim E \sim m_{\rm LSP} v^2$, 
and $\alpha_2m_\phi\sim v m_{\rm EW} \sim m_{\rm LSP} v^2$. A general 
MSSM parameter point, however, may also lead to heavy Higgs bosons, which 
though strongly suppressed in the potential, may give a large 
contribution to the last term in square brackets. The origin of 
this unphysical power-counting breaking contribution is the 
linearly divergent integral (\ref{eq:intlinear}). Since heavy Higgs-boson 
exchange causes a suppressed, local $(\chi^\dagger\chi)^2$ 
potential interaction, the simplest solution is to decouple the heavy 
Higgs bosons by not including them into the MSSM potential of 
Section~\ref{sec:potMSSM}. 
After this decoupling, the generated local interaction is a 
${\cal O}(v^2)$ correction to the long-distance part, which we 
consistently neglect. 
In practice, we simply eliminate Higgs-exchange 
from the potential for Higgs masses $m_H/m_{\rm LSP} > 0.5$ 
unless $m_H < 100\,$GeV.

\subsection{Approximate treatment of heavy channels}
\label{sec:heavychanapprox}

For a generic MSSM parameter space point the two-particle states 
will span a certain mass range and not be degenerate. While 
the improved method for the Schr\"odinger-equation solution 
allows us to cover the non-degenerate case without numerical 
instabilities there are still practical limitations related 
to the increasing CPU time, as the number of states, and hence 
the dimensionality of the matrix increases. For example, if for 
a given MSSM model and scattering energy $E$, the computation 
of the $^1S_0$ Sommerfeld factor in the charge-neutral channel 
for a velocity close to the threshold of a nearly degenerate 
state takes $0.1\,$s when only two out of the total of 14 channels 
are included in the computation, the CPU time increases to 
$14\,$s for four channels, $5\,$min for 8 channels and reaches 
nearly three hours for the solution of the full $14\times 14$ 
matrix problem.\footnote{The numbers strongly depend on the model and 
velocity but provide an indication of the increase in CPU time as 
the number of channels increases.} 
Since even for a single MSSM parameter set, 
many Sommerfeld factors must be computed for a single 
scattering energy, and further the thermal average requires an 
integral over the scattering energy, CPU considerations restrict
the number of channels, for which the Schr\"odinger equation is 
solved exactly. 

However, an exact solution of the $14\times 14$ problem is only 
necessary if all 14 states are nearly degenerate, which is rarely 
the case. As soon as the mass splitting becomes large, the heavier 
two-particle states should have little effect on the Sommerfeld 
enhancement of the lighter states, since the propagator of the 
heavier state in one of the loops of the ladder diagram series is 
off-shell and does not lead to an enhancement. Yet there is the 
possibility that a heavy state couples more strongly to the 
annihilation process than the lighter states and hence effectively 
enhances the annihilation rate. To cover this case, we allow the 
heavy channels to appear in the last loop before the annihilation 
vertex, but not elsewhere in the ladder diagrams. 
Moreover, the non-relativistic power-counting shows that
there is a suppression factor of order $[E/(M_h-2m_{\rm LSP})]^a$
when a light channel is replaced by a heavy one,
with  $a=1/2$ for the contribution  in the last loop before the annihilation,
but $a=3/2$ when the heavy state appears inside the ladder, 
which justifies the omission. In the following, 
we describe how the effect of the heavy channels in the last 
loop can be absorbed into an effective annihilation matrix for the 
lighter channels.

We divide the total number $N$ of two-particle states 
(14 in method-2 in the charge-neutral 
sector) into $n$ light states, which will be treated exactly, 
and $N-n$ heavy states, which we include only in the last 
loop.\footnote{We also include them as external states, but 
only at tree-level. Heavy external states are not relevant for 
the relic-density calculations, since they are Boltzmann-suppressed, 
except for special cases such as resonant annihilation of the 
heavy state, in which the resonant enhancement of the 
cross section compensates the Boltzmann suppression.} 
Suppose that $h=\{h_1 h_2\}$ refers to one of the heavy states, then
by extension of (\ref{eq:meladdersum}) we obtain 
\begin{eqnarray}
&& \langle 0| \chi^{c \dagger}_{h_2}\Gamma 
K\big[-\frac{i}{2}\overleftrightarrow{\bff{\partial}}\big]\chi^{}_{h_1}
| \chi_i \chi_j \rangle = 
\langle \xi^{c \dagger}_{j} \Gamma \xi_{i} \rangle 
\,\lim_{\hat E\to E} (-1)\left(\hat E-\frac{\vec{p}^{\,2}}{2\mu_{ij}}\right)
\nonumber\\
&&
\times\, 
\int\frac{d^3\vec{q}}{(2\pi)^3} \int\frac{d^3\vec{k}}{(2\pi)^3}
\,K[\vec{q}\,]\,
\hat{V}^{lh}(\vec{q}-\vec{k}) \,
\frac{1}{E-[M_h-2 m_{\rm LSP}]-\frac{\vec{q}^{\,2}}{2\mu_{h}}}\,
\tilde{G}^{il}(\vec{p},\vec{k};\hat E)
\nonumber\\
&&\hspace*{0.0cm}
+\, (-1)^{L+S}\,\{h\to \bar{h} \}
\,,
\label{eq:meladdersumheavy}
\end{eqnarray}
where $l$ refers to the light channels summed over, and 
$\tilde{G}^{il}$ is the Green function of the Sch\"odinger 
operator for the $n\times n$ problem for the light-states. 
The potential interaction 
$\hat{V}^{lh}$ converts the light channel into the annihilating 
heavy channel $h$.

The annihilation cross sections matrix elements such as above 
are multiplied by annihilation matrices 
$\hat f_{e e^\prime}\equiv \hat f^{ \chi \chi \to \chi \chi }_
{ \lbrace e_1 e_2\rbrace \lbrace e_4 e_3\rbrace }$ and summed 
over all two-particle channels $e$, see 
(\ref{eq:notation}). Dividing the sum over $e$ into the sum over light and 
the sum over heavy channels and proceeding for the heavy channels 
from (\ref{eq:meladdersumheavy}) to the integral involving the 
wave-function similar to (\ref{eq:sfpsi}), we find
\begin{eqnarray}
&& 
\hat f_{e e^\prime}\, \langle 0| \chi^{c \dagger}_{e_2}\Gamma 
K\big[-\frac{i}{2}\overleftrightarrow{\bff{\partial}}\big]\chi^{}_{e_1}
| \chi_i \chi_j \rangle = 
2 \,\langle \xi^{c \dagger}_{j} \Gamma \xi_{i} \rangle 
\int\frac{d^3\vec{q}}{(2\pi)^3}\,[\tilde{\psi}_E]^*_{li}(\vec{q}\,)
\nonumber\\
&&\hspace*{0.5cm}
\times \left[\hat f_{l e^\prime} K[\vec{q}\,]+\hat f_{h e^\prime}
\int\frac{d^3\vec{k}}{(2\pi)^3} K[\vec{k}\,]\,
\hat{V}^{lh}(\vec{k}-\vec{q}) \,
\frac{1}{E-[M_h-2 m_{\rm LSP}]-\frac{\vec{k}^{\,2}}{2\mu_{h}}}
\right],
\qquad
\label{eq:sfpsiheavy}
\end{eqnarray}
where the sum over $l$ ($h$) runs only over the light (heavy) channels,
and the equality in~(\ref{eq:sfpsiheavy}) holds when heavy channels 
contribute only in the last loop before the annihilation of the perturbative 
expansion of the matrix element. 

Our aim is to write the square bracket as an effective annihilation 
matrix $\hat f_{l e^\prime}^{\rm eff} K[\vec{q}\,]$ for the light channels.
We write the potential as in (\ref{eq:yukawasum}) and perform the 
integration over loop momentum $\vec{k}\,$ to obtain 
\begin{eqnarray}
\hat f_{l l^\prime}^{\rm eff, \,right} = \hat f_{l l^\prime} + 
I_{lh} \,\hat f_{h l^\prime}\,,
\label{eq:effectivematrix}
\end{eqnarray}
with 
\begin{eqnarray}
I_{lh} =\alpha_2 \sum_a 
c^{(a)}_{lh} \,(-2\mu_{h})\,I_L(2\mu_{h} 
(M_h-[2 m_{\rm LSP}+E]-i\epsilon),m_{\phi_a},\vec{q}^{\;2})\,.
\label{eq:effectivematrixI}
\end{eqnarray}
for $S$-wave matrix elements ($L=0$, $K[\vec{q}\,]=1$) 
\begin{eqnarray}
I_0(y,m,\vec{q}^{\;2}) &=& \frac{i}{2\sqrt{\vec{q}^{\;2}}}\,
\ln\frac{i (m^2-y+ \vec{q}^{\;2})+2 m \sqrt{\vec{q}^{\;2}}}
{i (m^2-y- \vec{q}^{\;2})+2 \sqrt{y\,\vec{q}^{\;2}}}\,,
\label{eq:I0}
\end{eqnarray}
and for the $P$-wave case ($L=1$, $K[\vec{q}\,]=\vec{q}\,$)
\begin{eqnarray}
I_1(y,m,\vec{q}^{\;2}) &=& 
\frac{\sqrt{y}-m}{2\vec{q}^{\;2}} + 
\frac{i (m^2-y+\vec{q}^{\;2})}{2(\vec{q}^{\;2})^{3/2}}\,
\ln\frac{i (m^2-y+ \vec{q}^{\;2})+2 m \sqrt{\vec{q}^{\;2}}}
{i (m^2-y- \vec{q}^{\;2})+2  \sqrt{y\,\vec{q}^{\;2}}}\,.
\label{eqI1}
\end{eqnarray}
The superscript ``right'' reminds us that (\ref{eq:meladdersumheavy}) 
represents only ``one half'' of the four-fermion operator that 
accounts for the square of the annihilation amplitude. Accounting for 
the fermion bilinear matrix element that multiplies the annihilation 
matrix from the left then gives 
\begin{eqnarray}
\hat f_{l l^\prime}^{\rm eff} = \hat f_{l l^\prime} 
+ I_{lh} \,\hat f_{h l^\prime}
+ I_{l^\prime h}^* \,\hat f_{l h} 
+ I_{lh} I_{l^\prime h^\prime}^*\,\hat f_{h h^\prime}
\label{eq:effectivematrixfull}
\end{eqnarray}
for the effective annihilation matrix in the space of light states 
including the heavy states in the last loop of the ladder.

We wish to note that the result~(\ref{eq:effectivematrix}) 
taking channel index $h$ to run over all relevant two-particle  
states regardless of their mass provides the leading-order
non-relativistic approximation to the 
1-loop annihilation amplitude of the state $l$, which has also been obtained 
by direct expansion
in~\cite{Drees:2013er}. It can be checked that the loop integrals $I_{S,P}$ 
defined in the latter reference, are equal to  
$(2\pi |\vec{q}\,|)I_{0,1}$.\footnote{The non-relativistic
expansion performed in~\cite{Drees:2013er} keeps in addition terms  
proportional to mass differences between the incoming $\chi$'s
and the virtual $\chi$'s inside the loop
that originate from the numerator of the full 1-loop amplitude. 
In our approach,
the latter can correspond to corrections of ${\cal O}(v^2)$ to the long-distance 
part, which we do not consider, as well as to 
the short-distance annihilation, where we have kept all $(m_i-m_{\rm LSP})/m_{\rm LSP}\sim {\cal O}(v^2)$ terms
in the Wilson coefficients of the annihilation operators. We note that similar
$(m_i-m_{\rm LSP})/m_{\rm LSP}$
terms arise as well from other sources, for instance
from the anti-particle 
pole contribution in the $q^0$-integration of the full amplitude, which 
have been neglected in~\cite{Drees:2013er} and also in our approach,
since they are a class of ${\cal O}(v^2)$ long-distance corrections.}

The expressions for $I_L$ above 
are not yet useful, since they define functions 
of $\vec{q}^{\;2}$, hence the evaluation of (\ref{eq:sfpsiheavy}) 
would require the full momentum-space wave-function 
$[\tilde{\psi}_E]_{li}(\vec{q}\,)$, which is not available, 
and not just the value and derivative at $r=0$ in coordinate space. 
However, we would like to make use of these expressions only for 
heavy channels when the mass splitting $M_h-2 m_{\rm LSP}$ is 
large compared to the scattering energy $E$. Since the typical 
relative momentum scales as $\vec{q}^{\;2} \sim m_{\rm LSP} E$, 
we can expand the integrals $I_L$ in $\sqrt{\vec{q}^{\;2}}/
(M_h-2 m_{\rm LSP})$. Keeping only the leading term in this 
expansion, we approximate
\begin{eqnarray}
&& I_0(y,m,\vec{q}^{\;2}) \to \frac{1}{\sqrt{y}+m} 
\,\qquad 
I_1(y,m,\vec{q}^{\;2}) \to \frac{2 \sqrt{y}+m}{3 (\sqrt{y}+m)^2}\,.
\label{eq:I1approx}
\end{eqnarray}
Inserting this into (\ref{eq:effectivematrixI}) gives
\begin{eqnarray}
I_{lh|L=0} &=& -2\mu_{h} \alpha_2 \sum_a 
\frac{c^{(a)}_{lh}}{\sqrt{y_h}+m_{\phi_a}}
\\
I_{lh|L=1} &=& 
-2\mu_{h} \alpha_2 \sum_a c^{(a)}_{lh} \,
\frac{2 \sqrt{y_h}+m_{\phi_a}}{3(\sqrt{y_h}+m_{\phi_a})^2}
\end{eqnarray}
with $y_h= 2\mu_{h} 
(M_h-[2 m_{\rm LSP}+E]-i\epsilon)$, from which the 
effective annihilation matrix (\ref{eq:effectivematrixfull}) follows.

The case of the second-derivative $S$-wave operator is somewhat more 
complicated, since one has to apply the equation-of-motion relation 
discussed in the previous section to the factor $K[\vec{k}\,]=\vec{k}^{\,2}$ 
in (\ref{eq:sfpsiheavy}). This is done by writing 
\begin{equation} 
\vec{k}^{\,2} = 2\mu_{h} \left(
2 m_{\rm LSP}+E - M_h - \left[2 m_{\rm LSP}+E - M_h - 
\frac{\vec{k}^{\,2}}{2\mu_{h}}\right] \right)
\end{equation}
which leads to the same factor as in (\ref{eq:d2me}).
The result is that the effective annihilation matrix for the 
second-derivative operators reads
\begin{eqnarray}
\hat{g}_{\kappa,ll^\prime}^{\rm eff} &=& \hat{g}_{\kappa,ll^\prime} 
+ \frac{1}{2}\left[I_{lh,D^2} \,\frac{\hat g_{h l^\prime}}{M^2}
+ I_{l^\prime h,D^2}^* \,\frac{\hat g_{l h}}{M^2} 
+ I_{lh|L=0} I_{l^\prime h^\prime,D^2}^*\,\frac{\hat g_{h h^\prime}}{M^2}
+ I_{lh,D^2} I_{l^\prime h^\prime|L=0}^*\,\frac{\hat g_{h h^\prime}}{M^2}\right]
\,,
\nonumber\\
\label{eq:gkappaeff}
\end{eqnarray}
where  
\begin{eqnarray}
I_{lh,D^2} =  2\mu_{h} \alpha_2 \left(
(2 m_{\rm LSP}+E - M_h) (-2\mu_{h}) 
\sum_a \frac{c^{(a)}_{lh}}{\sqrt{y_h}+m_{\phi_a}} + 
\sum_a m_{\phi_a}\,c^{(a)}_{lh}\right)\,.
\end{eqnarray}
The factor $M$ in each term in~(\ref{eq:gkappaeff}) is built from the 
particle species that enter the two-particle states specified by the 
indices of the accompanying $g$ matrix, as defined in~(\ref{eq:M}). 
For instance $\hat g_{lh}/M^2\equiv \hat g_{lh}/
[(m_{l_1}+m_{l_2}+m_{h_1}+m_{h_2})^2/4]$. 

\begin{figure}[t]
\begin{center}
\includegraphics[width=0.6\textwidth]{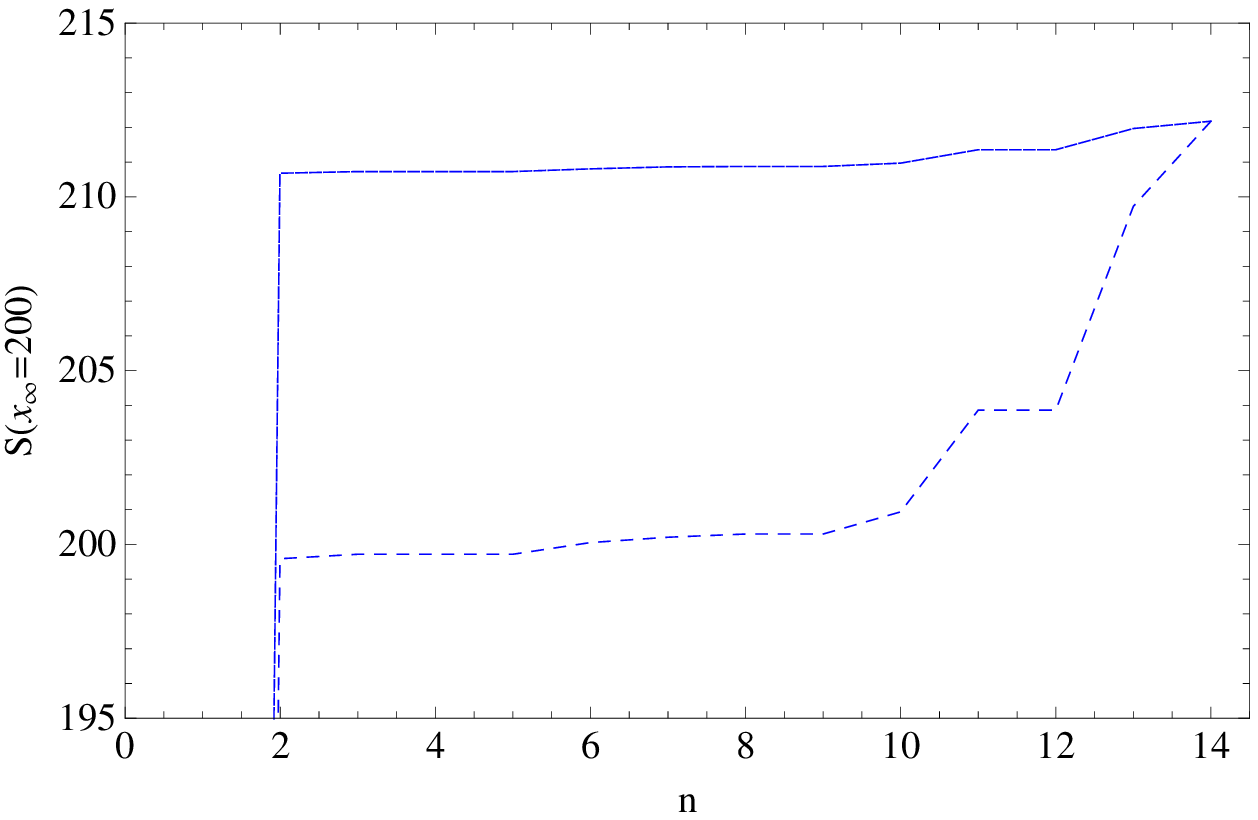}
\vskip0.4cm
\includegraphics[width=0.6\textwidth]{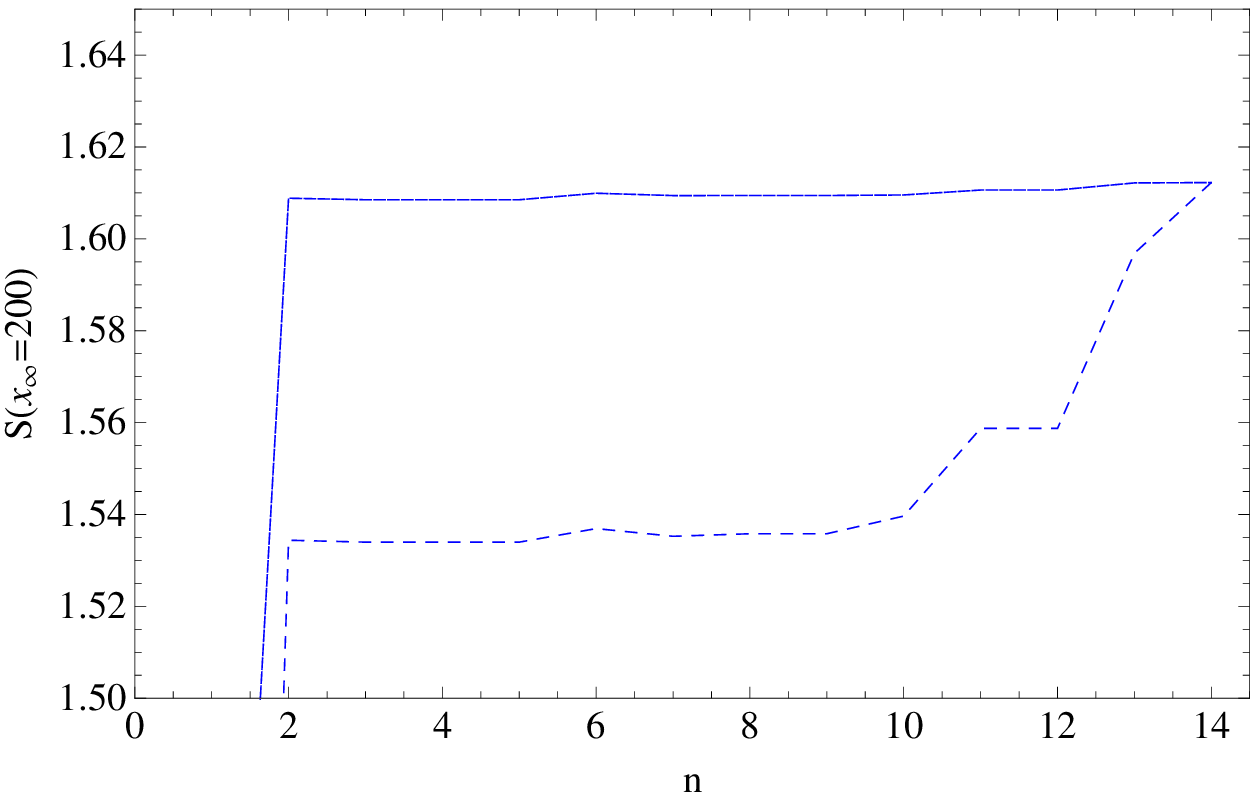}
\caption{Dependence of the Sommerfeld factor 
$S\equiv S_{\chi_1^0\chi_1^0}[\hat f(^{1}S_0)]$ in the 
charge-neutral sector on the number of 
states included in the computation. As $n$ increases from 1 to 14, 
the dashed curve shows $S(x_\infty=200)$, when only the $n$ 
lightest states are 
included and the remaining ones are ignored. 
The solid curve shows the result, when the $n$ 
lightest states are included in the solution of the 
Schr\"odinger equation, and the remaining $14-n$ heavier 
channels are treated approximately by including them 
in the last loop before the annihilation. 
We show the result for two velocity points, 
$v = 0.012$ (upper plot), slightly below the threshold for the 
nearly degenerate $\chi_1^+\chi_1^-$ state, and $v=0.15$ 
(lower plot).}
\label{fig:heavychannelsapprox}
\end{center}
\end{figure}

To illustrate the quality of approximation to the heavy channels 
described in this section, we employ the same wino-LSP model as 
discussed at the end of Section~\ref{sec:schrsolution}. The spectrum 
of neutralino masses is $m_{\chi^0} = \{2749.4, 2950.25, 3062.98, 
3083.54\}\,$GeV, the chargino masses are $m_{\chi^+} = 
\{2749.61, 3074.26\}\,$GeV. 
We consider the charge-neutral two-particle state sector with 
method 2, in which case there are 14 two-particle states of 
increasing mass. We study the approach to the full treatment 
of the $14\times 14$ matrix problem by plotting in 
Fig.~\ref{fig:heavychannelsapprox} the 
dependence of the Sommerfeld factor 
$S\equiv S_{\chi_1^0\chi_1^0}[\hat f(^{1}S_0)]$ of the lightest 
$\chi_1^0\chi_1^0$ state in the $S$-wave, spin-0 configuration 
on the number of 
states included in the computation. As $n$ increases from 1 to 14, 
the dashed curve shows 
$S(x_\infty=200)$, when only the $n$ lightest states are 
included and the remaining ones are ignored. 
The solid curve shows the result, when the $n$ 
lightest states are included in the solution of the 
Schr\"odinger equation, and the remaining $14-n$ heavier 
channels are treated approximately by including them 
in the last loop before the annihilation. For the purpose of 
this comparison we keep the $\delta m_{e_4 e_1} 
\delta m_{e_4 e_1}/M_{Z,W}^2$ terms in the potential for all 
14 two-particle states, not only for those $n$ treated 
exactly, such that the potential is independent of the number of 
states $n$ treated exactly.\footnote{This is explains why $S(x_\infty=200)$ 
for $v=0.012$ shown in the upper plot approaches 
the value 212.18 rather than the previously quoted value 
$S(\infty)\approx 199.59$, which corresponds to $n=2$ on 
the dashed curve. If, one the other hand the mass-difference 
terms in the potential are neglected entirely, we obtain 
$S(x_\infty=200) \approx 206.06$,
hence their effect is small. 
This is expected, since the Sommerfeld enhancement is 
mainly generated by the two lowest, nearly degenerate states, 
for which the mass difference terms are completely negligible. 
The rise of $S$ in Fig.~\ref{fig:heavychannelsapprox} when adding 
states 11, 13 and 14, which suggests that these states are 
comparatively more important than the lower ones (except the first 
and second), should 
be taken with a grain of salt, since it originates from the 
$(\delta m/M_{\rm EW})^2$ terms of the potential, which here are 
included for all states.}

%
%
We show the result for two velocity points, 
$v = 0.012$ (upper plot), slightly below the threshold for the 
nearly degenerate $\chi_1^+\chi_1^-$ state, and $v=0.15$ 
(lower plot). The figure shows that $n=1$ is always a 
poor approximation. This says that it is necessary 
to treat the nearly degenerate $\chi_1^+\chi_1^-$ state 
exactly, as is of course expected, since the bulk of the 
enhancement comes from the mixing of the  
$\chi_1^0\chi_1^0$ state with the charged state in models 
where the LSP is almost pure wino. Once $n>1$, the 
dependence of the Sommerfeld enhancement on the remaining 
states is no longer large, resulting in only another 5\% increase. 
We observe that the results that include the heavy channels 
approximately have a smoother approach to the full result 
attained for $n=14$ than the ones that leave out the heavier 
channels completely. 
This demonstrates that it is a good approximation to 
include the heavy channels only in last loop before the annihilation 
process, but to neglect them in 
the Schr\"odinger equation. In the two cases shown, 
it is essentially sufficient to treat only two states exactly 
and all other states approximately, as the solid curve for 
$n=2$ is already very close to the $n=14$ value.

While this shows that the one-loop approximation for heavy 
states is often a good approximation to the full, resummed result, 
it does not prove that either of the two, the one-loop or full 
resummed treatment of very heavy channels in the non-relativistic 
approximation is a reasonable approximation to the true 
loop corrections, since the non-relativistic approximation 
in the potential region is not expected to be a good 
approximation to the exact loop integral for very heavy 
channels. Since such states have little influence on the 
Sommerfeld factor anyway, in practice, we simply include 
them by the approximate procedure described in this section.

\section{Summary}
\label{sec:summary}

With this paper we conclude the presentation of the non-relativistic MSSM 
effective theory formalism, designed to calculate the enhanced 
radiative corrections to pair-annihilation of neutralinos and charginos 
at small relative velocity. The construction of the NRMSSM is similar 
to the non-relativistic EFT for heavy quarkonium annihilation. 
Characteristic features of the NRMSSM are the presence of more 
than one heavy particle species; massive mediators exchanged among the 
heavy particles, which generate electroweak Yukawa (rather than 
colour Coulomb) potentials; and weak couplings that allow one to 
compute the long-distance matrix elements; these contains the so-called 
Sommerfeld enhancements. The framework developed here applies to 
general MSSM scenarios, where the LSP
is an arbitrary admixture of the electroweak gauginos, and accounts for the
co-annihilations with all the neutralino and charginos, which are nearly
mass degenerate with the LSP. 

Papers I~\cite{Beneke:2012tg} and II~\cite{Hellmann:2013jxa} 
focused on the short-distance part of the annihilation process,
providing analytic results for the Wilson coefficients that
reproduce the tree-level annihilation rates of neutralinos and charginos 
including up to $P$- and next-to-next-to-leading order $S$-wave 
effects, {\it i.e.} up corrections of ${\cal O}(v_\text{rel}^2)$. 
The present paper describes all the necessary 
ingredients to account for the long-range interactions that produce 
the Sommerfeld enhancements in the annihilation rates
at the leading order in the non-relativistic power-counting, which 
treats $\alpha_2\sim v_\text{rel}\sim M_{Z,W}/m_{\rm LSP}$.
The calculated enhancements thus sum up large 
${\cal O}(\alpha_2/v_\text{rel})^n$, $n=1,2,\dots$, quantum corrections 
to the tree-level annihilation rates to all orders. 
The main results of this this work are:
\begin{itemize}
 \item A general formula for the Sommerfeld-corrected annihilation 
cross-section for a given channel $\chi_i\chi_j$ 
that includes ${\cal O}(v_\text{rel}^2)$ corrections in the short-distance part
has been written down in~(\ref{eq:SFenhancedsigma}). The formula is a 
generalization of the tree-level
approximation $\sigma_{\chi_{i} \chi_{j} } \, v_\text{rel} = 
a + b \, v_\text{rel} ^2$. Since the Sommerfeld
factors depend on the partial-wave configuration of the annihilating state,
$\sigma^{\chi_{i} \chi_{j} } \, v_\text{rel}$
has been decomposed in~(\ref{eq:SFenhancedsigma}) as a sum over partial waves, 
where the annihilation coefficients for each partial wave were computed
in papers I and II.
 \item Compact analytic results for the potential interactions among the 
chargino and neutralino pairs for a general MSSM parameter point, which 
feed into the multi-channel Schr\"odinger equation that 
has to be solved to obtain  the Sommerfeld correction factors. Transitions 
among different states are accounted
for by the off-diagonal entries in the potential.
 \item A novel method to solve the multi-channel Schr\"odinger equation. 
The standard method does provide an accurate result for the Sommerfeld 
factors when kinematically closed channels are
included due to numerical instabilities caused by the exponential growth of 
some of the solutions. The new method solves this problem by reformulating 
the Schr\"odinger equation for the quantity relevant for the 
Sommerfeld factor instead of solving for the scattering wave functions. 
This important improvement yields accurate numerical solutions when 
many non-degenerate two-particle channels are included, up to CPU time 
limitations. 
 \item An approximate treatment of heavy channels, which can includes them 
in an effective annihilation matrix for the Schr\"odinger equation of 
a lower-dimensional problem. We demonstrated that including heavy channels 
only in the last loop before annihilation provides a good approximation to 
the full resummed result,  and thus reduces the number of channels to be 
treated exactly.
\end{itemize}

While this paper, as well as papers I and II, have been dedicated to the 
technical details describing the computation of the Sommerfeld-enhanced 
annihilation cross sections, the investigation of the effect 
on the relic abundance in well-motivated MSSM scenarios with heavy 
neutralino LSP is the subject of an accompanying 
publication~\cite{paperIV}, which further illustrates the general use 
and potential of the framework.

\subsubsection*{Acknowledgements}
The work of M.B. was supported in part by the DFG
Sonder\-for\-schungs\-bereich/Trans\-regio~9 ``Computergest\"utzte
Theoreti\-sche Teilchenphysik''. 
C.H. thanks the ``Deutsche Telekom Stiftung'' for its support.
This work is supported in part by the Gottfried Wilhelm Leibniz programme of the
Deutsche Forschungsgemeinschaft (DFG) and the DFG
Sonder\-for\-schungs\-bereich/Trans\-regio~9 ``Computergest\"utzte
Theoreti\-sche Teilchenphysik''.
The work of P.~R. 
has been supported in part by the Spanish
Government and ERDF funds from the EU Commission
[Grants No. FPA2011-23778, No. CSD2007-00042
(Consolider Project CPAN)] and by Generalitat
Valenciana under Grant No. PROMETEOII/2013/007.
P.~R. also thanks
the ``Excellence Cluster Universe'' at TU Munich 
for its hospitality and support during completion of this work.
Feynman diagrams have been drawn with the packages 
{\sc Axodraw}~\cite{Vermaseren:1994je} and 
{\sc Jaxo\-draw}~\cite{Binosi:2008ig}.

\appendix

\section{Explicit expressions for the MSSM potential}
\label{sec:appendixpot}

\renewcommand{\arraystretch}{1.8}
 \setlength{\tabcolsep}{0.27em}
\begin{table}[tbp]
\begin{tabular}{| c | c c c c c |}
 \hline
&
$ \frac{\alpha_{2}\,e^{- M_{Z} \, r} }{r} $
&
$ \frac{\alpha_{2}\,e^{- M_{W} \, r} }{r} $
&
$ \frac{\alpha_{2}}{r} $
&
$ \frac{\alpha_{2}\,e^{- m_\phi \, r} }{r} $
&
$ \frac{\alpha_{2}\,e^{- M_{H^+} \, r} }{r} $
  \\
 \hline\hline 
 $\chi^0_{e_1}\chi^0_{e_2} \to \chi^0_{e_4}\chi^0_{e_3}$
&
$
\begin{array}{c}
\lambda^{Z} v^{(0), Z}_{e_4 e_1}  v^{(0), Z}_{e_3 e_2} \\[-2mm]
+  \lambda_S\, a^{(0), Z}_{e_4 e_1}  a^{(0), Z}_{e_3 e_2}
\end{array} 
$
&
0
&
0
&
$-s_{e_4 e_1}^{(0),\phi} \, s_{ e_3 e_2}^{(0),\phi}$
&
0
  \\
 \hline
$\chi^+_{e_1}\chi^-_{e_2} \to \chi^+_{e_4}\chi^-_{e_3}$
&
$
\begin{array}{c}
-\lambda^{Z} v^{Z}_{e_4 e_1}  v^{Z}_{e_3 e_2} \\[-2mm]
+ \lambda_S\, a^{Z}_{e_4 e_1}  a^{Z}_{e_3 e_2} 
\end{array}
$
&
0
&
$-v^{\gamma}_{e_4 e_1}  v^{\gamma}_{e_2 e_3}$
&
$-s_{e_4 e_1}^\phi \, s_{e_2 e_3}^{\phi}$
&
0
  \\
 \hline
 $\chi^0_{e_1}\chi^0_{e_2} \to \chi^+_{e_4}\chi^-_{e_3}$
&
0
&
$
\begin{array}{c}
- \lambda^{W} v^{W}_{e_4 e_1}  v^{W*}_{e_3 e_2} \\[-2mm]
+  \lambda_S\, a^{W}_{e_4 e_1}  a^{W^*}_{e_3 e_2}
\end{array} 
$
&
0
&
0
&
$-s_{e_4 e_1}^{H^+} \, s_{e_3 e_2}^{H^+*}$
 \\
  \hline
 $\chi^0_{e_1}\chi^0_{e_2} \to \chi^-_{e_4}\chi^+_{e_3}$
&
0
&
$
\begin{array}{c}
- \lambda^{W} v^{W*}_{e_4 e_1}  v^{W}_{e_3 e_2} \\[-2mm]
+  \lambda_S\, a^{W*}_{e_4 e_1}  a^{W}_{e_3 e_2}
\end{array} 
$
&
0
&
0
&
$-s_{e_4 e_1}^{H^+*} \, s_{e_3 e_2}^{H^+}$
 \\
  \hline
 $\chi^0_{e_1}\chi^+_{e_2} \to \chi^0_{e_4}\chi^+_{e_3}$
&
$
\begin{array}{c}
 \lambda^{Z} v^{(0), Z}_{e_4 e_1}  v^{Z}_{e_3 e_2} \\[-2mm]
+ \lambda_S\, a^{(0), Z}_{e_4 e_1}  a^{Z}_{e_3 e_2}
\end{array} 
$
&
0
&
0
&
$-s_{e_4 e_1}^{(0),\phi} \, s_{ e_3 e_2}^{\phi}$
&
0
  \\
 \hline
 $\chi^0_{e_1}\chi^-_{e_2} \to \chi^0_{e_4}\chi^-_{e_3}$
&
$
\begin{array}{c}
- \lambda^{Z} v^{(0), Z}_{e_4 e_1}  v^{Z}_{e_2 e_3} \\[-2mm]
+  \lambda_S\, a^{(0), Z}_{e_4 e_1}  a^{Z}_{e_2 e_3}
\end{array} 
$
&
0
&
0
&
$-s_{e_4 e_1}^{(0),\phi} \, s_{ e_2 e_3}^{\phi}$
&
0
  \\
 \hline
 $\chi^0_{e_1}\chi^+_{e_2} \to \chi^+_{e_4}\chi^0_{e_3}$
&
0
&
$
\begin{array}{c}
 \lambda^{W} v^{W}_{e_4 e_1}  v^{W*}_{e_2 e_3} \\[-2mm]
+  \lambda_S\, a^{W}_{e_4 e_1}  a^{W*}_{e_2 e_3}
\end{array} 
$
&
0
&
0
&
$-s_{e_4 e_1}^{H^+} \, s_{e_2 e_3}^{H^+*}$
 \\
  \hline
 $\chi^0_{e_1}\chi^-_{e_2} \to \chi^-_{e_4}\chi^0_{e_3}$
&
0
&
$
\begin{array}{c}
\lambda^{W} v^{W*}_{e_4 e_1}  v^{W}_{e_2 e_3} \\[-2mm]
+ \lambda_S\, a^{W*}_{e_4 e_1}  a^{W}_{e_2 e_3}
\end{array} 
$
&
0
&
0
&
$-s_{e_4 e_1}^{H^+*} \, s_{e_2 e_3}^{H^+}$
 \\
  \hline
 $\chi^+_{e_1}\chi^+_{e_2} \to \chi^+_{e_4}\chi^+_{e_3}$
&
$
\begin{array}{c}
\lambda^{Z} v^{Z}_{e_4 e_1}  v^{Z}_{e_3 e_2} \\[-2mm]
+  \lambda_S\, a^{ Z}_{e_4 e_1}  a^{ Z}_{e_3 e_2}
\end{array} 
$
&
0
&
$v^{\gamma}_{e_4 e_1}  v^{\gamma}_{e_3 e_2}$
&
$-s_{e_4 e_1}^{\phi} \, s_{ e_3 e_2}^{\phi}$
&
0
  \\
 \hline
 $\chi^-_{e_1}\chi^-_{e_2} \to \chi^-_{e_4}\chi^-_{e_3}$
&
$
\begin{array}{c}
\lambda^{Z} v^{ Z}_{e_1 e_4}  v^{Z}_{e_2 e_3} \\[-2mm]
+  \lambda_S\, a^{Z}_{e_1 e_4}  a^{Z}_{e_2 e_3}
\end{array} 
$
&
0
&
$v^{\gamma}_{e_1 e_4}  v^{\gamma}_{e_3 e_2}$
&
$-s_{e_1 e_4}^{\phi} \, s_{ e_2 e_3}^{\phi}$
&
0
  \\
 \hline
\end{tabular}
\caption{Potentials describing the non-relativistic interactions
among chargino and neutralino pairs in the MSSM. The potential from
neutral scalar exchange $\phi$ has to be summed over the ``physical''
neutral Higgs bosons, $\phi=H^0,h^0,A^0$. The expressions obtained from
the table correspond to the potential entries in method-1.
The potentials for the channels not shown are
obtained trivially by interchanging indices 
(like $\chi^-\chi^+\to \chi^-\chi^+$ or  $\chi^+\chi^0\to \chi^+\chi^0$) or 
are vanishing (like $\chi^-\chi^+\to \chi^+\chi^-$). We have introduced $\lambda_S\equiv (3-4S)$
and
$\lambda^{Z/W}= 1+\delta m_{e_4e_1}\delta m_{e_3 e_2}/ M_{Z/W}^2$. }
\label{tab:potentials}
\end{table}

The leading-order non-relativistic potential interactions among MSSM 
neutralino and chargino two-particle states 
are given in this appendix. Tab.~\ref{tab:potentials} provides the coefficients
of the Yukawa terms $e^{-m_{X_i} r}/r$ generated by the exchange of boson 
$X_i\,(X_i=Z,W^\pm,\gamma,H^0_m,A^0_1, H^\pm)$ for 
each process. The complete potential is obtained as the sum of the coefficients multiplied
by the corresponding potential terms. The contribution from the pseudoscalar Goldstone
boson ($A^0_2$) does not have to be considered, since it cancels
against a piece of the $Z$-exchange potential, as explained in Sec.~\ref{sec:potMSSM}. 
A similar cancellation occurs between the charged pseudo-Goldstone bosons $G^\pm = H_2^\pm$ 
and part of the $W^\pm$-exchange potential.
The coefficients have been written in terms of the coupling factors as defined in~(\ref{eq:gaugecoup},\ref{eq:higgscoup}), whose explicit
expressions are given next.

The (axial-) vector and \mbox{(pseudo-)} scalar couplings, that encode 
chargino interactions 
with the neutral scalar and pseudo-scalar Higgs particles ($H^0_m,A^0_m$),
the $Z$-boson or the photon read
\begin{align}
\nonumber
 s^{H^0_m}_{ij} \, ( p^{H^0_m}_{ij} ) \ = \
&
            - \frac{1}{2 \sqrt 2} \Bigl[
                   \, Z_R^{1m} \Bigl( \widetilde Z_-^{2j *} \widetilde Z_+^{1i *} \pm \widetilde Z_-^{2i} \widetilde Z_+^{1j} \Bigr)
                    + Z_R^{2m} \Bigl( \widetilde Z_-^{1j *} \widetilde Z_+^{2i *} \pm \widetilde Z_-^{1i} \widetilde Z_+^{2j} \Bigr)
                               \, \Bigr] \ ,
\\\nonumber
 s^{A^0_m}_{ij} \, ( p^{A^0_m}_{ij} ) \ = \
&
            - \frac{i}{2 \sqrt 2} \Bigl[
                   \, Z_H^{1m} \Bigl( \widetilde Z_-^{2j *} \widetilde Z_+^{1i *} \mp  \widetilde Z_-^{2i} \widetilde Z_+^{1j} \Bigr)
                    + Z_H^{2m} \Bigl( \widetilde Z_-^{1j *} \widetilde Z_+^{2i *} \mp  \widetilde Z_-^{1i} \widetilde Z_+^{2j} \Bigr)
                              \,   \Bigr] \ ,
\\\nonumber
 v^Z_{ij} \ = \
&
            - \frac{1}{4 c_W} \Bigl(
                    \widetilde Z_-^{1i} \widetilde Z_-^{1j *}
                  + \widetilde Z_+^{1i *} \widetilde Z_+^{1j} 
                  + 2  ( c_W^2 - s_W^2 ) \delta_{ij}
                        \Bigr) \ ,
\\\nonumber
 a^Z_{ij} \ = \
&
            \phantom{-} \frac{1}{4 c_W} \Bigl(
                    \widetilde Z_+^{1i *} \widetilde Z_+^{1j} 
                  - \widetilde Z_-^{1i} \widetilde Z_-^{1j *}
                        \Bigr) \ ,
\\\nonumber
 v^\gamma_{ij} \ = \
&
           - s_W \delta_{ij} \ ,
\\
 a^\gamma_{ij} \ = \
&
           \phantom{-} 0 \ ,
\label{eq:app_couplings_ccX}
\end{align}
where $H_1^0 \equiv H^0$, $H_2^0 \equiv h^0$ and $A_1^0\equiv A^0$, $A_2^0\equiv G^0$.
We have used the abbreviations $s_W=\sin \theta_W$ and $c_W=\cos \theta_W$, with $\theta_W$ 
denoting the Weinberg angle.
The   coupling factors for the 
three-point interaction
of an incoming neutralino $\chi^0_j$, an outgoing chargino $\chi^+_i$ and
either an incoming charged Higgs particle $G^+$ or $H^+$ or an incoming
$W^+$-boson are given by
\begin{align}
\nonumber
 s^{H_m^+}_{ij} \, ( p^{H^+_m}_{ij}   ) \ = \
&
           - \frac{1}{2} \Bigl[
              \phantom{-} Z_H^{2m} \left(
                     \widetilde Z_N^{4j *} \widetilde Z_+^{1i *}
                   + \frac{1}{\sqrt 2} \widetilde Z_+^{2i *}
                     ( \widetilde Z_N^{2j *} + \tan \theta_W \widetilde Z_N^{1j *})
                                 \right)
\\\nonumber
&
         \phantom{- \frac{1}{2} \ \ \Bigl[}
                       \pm Z_H^{1m} \left(
                     \widetilde Z_N^{3j} \widetilde Z_-^{1i}
                   - \frac{1}{\sqrt 2} \widetilde Z_-^{2i}
                     ( \widetilde Z_N^{2j} + \tan \theta_W \widetilde Z_N^{1j})
                                 \right)
                        \ \Bigr] \ ,
\\\nonumber
 v^W_{ij} \ = \
&
 \phantom{-} \frac{1}{2} \left(
                      \widetilde Z_N^{2j *} \widetilde Z_-^{1i}
                   + \widetilde Z_N^{2j} \widetilde Z_+^{1i *}
                   + \frac{1}{\sqrt 2} \widetilde Z_N^{3j *} \widetilde Z_-^{2i}
                   - \frac{1}{\sqrt 2} \widetilde Z_N^{4j} \widetilde Z_+^{2i *}
                        \right) \ ,
\\
 a^W_{ij} \ = \
&
 \phantom{-} \frac{1}{2} \left(
                      \widetilde Z_N^{2j *} \widetilde Z_-^{1i}
                    - \widetilde Z_N^{2j} \widetilde Z_+^{1i *}
                    + \frac{1}{\sqrt 2} \widetilde Z_N^{3j *} \widetilde Z_-^{2i}
                    + \frac{1}{\sqrt 2} \widetilde Z_N^{4j} \widetilde Z_+^{2i *}
                        \right) \ ,
\label{eq:app_couplings_ncX}
\end{align}
where $H_1^+\equiv H^+$ and $H_2^+\equiv G^+$, and the mixing matrices are
defined as in Ref.~\cite{Rosiek:1989rs}.
Finally, three-point interactions of an incoming $\chi^0_j$ and an
outgoing $\chi^0_i$ with a (pseudo-) scalar Higgs particle or the $Z$-boson
involve the following (axial-) vector or (pseudo-) scalar coupling factors:
\begin{align}
\nonumber
 s^{(0) , H^{0}_m}_{ij} \, ( p^{(0), H^{0}_m}_{ij}   )  = \
&
            \frac{1}{4} \,\Bigl[ \,
                         \left(
                    Z_R^{2m} \, \widetilde Z_N^{4i *}
                   - Z_R^{1m} \, \widetilde Z_N^{3i *}
                         \right)
                         \left(
                  \widetilde Z_N^{2j *} - \tan\theta_W \, \widetilde Z_N^{1j *}
                         \right)
                       + \left( i \leftrightarrow j \right)\,
                    \Bigr]
                    \pm c.c.\,,
\\\nonumber
 s^{(0), A^{0}_m}_{ij}  (  p^{(0), A^{0}_m}_{ij}  )  = \
&
            \frac{i}{4} \,\Bigl[\,
                         \left(
                  \, Z_H^{2m} \, \widetilde Z_N^{4i *}
                   - Z_H^{1m} \, \widetilde Z_N^{3i *}
                         \right)
                         \left(
                  \widetilde Z_N^{2j *} - \tan\theta_W \, \widetilde Z_N^{1j *}
                          \right)
                       + \left( i \leftrightarrow j \right)\,
                    \Bigr]
                    \pm c.c. \, ,
\\
 v^{(0), Z}_{ij}  ( a^{(0), Z}_{ij})  = \
&
            \frac{1}{4 c_W} \Bigl(
                 \, \widetilde Z_N^{3i} \widetilde Z_N^{3j *} 
                  - \widetilde Z_N^{4i} \widetilde Z_N^{4j *}
                         \mp (i \leftrightarrow j) \
                        \Bigr) \ .
\label{eq:app_couplings_nnX}
\end{align}
The (axial-) vector and (pseudo-) scalar coupling factors in
(\ref{eq:app_couplings_ccX}) and (\ref{eq:app_couplings_nnX}), which are all
related to interactions with neutral SM and Higgs particles, satisfy
\begin{align}
 \nonumber
v^*_{ij} \ &=  \phantom{-} v_{ji} \ , 
&a^*_{ij} \ =  \phantom{-} a_{ji} \ ,
\\
s^*_{ij} \ &=  \phantom{-} s_{ji} \ ,
&p^*_{ij} \ =  - p_{ji} \ .
\end{align}
as a consequence of the hermiticity of the underlying SUSY Lagrangian.

\section{Equivalence between method-1 and method-2}
\label{sec:appendixmethodIvsII}

We wish to show in this appendix by means of an example, that using any of the two basis of
two-particle states introduced in Sec.~\ref{sec:potMSSM},
named as method-1 and method-2, gives the same outcome for the Sommerfeld enhancement factors.

Let us consider the simple case of a system which in method-1 consist of three states labelled
with $a=11,12,21$, corresponding
to $\chi_1\chi_1$, $\chi_1\chi_2$ and $\chi_2\chi_1$, respectively.
We can think of $\chi_1$ and $\chi_2$ as two different neutralinos
(or two different charginos), and we note that only state 11 is composed of identical particles. 
Conversely, the basis of method-2 is built 
from only two states, $(\chi\chi)_{11}$ and $(\chi\chi)_{12}$.  

The Schr\"odinger equation in method-1 is a $3\times 3$ matrix where the
generic potential matrix
$V_{ab}^{(1)}$ ($a,b=1,2,3$) 
reads
\begin{align}
V^{(1)}(r)   &= 
    \left(\begin{array}{ccc}
  V_{11} &  V_{12} &  V_{12}  \\
  V_{21} &  V_{22} &  V_{23}  \\
  V_{21} &  V_{23} &  V_{22} 
  \end{array}\right) \ ,
\label{eq:potI}
\end{align}
and we have used that the potential 
for scattering of state $11$ to either $12$ or $21$ is the same, since the
exchange of state $12$ by $21$, which corresponds to crossing the lines of $\chi_1$
and $\chi_2$ in the diagram, gives the same amplitude  due to the identical particle
nature of the state $11$. In addition we have taken into account that
$V_{32}=V_{23}$ (as well as $V_{33}=V_{22}$), since the amplitudes that produce these potential
terms only differ by a relabeling of the internal vertices of the associated diagram.
Note that hermiticity requires as well that $V_{ab}=V_{ba}^*$.
The Schr\"odinger equation~(\ref{eq:schroedingerpartial}) in method-1 when applied to a generic 
multi-component wave function $\Psi=(u_{1},u_{2},u_{3})$ reads
\begin{align}
& D \,u_{1} + V_{11} \, u_{1} + V_{12}\,  (u_{2}+u_{3}) = 0 \ ,
\nonumber\\
& D \,u_{2} + V_{21}\, u_{1} + V_{22} \, u_{2} + V_{23}\,  u_{3} = 0 \ ,
\nonumber\\
& D \,u_{3} + V_{21}\, u_{1} + V_{23} \, u_{2} + V_{22}\,  u_{3} = 0 \ .
\label{eq:sch3states}
\end{align}
$D$ stands for the differential operator multiplying $\delta^{ab}$ 
in~(\ref{eq:schroedingerpartial}), whose specific form is irrelevant for the 
current discussion, and we also omit writing explicitly the dependence of 
the quantities $u_a$ and $V_{ab}$ on the
independent variable $r$.
We have to seek for three independent
solutions of the system~(\ref{eq:sch3states}). Let us first choose two solutions 
of the form 
$\Psi_{i}=(\sqrt{2} \, u_{1i},u_{2i},u_{2i})$, with $i=1,2$. 
Then the set of equations~(\ref{eq:sch3states}) applied to $\Psi_{1,2}$
reduce to two independent ones 
(since eqs.~(\ref{eq:sch3states}) are symmetric in $u_2$ and $u_3$), and
acquire the form
\begin{align}
& D \,u_{1i} + V_{11} \, u_{1i} + \sqrt{2}\,V_{12}\, u_{2i} = 0 \ ,
\nonumber\\
& D \,u_{2i} + \sqrt{2}\,V_{21}\, u_{1i} + ( V_{22} + V_{23})\,  u_{2i} = 0 \ ,
\label{eq:sch_sols12}
\end{align}
for $i=1,2$.
A third solution to~(\ref{eq:sch3states}) is chosen as
$\Psi_{3}=[u_L]_{a3}=(0,u_{23},-u_{23})$, in which case only one equation remains:
\begin{align}
D \,u_{23} + ( V_{22} - V_{23})\,  u_{23} = 0 
\ .
\label{eq:sch_sol3}
\end{align}
In order to compute the Sommerfeld enhancement factors we need the
matrix $T$ and the annihilation matrix, see~(\ref{eq:SFwithT}). The entries
of the annihilation matrix in method-1 read
\begin{align}
[\hat{f}({}^{2S+1}L_J)]^{(1)}  &= 
    \left(\begin{array}{ccc}
  \frac{1+(-1)^{S+L}}{2}\,  \hat{f}_{11} &  \frac{1+(-1)^{S+L}}{2}\,  \hat{f}_{12} &  \frac{1+(-1)^{S+L}}{2}\,\hat{f}_{12}  \\
   \frac{1+(-1)^{S+L}}{2}\, \hat{f}_{21} &  \hat{f}_{22} &  (-1)^{S+L}\, \hat{f}_{22}  \\
 \frac{1+(-1)^{S+L}}{2}\, \hat{f}_{21} & \; (-1)^{S+L}\, \hat{f}_{22} \; &  \hat{f}_{22}
  \end{array}\right) \ ,
\label{eq:anmatI}
\end{align}
where we have used the symmetry relations under the exchange of the 
particle labels $e_1 \leftrightarrow e_2$ derived 
in~\cite{Beneke:2012tg,Hellmann:2013jxa}, namely 
$\hat{f}_{ \lbrace e_2 e_1\rbrace \lbrace e_4 e_3\rbrace }\left( {}^{2S+1}L_J \right)
= (-1)^{S+L} \  \hat{f}_{ \lbrace e_1  e_2\rbrace \lbrace e_4 e_3\rbrace }  \left( {}^{2S+1}L_J \right)
$ and likewise for the exchange $e_3 \leftrightarrow e_4$,
to write the annihilation matrix entries involving state $21$ in terms of
 those of $12$.
These same relations also tell us that the annihilation rates involving channel 11 vanish
for odd $L+S$, since such an state cannot be formed out of two identical spin-1/2 particles.
Note also that $\hat{f}_{21}=\hat{f}_{12}^*$.
On the other hand, the matrix $T$ is determined following~(\ref{eq:buildT}) from the matrices 
$[u_L^{(L+1)}(0)]_{ab}$ and  $M_{ab}$~(\ref{eq:Mdef}), which are both built
from the independent solutions of the Schr\"odinger equations. Given the solutions $\Psi_i$,
$i=1,2,3$, suggested above these matrices must have the structure
\begin{align}
[u_L^{(L+1)}(0)]  = 
    \left(\begin{array}{ccc}
  \sqrt{2}\, U_{11} & \sqrt{2}\, U_{12} &  0  \\
  U_{21} &  U_{22} &  U_{23}  \\
  U_{21} &  U_{22} &  -U_{23}
  \end{array}\right) \ , \quad
M  = 
    \left(\begin{array}{ccc}
  \sqrt{2}\, M_{11} & \sqrt{2}\, M_{12} &  0  \\
  M_{21} &  M_{22} &  M_{23}  \\
  M_{21} &  M_{22} &  -M_{23}
  \end{array}\right) \ .
\label{eq:u0andMI}
\end{align}
With the matrix $T$ built from~(\ref{eq:u0andMI}), we can readily compute the Sommerfeld factors for method-1 using~(\ref{eq:SFwithT}). For the case
$L+S={\rm even}$, and up to the global factor 
$[(2L+1)!!/(L+1)!\,k_a^{L+1}]^2$, we obtain 
\begin{align}
S_{11}^{\,{\rm even}}  \ = \ & \bigg| \frac{ M_{22}\, U_{11} -M_{21}\, U_{12} }{ M_{12}\,M_{21}-M_{11}\,M_{22}} \bigg|^2 
+ 2 \, \bigg| \frac{ M_{22}\, U_{21} -M_{21}\, U_{22} }{ M_{12}\,M_{21}-M_{11}\,M_{22}} \bigg|^2 
 \,\frac{\hat{f}_{22}}{\hat{f}_{11}}
 \nonumber\\[2mm]
&+2\sqrt{2}\,{\rm Re} 
\bigg[ \frac{ ( M_{22}\, U_{21} - M_{21}\, U_{22}) \, ( M_{22}\, U_{11} -M_{21}\, U_{12})^* }
                      { | M_{12}\, M_{21} - M_{11}\, M_{22} |^2 } \,\frac{\hat{f}_{12}}{\hat{f}_{11}}
\bigg] 
     \ ,
 \nonumber\\[2mm]
S_{12}^{\,{\rm even}}  \ = \ & \bigg| \frac{ M_{12}\, U_{21} -M_{11}\, U_{22} }{ M_{12}\,M_{21}-M_{11}\,M_{22}} \bigg|^2 
+ \frac{1}{2} \, \bigg| \frac{ M_{12}\, U_{11} -M_{11}\, U_{12} }{ M_{12}\,M_{21}-M_{11}\,M_{22}} \bigg|^2 
 \,\frac{\hat{f}_{11}}{\hat{f}_{22}}
 \nonumber\\[2mm]
&+\,\sqrt{2}\,{\rm Re} 
\bigg[ \frac{ ( M_{12}\, U_{21} - M_{11}\, U_{22}) \, ( M_{12}\, U_{11} -M_{11}\, U_{12})^* }
                      { | M_{12}\, M_{21} - M_{11}\, M_{22} |^2 } \,\frac{\hat{f}_{12}}{\hat{f}_{22}}
\bigg]
     \ ,
\label{eq:SfacevenI}
\end{align}
whereas for $L+S={\rm odd}$
\begin{align}
S_{11}^{\,{\rm odd}} \ = \ & 0
      \ ,
 \nonumber\\[2mm]
S_{12}^{\,{\rm odd}}  \ = \ & \bigg| \frac{  U_{23} }{ M_{23} } \bigg|^2 
     \ .
\label{eq:SfacoddI}
\end{align}
 The Sommerfeld factors for the state $21$ are found  to be equal to those of state $12$; this
 should be the case since both states are physically equivalent. We have used that 
 $\hat{f}_{21}=\hat{f}_{12}^*$ to write~(\ref{eq:SfacevenI}) in a more compact form.
 
 Let us now turn to method-2. As mentioned above in this  case the basis is given by
 the  two states $a=11,12$. Applying the rules described in Sec.~\ref{sec:eft} 
 (see~(\ref{eq:potmethods}) and the discussion following that equation), we can build the potential
 matrix entries corresponding to method-2 from those of method-1. Distinguishing 
 even and odd $S+L$, we find
\begin{align}
V^{(2),\,{\rm even}}(r)   = 
    \left(\begin{array}{ccc}
  V_{11} &  \sqrt{2} \, V_{12} \\
  \sqrt{2} \, V_{21} &  V_{22} + V_{23}   \end{array}\right) \ , \quad
  V^{(2),\,{\rm odd}}(r)   = 
    \left(\begin{array}{ccc}
 0 & 0 \\
  0 &  V_{22} - V_{23}   \end{array}\right) \ .
\label{eq:potII}
\end{align}
In particular, note that for odd $L+S$ the identical-particle state 11 does not exist, and
the entries of the potential matrix involving that state have to be set to zero, to 
avoid that a potential transition $11\to 12$ could give a non-vanishing 
annihilation amplitude for an incoming state $11$, as 
argued in Sec.~\ref{sec:eft}. In method-1 this is taken into account automatically 
and one gets the result $S_{11}^{\,\rm odd}=0$, as found in~(\ref{eq:SfacoddI}), at the expense
of using two different states (12 and 21) to describe the physical state built from $\chi_1$ and $\chi_2$.
Coming back to method-2, we then see that in the odd $L+S$ sector we deal with just a single 
annihilating state. We can therefore switch to a one-dimensional problem with  $V^{(2),\,{\rm odd}}(r)   = V_{22} - V_{23}$.

The Schr\"odinger equations which derive
from the potential  $V^{(2),\,{\rm even}}$ for the wave function 
$\widetilde{\Psi}=(\tilde{u}_1,\tilde{u}_2)$ of method-2 read
\begin{align}
& D \,\tilde{u}_{1} + V_{11} \, \tilde{u}_{1} + \sqrt{2}\,V_{12}\, \tilde{u}_{2} = 0 \ ,
\nonumber\\
& D \,\tilde{u}_{2} + \sqrt{2}\,V_{21}\, \tilde{u}_{1} + ( V_{22} + V_{23})\,  \tilde{u}_{2} = 0 \ .
\label{eq:sch_sols12met2}
\end{align}
Eqs.~(\ref{eq:sch_sols12met2}) agree with (\ref{eq:sch_sols12}) obtained 
previously for the components of the solutions $\Psi_{1,2}$ of method-1. 
Therefore we can choose $\widetilde{\Psi}_1=(u_{11},u_{21})$ and 
$\widetilde{\Psi}_2=(u_{12},u_{22})$ as linear independent solutions for 
method-2 in the case $L+S={\rm even}$. The matrices $[u_L^{(L+1)}(0)]$ and 
$M$ that derive from solutions $\widetilde{\Psi}_{1,2}$ follow immediately:
\begin{align}
[u_L^{(L+1)}(0)]^{\rm even}  = 
    \left(\begin{array}{cc}
  U_{11} &  U_{12}   \\
 U_{21} &  U_{22} 
  \end{array}\right) \ , \quad
M^{\rm even}  = 
    \left(\begin{array}{cc}
  M_{11} &  M_{12}   \\
  M_{21} &  M_{22} 
  \end{array}\right) \ .
\label{eq:u0andMIIeven}
\end{align}
For the case of odd $L+S$, there is  
just one equation,
\begin{align}
D \,\tilde{u}_{2} + ( V_{22} - V_{23})\, \tilde{u}_{2} = 0 
\ ,
\label{eq:sch_sol3met}
\end{align}
which is the same differential equation as~(\ref{eq:sch_sol3}), satisfied
by the component $u_{23}$ of $\Psi_{3}$. Hence,  
we can choose $\widetilde{\Psi}=u_{23}$ as solution for odd $L+S$. 
The (one-dimensional)
matrices $[u_L^{(L+1)}(0)]$ and $M$ for odd $L+S$ are simply given by
\begin{align}
[u_L^{(L+1)}(0)]^{\rm odd}  = 
  U_{23} \ , \quad
M^{\rm odd}  = 
   M_{23} \ .
\label{eq:u0andMIIodd}
\end{align}
Finally, we require the annihilation matrices in method-2 to compute 
the Sommerfeld factors. These are obtained 
from those of method-1 by adding a factor of $1/\sqrt{2}$ for each identical-particle
state involved in the entry, as explained in Sec.~\ref{sec:eft}:
\begin{align}
\hat{f}^{(2),\,{\rm even}}  = 
    \left(\begin{array}{ccc}
  \frac{1}{2}\,\hat{f}_{11} &   \frac{1}{\sqrt{2}}\, \hat{f}_{12} \\
  \frac{1}{\sqrt{2}}\,  \hat{f}_{21} &  \hat{f}_{22}  
  \end{array}\right) \ , \quad
\hat{f}^{(2),\,{\rm odd}}  =  \hat{f}_{22}
  \ .
  \label{eq:anmatII}
\end{align}
From the expressions for $[u_L^{(L+1)}(0)]$ and $M$ 
obtained for the even and the odd cases in~(\ref{eq:u0andMIIeven})
and~(\ref{eq:u0andMIIodd}), respectively, it is immediate to compute the 
corresponding $T$ matrices. The Sommerfeld factors for method-2 
then follow from~(\ref{eq:SFwithT}) by inserting the annihilation matrices~(\ref{eq:anmatII}). 
This straightforward computation
yields results for the Sommerfeld factor of states $11$ and $12$ for
even and odd $L+S$ that match the formulae~(\ref{eq:SfacevenI})
and~(\ref{eq:SfacoddI}) derived above using method-1.


\providecommand{\href}[2]{#2}\begingroup\raggedright\endgroup

\end{document}